\documentclass[AMA,STIX1COL]{WileyNJD-v2}

\usepackage{hyperref}
\usepackage{color}
\usepackage{mathtools}
\usepackage[mathscr]{eucal}
\usepackage{subfig}
\usepackage{epstopdf}
\usepackage{graphicx}
\usepackage{threeparttable}
\graphicspath{{./figures/}}
\usepackage{pgfplots}
\usepackage{tikz}
\usepackage{multirow}
\usepackage[separate-uncertainty]{siunitx}
\pgfplotsset{compat=1.13}
\usepackage{cases}
\usepgfplotslibrary{groupplots}
 
\newlength\figureheight
\newlength\figurewidth

\newcommand{\iema}[3]{{\overset{{#2}\rightarrow{#3}}{#1}}}
\newcommand{\tns}{\bs}
\newcommand{\bs}{\boldsymbol}
\newcommand{\bsd}[1]{\boldsymbol{ \mathsf{#1}}}
\newcommand{\grad}   { \boldsymbol{\nabla}    }
\renewcommand{\div} { \grad \! \cdot } 

\newcommand{\lbig}{\Lambda_{\mathrm{L}}}
\newcommand{\lbigdisc}{\Lambda_{{\mathrm{L}},h}}
\newcommand{\lsmall}{\Lambda_{\mathrm{S}}}
\newcommand{\sgn}{\mathop{\mathrm{sgn}}}

\newcommand{\porosity}{\varepsilon}
\DeclareMathOperator*{\argmax}{argmax}

\articletype{Research article - fundamental}%

\begin{document}

\title{Validation and parameter optimization of a hybrid embedded/homogenized solid tumor perfusion model}

\author[1]{Johannes Kremheller*}

\author[1]{Sebastian Brandstaeter}

\author[2,3]{Bernhard A. Schrefler}

\author[1]{Wolfgang A. Wall}

\authormark{KREMHELLER \textsc{et al}}

\address[1]{\orgdiv{Institute for Computational Mechanics}, \orgname{Technical University of Munich}, \orgaddress{Boltzmannstrasse 15, D-85748 Garching b. M\"unchen, Germany}}

\address[2]{\orgdiv{Institute for Advanced Study}, \orgname{Technical University of Munich}, \orgaddress{Lichtenbergstrasse 2a, D-85748 Garching b. M\"unchen, Germany}}

\address[3]{\orgdiv{Department of Civil, Environmental and Architectural Engineering}, \orgname{University of Padova}, \orgaddress{Italy}}

\corres{*Johannes Kremheller, Institute for Computational Mechanics, Technical University of Munich, Boltzmannstrasse 15, D-85748 Garching b. M\"unchen, Germany. \email{kremheller@lnm.mw.tum.de}}

\presentaddress{Institute for Computational Mechanics, Technical University of Munich, Boltzmannstrasse 15, D-85748 Garching b. M\"unchen, Germany}

\abstract[Summary]{The goal of this paper is to investigate the validity of a hybrid embedded/homogenized \textit{in-silico} approach for modeling perfusion through solid tumors. The rationale behind this novel idea is that only the larger blood vessels have to be explicitly resolved while the smaller scales of the vasculature are homogenized. As opposed to typical discrete or fully-resolved 1D-3D models, the required data can be obtained with \textit{in-vivo} imaging techniques since the morphology of the smaller vessels is not necessary. By contrast, the larger vessels, whose topology and structure is attainable non-in\-va\-sive\-ly, are resolved and embedded as one-dimension\-al inclusions into the three-dimensional tissue domain which is modeled as a porous medium. A sound mortar-type formulation is employed to couple the two representations of the vasculature. We validate the hybrid model and optimize its parameters by comparing its results to a corresponding fully-resolved model based on several well-defined metrics. These tests are performed on a complex data set of three different tumor types with heterogeneous vascular architectures. The correspondence of the hybrid model in terms of mean representative elementary volume blood and interstitial fluid pressures is excellent with relative errors of less than $\SI{4}{\percent}$. Larger, but less important and explicable errors are present in terms of blood flow in the smaller, homogenized vessels. We finally discuss and demonstrate how the hybrid model can be further improved to apply it for studies on tumor perfusion and the efficacy of drug delivery.
}


\jnlcitation{\cname{%
\author{J. Kremheller}, 
\author{S. Brandstaeter},
\author{B. A. Schrefler}, and 
\author{W. A. Wall}} (\cyear{2021}), 
\ctitle{Validation and parameter optimization of a hybrid embedded/homogenized solid tumor perfusion model}, \cjournal{Int J Numer Meth Biomed Engng}.}

\maketitle

\section{Introduction}
\label{sec:intro}
Mathematical modeling of blood flow and mass transport is of increasing importance to study a number of highly relevant biomedical questions in health and disease. Computational tools offer the possibility to gain new insight into physiologically relevant processes such as the transport of nutrients, oxygen or drugs across the vascular system and inside the tissue micro-environment. These methods can ultimately lead to a new rationale for developing and non-invasive testing of novel therapies.\cite{Dewhirst2017} Concurrently, such \textit{in-silico} models will make the design of drugs both cheaper and faster. 

In this paper, we are concerned with the simulation of blood flow and tissue perfusion at the scale of the microcirculation with a special focus on solid tumors where transport processes can be decisive for disease progression and treatment efficacy. This includes, first, the vasculature, which is embedded in the surrounding tissue, second, passage across the blood vessel walls into the surrounding extravascular space and, third, flow of the fluid filling this space, namely the interstitial fluid (IF). Subsequently, we will distinguish between three different modeling strategies for these transport processes, namely discrete, continuum and hybrid approaches. All strategies have been developed and implemented in the context of our vascular multiphase tumor growth model.\cite{Kremheller2018,Kremheller2019} For the discrete or fully-resolving variant, we follow a common modeling approach, where the vasculature is represented by a network of one-dimensional blood vessel segments embedded in the encompassing three-dimensional tissue domain which is modeled as a porous medium. A 1D partial differential equation (PDE) is employed to model mass transport in the vasculature while a corresponding 3D PDE governs the surrounding IF. Both domains are coupled via source terms which account for the exchange across the blood vessel wall. Such models are well-studied and first contributions include the so-called embedded multiscale method developed by D'Angelo and Quarteroni~\cite{DAngelo2007,DAngelo2008,DAngelo2012} and the Green's function method of Secomb et al.\cite{Secomb1994,Secomb2004} More recent approaches with such a philosophy include drug delivery\cite{Cattaneo2013,Cattaneo2014}, hyperthermia treatment\cite{Nabil2015,Nabil2016} and a combination of a numerical framework with optical imaging to predict fluid and species mass transport through whole tumors with heterogenous blood vessel architecture.\cite{Desposito2018,Sweeney2019} These approaches are commonly termed discrete models in distinction from continuum models which involve a homogenization procedure. Thereby, the vasculature is approximated as a homogeneous porous medium resulting in two distinct pore spaces which are the aforementioned interstitium and the homogenized vasculature. Flow in both domains is modeled via the Darcy equation and suitable exchange terms are defined.\cite{Kremheller2018,Chapman2008,Shipley2010,Tully2011,Penta2015,Penta2015a,Mascheroni2017b}

These two distinct approaches have different use cases: Discrete models can and should be applied when the entire structure of the vasculature including the smallest scales, i.e., the capillaries, is known and its resolution is needed for the question at hand. This is usually restricted to small domain sizes of an order of several $\mathrm{mm}^3$. By contrast, continuum models are used to simulate mass transport at larger scales, e.g., through whole organs. Both approaches have advantages and disadvantages: On the one hand, the computational cost of continuum models is usually smaller than for discrete ones which makes the application to larger domains possible in the first place. On the other hand, the information about the exact morphology of the vascular network is lost such that blood flow can only be described in an averaged sense. Discrete models, however, are computationally more expensive. Furthermore, they require the full structure of the part of the vasculature under consideration. This is usually realized via a graph whose edges are assigned the radius of the blood vessel segments between nodes. Such high-resolution data including blood vessel radii, connectivity and positions can at present only be acquired through \textit{ex-vivo} imaging.\cite{Shipley2019} In addition, the acquisition of high-quality data is still challenging and error-prone especially on the finest scales.\cite{Koeppl2020} By contrast, \textit{in-vivo} imaging is currently only possible for larger vessels and flow therein.\cite{Shipley2019,Li2020a} Therefore, discrete models rely on data which is not available via non-invasive imaging. An additional difficulty is the assignment of blood pressure or flow boundary conditions which can only be estimated for large networks.\cite{Sweeney2019,Fry2012} In any case, validation of these models is usually only performed on macroscopic quantities such as tissue perfusion\cite{Desposito2018} since measuring flow or pressures inside single micro-vessels is not possible.\cite{Sweeney2019}

This has motivated the development of hybrid methods which are especially suited for cases where the full vascular morphology is unknown or too large to be modeled with a discrete approach. The idea behind them is to explicitly resolve the larger vessels through a discrete model and to use a homogenized approach for the capillary bed. Next to our own work,\cite{Kremheller2019} such hybrid approaches have also been developed by Vidotto et al.\cite{Vidotto2018}, Shipley et al.\cite{Shipley2019} and Kojic et al.\cite{Kojic2017} Compared to pure homogenized formulations, their advantage is that the structure of the larger vessels is retained and, therefore, the heterogeneity of blood flow and pressure in the major vessel branches is better represented. Moreover, compared to discrete models, less anatomic data is needed since the morphology of the smallest vessels is not required. This could also have the additional advantage of a smaller computational cost and make them applicable to larger domains. Also quantities typically needed for validation such as tissue perfusion, blood flow or pressures at the resolution of current imaging techniques can equally be acquired from hybrid models. A related approach, where no homogenization of the capillaries is needed, is to generate a discrete surrogate network of the smaller scales based on the oxygen demand of the tissue.\cite{Koeppl2020} 

We have previously incorporated a hybrid method for coupling discrete, one-dimensional blood vessels with a homogenized representation of the vasculature\cite{Kremheller2019} into our vascular multiphase tumor growth model.\cite{Kremheller2018} Therein, we couple a discrete representation of the pre-existing vasculature with a homogenized representation of the neovasculature which is formed during angiogenesis. For that, we employ constraint enforcement strategies which are well-known from solid mechanics. In our previous paper, the main focus was on modeling vascular tumor growth. In this contribution, we validate the applicability of the hybrid embedded/homogenized approach for the study of perfusion through solid tumors. Here, accurate models of fluid mass transport are of high relevance since efficient drug delivery to cancerous tissue relies on the fact that the drug reaches a large fraction of the tumor cells. Therefore, physiological characteristics such as microvascular flow, the structure of the extracellular matrix or the IF pressure profile may influence the transport of drugs through tumor tissue and, hence, ultimately the success of treatment.\cite{Dewhirst2017} For instance, increased interstitial pressure due to highly permeable vessels and inefficient lymphatic drainage has been identified as an obstacle for successful drug delivery.\cite{Baxter1989,Jain1994,Heldin2004} Novel nanoparticle-based therapies aim to exploit these properties of the tumor vasculature for more specific targeting of tumor sites.\cite{Matsumura1986} Appropriate models of these transport phenomena can provide additional insight into and guidelines for drug design. In the context of cancer, this paradigm shift is described by the concept of transport oncophysics with the objective to engineer drugs with optimized transport properties.\cite{Ferrari2010,Nizzero2018}

Sound computational models are required to achieve this goal. We therefore took great care in the development of our hybrid model, i.e., both in the theoretical basis and its implementation. In this paper the main focus is on the validation of our hybrid embedded/homogenized scheme with three complex tumor-specific vascular networks based on large tissue samples containing more than $\num{100000}$ blood vessels.\cite{Desposito2018,Sweeney2019} We put a special emphasis on the extraction of the larger vessels from the fully-resolved network data such that it qualitatively matches the topology and distribution of larger vascular structures inside tumors available via \textit{in-vivo} imaging. Thus, we make sure that the hybrid approach is investigated for cases which closely resemble real-life scenarios where the structure of the considered part of the microcirculation is not entirely known. Here, the complete topology of the vasculature in the given tissue domain is available which allows us to generate reference solutions with a fully-resolved model and to quantify the error introduced by the homogenization in the hybrid model. We evaluate the error by means of several well-defined metrics involving the agreement of pressures and flow between the two models. Concurrently, the parameters of the hybrid model are identified such that the correspondence of the models is maximized. Evaluating the error of the hybrid model in comparison to a fully-resolved one is a first and indispensable step towards realistic hybrid models of tumor perfusion relying only on non-invasively available physiological data. For a full validation and parameter optimization, similar methods as applied herein need to be combined with advanced \textit{in-vivo} imaging techniques. The comparison of two purely numerical approaches and the inverse identification of the optimal parameters allows us to investigate the hybrid model in a controlled environment unaffected by any further influences such as uncertainties in experimental or clinical data.

The remainder of this work is structured as follows: We introduce both the hybrid and the fully-resolved model in Section~\ref{sec:models}. The employed tumor vasculature data sets as well as the setup of the models including the assignment of boundary conditions and the extraction of the hybrid model from the fully-resolved one are described in Section~\ref{sec:setup}. Numerical experiments to compare the accuracy of the hybrid model w.r.t.\ the full model and to evaluate its main errors are conducted in Section~\ref{sec:num_ex}. We illustrate some possible improvements of the hybrid model in Section~\ref{sec:poss_improvements} before summarizing our findings in Section~\ref{sec:concl}.

\section{Mathematical models and numerical methods}
\label{sec:models}
\definecolor{myred}{RGB}{220,0,0}
\def\svgwidth{0.35\textwidth}
\begin{figure}
\centering
     \subfloat[Fully-resolved\label{fig:notation_boundary_full}]{
     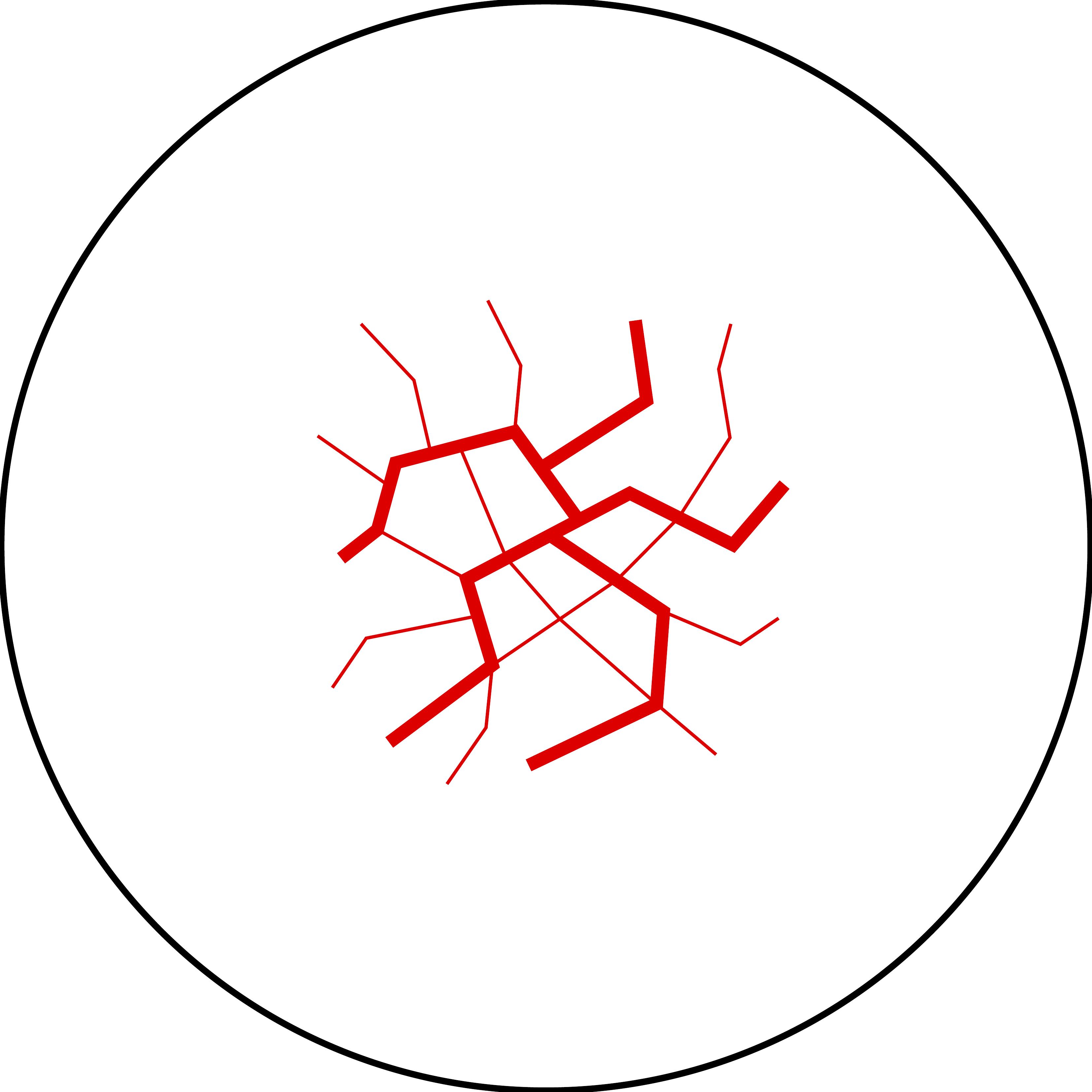}
     \def\svgwidth{0.35\textwidth}\hspace{1cm}
     \subfloat[Hybrid\label{fig:notation_boundary_red}]{
     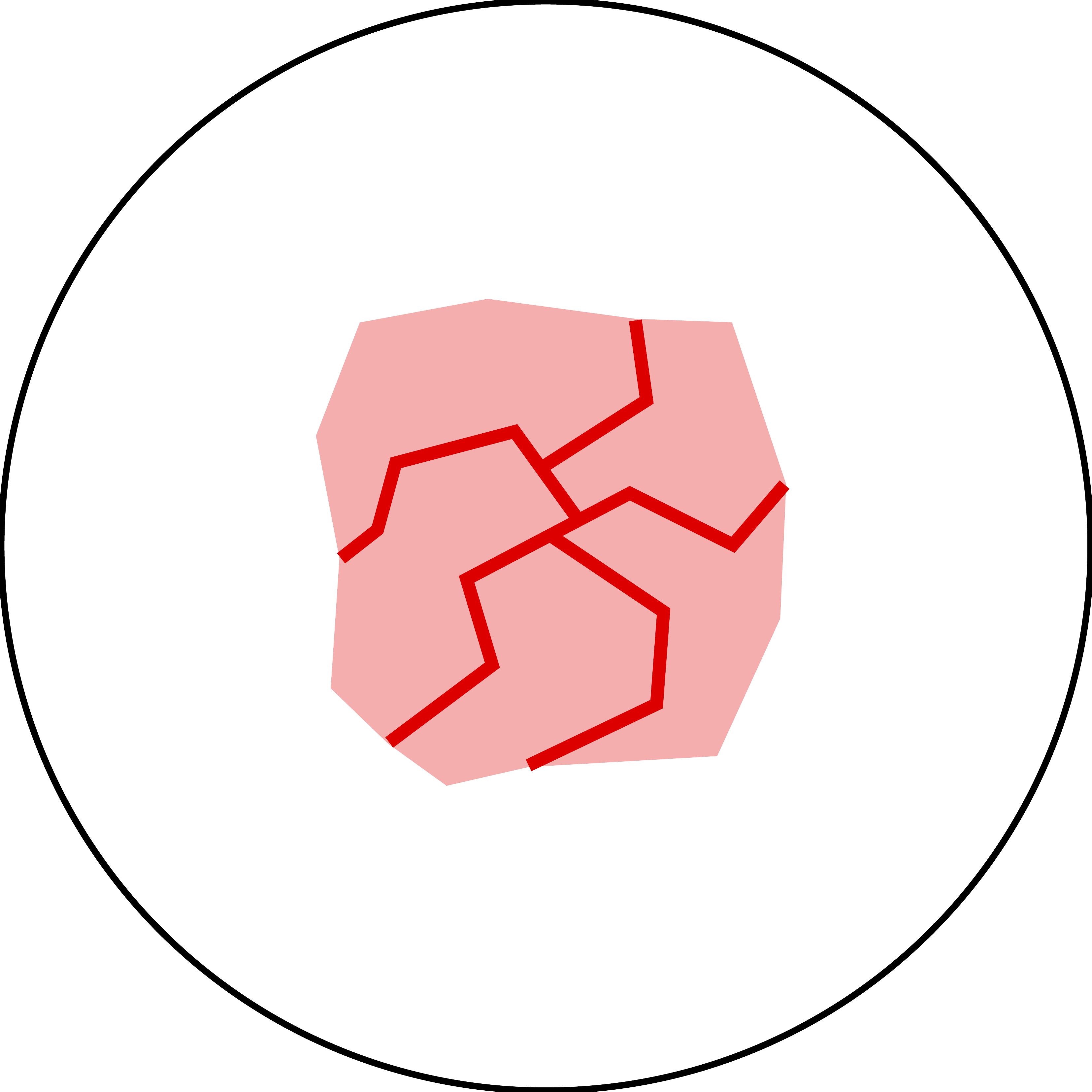}
     \caption{Notation for domains and boundaries}
\label{fig:notation_boundary}
\end{figure}
In this section, we describe the mathematical models we employ to solve the interaction between microcirculation and interstitial tissue perfusion including their main simplifications. We outline both our fully-resolved and our hybrid approach and their discretization by means of the finite element method (FEM).
\subsection{Problem setting}
\label{sec:setting}
As in other publications\cite{Kremheller2019,DAngelo2007,DAngelo2008,DAngelo2012,Secomb1994,Secomb2004,Cattaneo2013,Cattaneo2014,Nabil2015,Nabil2016,Desposito2018,Sweeney2019}, topology and structure of the microcirculation is described by a graph with straight edges, i.e., blood vessel segments. The segments connect the nodes of the network. A radius $R_k$ is assigned to each segment $\Lambda_k$. Available experimental data including the one employed here is also commonly provided in this format. Therefore, the vascular domain is given as the union of straight cylinders which are embedded in the three-dimensional tissue domain $\Omega$, see also Figure~\ref{fig:notation_boundary_full}. Based on the flow physics and the quantities of interest, it is obvious to employ a one-dimensional blood flow model for this part of the vascular system. In the following, we will denote the 1D embedded blood vessel network with $\Lambda$. Similar to Vidotto et al.\cite{Vidotto2018}, we further divide it into two subsets $\lbig$ and $\lsmall$ which correspond to the larger and smaller vessels in the network such that
\begin{equation}
\label{eq:sets_small_large}
\lbig:=\bigcup_{k\in I_{\mathrm{L}}} \Lambda_k,\;\;\lsmall:=\bigcup_{k\in I_{\mathrm{S}}} \Lambda_k\;\;{\mathrm{and}}\;\;\Lambda=\lbig\cup\lsmall
\end{equation}
with the index sets of large and small blood vessel segments $I_{\mathrm{L}}$ and $I_{\mathrm{S}}$, respectively. We will show in detail how this partition is realized in Section~\ref{sec:diff_fully_hyb}. Whereas the larger vessels are kept in the hybrid model, the smaller scale vessels are replaced by an appropriate homogenized representation as a porous network occupying the domain $\Omega_v\subset \Omega$, cf.~Figure~\ref{fig:notation_boundary_red}.
\subsection{Fully-resolved 1D-3D model}
\label{sec:resolved}
Gravity, inertial effects and the pulsatility of blood flow are neglected which are valid assumptions since we deal with microcirculatory flow. The balance of mass in the 1D vasculature domain $\Lambda$ is then given by the following equation
\begin{equation}
\label{eq:1D_blood_flow}
-\frac{\partial}{\partial s}\left( \frac{\pi R^4}{8\mu^{\hat{v}}}\frac{\partial p^{\hat{v}}}{\partial s} \right)= - \frac{\iema {M_{\mathrm{leak}}}{\hat{v}}{l\;\;\;\;}}{\rho^{\hat{v}}}
\quad\quad{\mathrm{on}}\;\;\Lambda,
\end{equation}
where we have applied the Hagen-Poiseuille law for flow in cylindrical pipes and assumed constant blood density $\rho^{\hat{v}}$. Here and in the following, quantities defined on the 1D vasculature domain are denoted by superscript $\hat{v}$. In the previous equation, $R$ is the blood vessel radius, $p^{\hat{v}}$ the pressure inside the vasculature, $\mu^{\hat{v}}$ the blood viscosity and $s$ the arc length coordinate along the 1D blood vessel segment. To account for the non-Newtonian behaviour of blood, we employ the algebraic relationship developed by Pries and Secomb~\cite{Pries2005} for \textit{in-vivo} blood viscosity depending on blood vessel diameter and hematocrit. As in~\cite{Sweeney2019}, we fix the hematocrit to $0.45$, thus, the blood viscosity $\mu^{\hat{v}}$ in each individual blood vessel segment depends only on its diameter. Finally, the right-hand-side term
\begin{equation}
\label{eq:starling}
\iema {M_{\mathrm{leak}}}{\hat{v}}{l\;\;\;\;}=\rho^{l}\cdot 2\pi R\cdot L_{p,\hat{v}} \cdot \left( p^{\hat{v}}- p^l-\sigma \left( \pi^b - \pi^l \right)  \right)
\end{equation}
is employed to model leakage of fluid across the blood vessel wall into the interstitium. For that, we use Starling's law with hydraulic conductance $L_{p,\hat{v}}$, density of blood plasma $\rho^l$, oncotic reflection coefficient $\sigma$ and the oncotic pressures of blood $\pi^b$ and the interstitial fluid (IF) $\pi^l$. In summary, the transvascular flux from the vascular network into the interstitial fluid is proportional to the pressure difference between vasculature and IF whose pressure in~\eqref{eq:starling} is denoted as $p^l$. It has long been  known that blood vessels inside tumors are leakier than normal ones which, in combination with a non-functional lymphatic system, leads to increased interstitial pressure inside solid tumors and, concurrently, resistance to efficient drug delivery.\cite{Baxter1989,Jain1994,Heldin2004} Note that our data sets are whole-tumor blood vessel networks where also larger vessels are leaky\cite{Desposito2018,Sweeney2019} which is why we apply the transvascular exchange term~\eqref{eq:starling} also on the subset of larger vessels $\lbig$.

As in related works, the tissue domain $\Omega$ is modeled as a porous medium. Therefore, flow in the interstitial fluid is accounted for by the following Darcy equation
\begin{equation}
\label{eq:mass_IF}
 -  \div \left(\frac{\tns{k}^l}{\mu^{l}} \grad p^{l} \right) = \delta_\Lambda\cdot\frac{\iema {M_{\mathrm{leak}}}{\hat{v}}{l\;\;\;\;} }{\rho^l}\quad\quad{\mathrm{in}}\;\;\Omega
\end{equation}
with (isotropic) permeability $\tns{k}^l = k^l\cdot \tns{I}$ and IF viscosity $\mu^l$. Hence, the primary variable for fluid flow through the tissue is the IF pressure $p^l$. The right hand side represents the counterpart of the leakage of fluid from the vasculature into the IF from~\eqref{eq:1D_blood_flow}. As proposed by D'Angelo and Quarteroni\cite{DAngelo2007,DAngelo2008,DAngelo2012} this mass transfer term is concentrated as a Dirac measure $\delta_\Lambda$ along the centerline $\Lambda$ of the vasculature resulting in a 1D-3D coupled problem. The mathematical properties including reduced convergence rates due to the introduced singular line source in the 3D pressure field are extensively studied in\cite{DAngelo2008,DAngelo2012,Koppl2014}. Alternative 2D-3D coupled approaches, where the mass exchange is evaluated at the lateral surfaces of the cylindrical blood vessel segments, have been proposed to increase the regularity of the solution.\cite{Koeppl2020,Koeppl2018} However, in our data sets the diameter $D$ is smaller than the element size $h$ in the 3D domain, see also Table~\ref{tab:analysis_datasets}. Therefore, we modify the approach of \cite{DAngelo2008,DAngelo2012}, which would involve taking the average value $\overline{p^l}$ of the pressure in the IF at the outer surface of the cylindrical vessels in the exchange term~\eqref{eq:starling}, and instead take the IF pressure value at the centerline $\Lambda$, which is a reasonable approach for the case $h>D$.\cite{DAngelo2007,Kremheller2019} This has recently also been investigated in the analogous solid mechanics problem of embedding thin 1D structures, i.e., beams, into 3D solid volumes.\cite{Steinbrecher2020} The weak form of the 1D-3D coupled problem may be written as
\begin{subnumcases}{}
\left(\frac{\partial \delta p^{\hat{v}}}{\partial s}, \frac{\pi R^4}{8\mu^{\hat{v}}}\frac{\partial p^{\hat{v}}}{\partial s}\right)_{\Lambda}+\left(\delta p^{\hat{v}}, \frac{\iema {M_{\mathrm{leak}}}{\hat{v}}{l\;\;\;\;}}{\rho^{\hat{v}}}\right)_{\Lambda} &= 0 \label{eq:full_weak_1D} \\
\left(\grad \delta p^{l}, \frac{\tns{k}^l}{\mu^{l}} \grad p^{l} \right)_{\Omega}-\left(\delta p^{l}, \frac{\iema {M_{\mathrm{leak}}}{\hat{v}}{l\;\;\;\;}}{\rho^{l}}\right)_{\Lambda} &= 0 \label{eq:full_weak_3D}
\end{subnumcases}
with test functions $\delta p^{\hat{v}}$ defined on the 1D domain and $\delta p^{l}$ defined on the 3D domain. Our approach allows for non-matching 1D discretizations $\Lambda_h$ and 3D discretizations $\Omega_h$ such that the two domains can be meshed independently of each other. This requires the numerical integration of products of 1D shape functions with 3D shape functions and products of 3D shape functions with 3D shape functions along the one-dimensional discretization $\Lambda_h$ which we realize via a segment-based line integration approach\cite{Kremheller2019,Steinbrecher2020} to avoid integration over kinks of shape functions, see Appendix~\ref{sec:app1}. After space discretization, the nodal primary variables of both domains are
\begin{equation}
\label{eq:primvars_full}
\bsd{p}^{\hat{v}}\in\mathbb{R}^{n_{{\mathrm{nodes}},\Lambda}}\;\;{\mathrm{and}}\;\;\bsd{p}^{ l}\in\mathbb{R}^{n_{{\mathrm{nodes}},\Omega}},
\end{equation}
that is, the nodal blood pressure in the discretized 1D domain and the nodal IF pressure in the discretized 3D domain, which consist of $n_{{\mathrm{nodes}},\Lambda}$ and $n_{{\mathrm{nodes}},\Omega}$, respectively. Details on the employed boundary conditions are given in Section~\ref{sec:bc_full}. 

Finally, we arrive at the global system of equations, which may be written as a $2\times 2$ block matrix
\begin{equation}
\label{eq:mat_vec_full}
\begin{bmatrix}
\bsd{K}^{\hat{v}\hat{v}} & \bsd{G}^{\hat{v}l} \\
\bsd{H}^{l\hat{v}} & \bsd{K}^{l l}
\end{bmatrix}
\begin{bmatrix}
\bsd{p}^{\hat{v}} \\
\bsd{p}^{l}
\end{bmatrix}
=
\begin{bmatrix}
\bsd{F}^{\hat{v}} \\
\bsd{F}^{l}
\end{bmatrix}.
\end{equation}
Herein, the main diagonal blocks $\bsd{K}^{ii}$ comprise contributions from the diffusive term and the exchange term in~\eqref{eq:full_weak_1D} and~\eqref{eq:full_weak_3D} while the off-diagonal submatrices $\bsd{G}^{\hat{v}l}$ and $\bsd{H}^{l\hat{v}}$ contain the "mixed" contributions from the exchange term. The right hand side terms $\bsd{F}^i$ represent the constant contribution to the exchange term stemming from the oncotic pressures in~\eqref{eq:starling}. To solve the coupled linear system~\eqref{eq:mat_vec_full} we employ a GMRES iterative solver in combination with the AMG(BGS) block preconditioner presented in Verdugo and Wall.\cite{Verdugo2016}
\subsection{Hybrid 1D-3D model}
\label{sec:hybrid}
The main idea behind our hybrid 1D-3D model, based on our previous work introduced in\cite{Kremheller2019} in the context of our vascular multiphase tumor growth model\cite{Kremheller2018}, is the following: The full resolution of the larger vessels $\lbig$ is kept, i.e., these are still modeled as a 1D embedded vasculature. Consequently, the hierarchy, topology and vascular properties such as individual blood vessel radii and viscosities of each segment are retained, see also Figure~\ref{fig:notation_boundary_red}. The smaller vessels $\lsmall$, for which this high-resolution data might either not be available through non-invasive imaging techniques or susceptible to errors, are instead represented as an additional porous network. This results in a double-porosity formulation where the first porous network is, as before, the interstitial space and the second one the smaller vessels occupying the domain $\Omega_v$. In the following, we will present the governing equations and the space discretization of this formulation.

As stated above, the model for the larger vessels does not change. Therefore, the mass balance equation inside the large vessels is given by
\begin{equation}
\label{eq:1D_blood_flow_large}
-\frac{\partial}{\partial s}\left( \frac{\pi R^4}{8\mu^{\hat{v}}}\frac{\partial p^{\hat{v}}}{\partial s} \right)= - \frac{\iema {M_{\mathrm{leak}}}{\hat{v}}{l\;\;\;\;}}{\rho^{\hat{v}}}
\quad\quad{\mathrm{on}}\;\;\lbig
\end{equation}
with the only difference to~\eqref{eq:1D_blood_flow} being that it holds only on the subset $\lbig\subset\Lambda$ of bigger vessels. The mass balance equation for the smaller vessels $\lsmall$ is replaced by a homogenized Darcy equation in the vascular domain $\Omega_v$, which we formulate as
\begin{equation}
\label{eq:mass_homo_vasc}
 -  \div \left(\frac{\tns{k}^v}{\mu^{v}} \grad p^{v} \right) =-\frac{\iema {M_{\mathrm{leak}}}{v}{l\;\;\;\;}}{\rho^{v}}\quad\quad{\mathrm{in}}\;\;\Omega_v.
\end{equation}
The unknown in this equation is the blood pressure $p^v$ in the homogenized part of the vasculature which is now defined in the entire 3D domain $\Omega_v$, thereby replacing the blood pressure of the smaller vessels in the 1D domain $\lsmall$ as the variable governing flow inside the smaller vessels. For simplicity, in a first step we consider an isotropic permeability tensor $\tns{k}^v= k^v\cdot \tns{I}$ for the additional porous network. This permeability and the averaged blood viscosity $\mu^v$ are the two model parameters governing this equation together with the right hand side term
\begin{equation}
\label{eq:M_leak}
\iema {M_{\mathrm{leak}}}{v}{l\;\;\;\;}=
\begin{cases}
\rho^l\cdot L_{p,v} (S/V)_{\lsmall} \cdot \left( p^{v}- p^l-\sigma \left( \pi^b - \pi^l \right)  \right) &\quad\quad{\mathrm{in}}\;\;\Omega_v \\
0 &\quad\quad{\mathrm{in}}\;\;\Omega\backslash\Omega_v
\end{cases}.
\end{equation}
This term replaces the outflow of fluid from the smaller vessels into the IF by a homogenized representation of the Starling equation~\eqref{eq:starling} involving the surface-to-volume ratio of the smaller blood vessels $(S/V)_{\lsmall}$ as an additional parameter. The mass balance equation of the IF for the fully-resolved model~\eqref{eq:mass_IF} is adapted as 
\begin{equation}
\label{eq:mass_IF_homo}
 -  \div \left(\frac{\tns{k}^l}{\mu^{l}} \grad p^{l} \right) = \frac{1}{\rho^l}\left(\delta_{\lbig}\cdot\iema {M_{\mathrm{leak}}}{\hat{v}}{l\;\;\;\;}+\iema {M_{\mathrm{leak}}}{v}{l\;\;\;\;}\right)\quad\quad{\mathrm{in}}\;\;\Omega
\end{equation}
in the homogenized formulation. Comparing the two equations, it is obvious that leakage from the large vessels is still treated equivalently, i.e., the large vessels are still embedded as 1D inclusions in the tissue with a Dirac measure (now defined only on $\lbig$). By contrast, leakage from the smaller blood vessels is replaced by the homogenized mass transfer term~\eqref{eq:M_leak} from the vascular domain $\Omega_v$ into the interstitium, i.e., from~\eqref{eq:mass_homo_vasc} into~\eqref{eq:mass_IF_homo}. 

So far, this procedure is analogous to other hybrid approaches.\cite{Shipley2019,Vidotto2018} The main difference to our methodology lies in the coupling between the larger vessels $\lbig$ and the homogenized vasculature $\Omega_v$. In the aforementioned publications, this was realized at the free ends of the larger vessels, i.e., as an outflow at the tips of the 1D discretization into the homogenized 3D vasculature domain. This was possible since the employed data sets had a clear vascular hierarchy with larger arterioles and venules connected to smaller capillaries. Our vascular networks, which we will describe in detail in Section~\ref{sec:analysis_datasets}, have been segmented from solid tumors and, therefore, have a much more complex, disorganized structure including variable vessel lengths and diameters as well as dead ends. All this is typical for tumor-specific vasculature.\cite{Carmeliet2000,Baluk2005} As shown in detail in Section~\ref{sec:diff_fully_hyb} for our data and the employed methodology to distinguish between large, flow-carrying vessels and smaller ones, another approach is more sensible: We enforce the coupling between larger vessels and the homogenized vasculature along the entire 1D representation of larger vessels $\lbig$ with a line-based coupling instead of a point-based coupling at the tips of the larger vessels flowing into the capillary bed as described before. Compared to these hybrid approaches, our proposed method has the advantage that no additional parameter -- apart from the penalty parameter -- is involved for the coupling of the two representations.

For that, we formulate a constraint of equal pressures in $\lbig$ and $\Omega_v$ as
\begin{equation}
\label{eq:const_simpl}
g = p^{\hat{v}}- p^{v}=0\quad\quad{\mathrm{on}}\;\;\lbig,
\end{equation}
which enforces a coupling between pressures $p^{\hat{v}}$ in the one-dimensional, large vessel domain $\lbig$ and homogenized pressures $p^v$ in the 3D domain $\Omega_v$. We aim to reproduce the fact that the pressure in a smaller vessel branching from a larger vessel at a specific node is equal to the pressure at the same node. If this smaller vessel is homogenized and, thus, removed from the 1D representation, we want to enforce these equal pressures between the resolved part and the homogenized part of the vasculature along the 1D vessel domain $\lbig$. In Section~\ref{sec:diff_fully_hyb}, we justify formulating this constraint along the entire 1D domain $\lbig$ considering the connectivity between larger and smaller vessels in our cases. We have previously employed a similar strategy in our hybrid treatment of the vasculature in a multiphase tumor growth model\cite{Kremheller2019} and the related solid mechanics problem of beam-to-solid mesh tying.\cite{Steinbrecher2020} We follow the same approach as in the two aforementioned publications and incorporate the constraint with an additional Lagrange multiplier (LM) field into the weak form of our hybrid model, which reads as
\begin{subnumcases}{}
	\left(\frac{\partial \delta p^{\hat{v}}}{\partial s}, \frac{\pi R^4}{8\mu^{\hat{v}}}\frac{\partial p^{\hat{v}}}{\partial s}\right)_{\lbig}+\left(\delta p^{\hat{v}}, \frac{\iema {M_{\mathrm{leak}}}{\hat{v}}{l\;\;\;\;}}{\rho^{\hat{v}}}\right)_{\lbig}+\left(\delta p^{\hat{v}}, \lambda\right)_{\lbig} &= 0 \label{eq:homo_weak_1D} \\
	\left(\grad \delta p^{v}, \frac{\tns{k}^v}{\mu^{v}} \grad p^{v} \right)_{\Omega_v}+\left(\delta p^{v}, \frac{\iema {M_{\mathrm{leak}}}{v}{l\;\;\;\;}}{\rho^{v}}\right)_{\Omega_v}-\left(\delta p^{v}, \lambda\right)_{\lbig} &= 0 \label{eq:homo_weak_vasc_3D} \\
	\left(\grad \delta p^{l}, \frac{\tns{k}^l}{\mu^{l}} \grad p^{l} \right)_{\Omega}-\left(\delta p^{l}, \frac{\iema {M_{\mathrm{leak}}}{\hat{v}}{l\;\;\;\;}}{\rho^{l}}\right)_{\lbig}-\left(\delta p^{l}, \frac{\iema {M_{\mathrm{leak}}}{v}{l\;\;\;\;}}{\rho^{l}}\right)_{\Omega_v} &= 0 \label{eq:homo_weak_IF_3D} \\
	\left(\delta\lambda ,\left(p^{\hat{v}}-p^{v}\right)\right)_{\lbig} &= 0 \label{eq:homo_lagrange_weak}
\end{subnumcases}
Therein, the first line is the weak form of flow in the larger vessels~\eqref{eq:1D_blood_flow_large} which is coupled to the weak form of flow in the homogenized vasculature domain, i.e., the second line~\eqref{eq:homo_weak_vasc_3D} with a continuous LM field $\lambda$ defined along the blood vessel center line. The third line is the weak form of flow in the IF. Compared to the fully-resolved model, conf. eqn.~\eqref{eq:full_weak_3D}, the additional mass transfer term arises due to leakage from the homogenized part of the vasculature into the IF. The fourth line represents the variational form of the coupling constraint~\eqref{eq:const_simpl}. Conveniently, the LM field employed to enforce this constraint can then be interpreted as a mass transfer term from the 1D resolved bigger vessels into the 3D homogenized vasculature, i.e., $\lambda=\iema M{\hat{v}}{v}$. Alternatively, a Gauss-point-to-segment scheme could also be employed but suffers from over-constraining of the system for large penalty parameters.\cite{Kremheller2019,Steinbrecher2020} Spatial discretization of the weak form~\eqref{eq:homo_weak_1D}-\eqref{eq:homo_lagrange_weak} leads to a saddle-point problem with nodal primary variables
\begin{equation}
\label{eq:primvars_homo}
\bsd{p}^{\hat{v}}\in\mathbb{R}^{n_{{\mathrm{nodes}},\lbig}},\bsd{\lambda}\in\mathbb{R}^{n_{{\mathrm{nodes}},\lbig}},\bsd{p}^{l}\in\mathbb{R}^{n_{{\mathrm{nodes}},\Omega}}\;\;{\mathrm{and}}\;\;\bsd{p}^{v}\in\mathbb{R}^{n_{{\mathrm{nodes}},\Omega_v}},
\end{equation}
that is, nodal pressures and nodal LMs in $\lbigdisc$, nodal IF pressures in $\Omega_h$ and nodal blood pressures of the homogenized vasculature in $\Omega_{v,h}$. In the following, we will specifically focus on the discretization of the terms arising due to the LM method. Approximating those contributions with a finite element interpolation yields a mortar-type formulation where the nodal LMs are additional degrees of freedom, condensed out with a dual approach\cite{Wohlmuth2000,Popp2010} or a penalty regularization of the mortar method is employed to remove the additional degrees of freedom and the saddle-point structure.\cite{Yang2005} Here, we follow the latter approach just as in our previous work on 1D-3D type couplings.\cite{Kremheller2019,Steinbrecher2020} The contributions to the weak form of the mass balance equations, i.e., the two last terms in~\eqref{eq:homo_weak_1D} and~\eqref{eq:homo_weak_vasc_3D} can be written as
\begin{equation}
\label{eq:lm_pot_weak_disc}
\delta\Pi_{{\mathrm{LM}},h}=\sum_{j=1}^{n_{{\mathrm{nodes}},\lbig}}\sum_{k=1}^{n_{{\mathrm{nodes}},\lbig}}\lambda_j D_{jk} \delta p^{\hat{v}}_k-\sum_{j=1}^{n_{{\mathrm{nodes}},\lbig}}\sum_{l=1}^{n_{{\mathrm{nodes}},\Omega_v}}\lambda_j M_{jl}\delta p^{v}_l
\end{equation}
with the so-called mortar matrices
\begin{align}
\label{eq:D}
\bsd{D}\left[j,k\right]=D_{jk}&=\int_{\Lambda_{{\mathrm{L}},h}}\hat{\Phi}_j \hat{N}_k\,ds \\
\intertext{and}
\label{eq:M}
\bsd{M}\left[j,l\right]=M_{jl}&=\int_{\Lambda_{{\mathrm{L}},h}}\hat{\Phi}_j N_l\,ds.
\end{align}
The entries of these matrices involve integrals of products of LM shape functions $\hat{\Phi}_j$ defined on the discretized 1D domain $\lbigdisc$ with 1D shape functions $\hat{N}_k$ and with 3D shape functions $N_l$ defined in the 3D domain $\Omega_v$. Hence, these terms are again evaluated using a segment-based approach, see Appendix~\ref{sec:app1}. We choose linear shape functions for both primary variables and the LM interpolation, i.e., $\hat{\Phi}_j = \hat{N}_j$. The weak form of the constraint~\eqref{eq:homo_lagrange_weak} may then be written in discretized form as
\begin{equation}
\label{eq:constraint_weak_disc}
\delta\Pi_{{\mathrm{\lambda}},h}=\delta\bsd{\lambda}^T\left(\bsd{D}\bsd{p}^{\hat{v}}-\bsd{M}\bsd{p}^{v}\right)=:
\delta\bsd{\lambda}^T\bsd{g}\left(\bsd{p}^{\hat{v}},\bsd{p}^{v}\right),
\end{equation}
where we have defined a weighted pressure gap $\bsd{g}$ at each node in $\lbigdisc$. This gap is then further used for the penalty regularization of the mortar method to explicitly define the nodal LMs in terms of 1D and 3D nodal blood pressures as
\begin{equation}
\label{eq:lambda}
\bsd{\lambda}=\epsilon\bsd{\kappa}^{-1}\bsd{g}\left(\bsd{p}^{\hat{v}},\bsd{p}^{v}\right).
\end{equation}
Hence, the LMs are no longer independent variables in the system but depend on the primary variables $\bsd{p}^{\hat{v}}$ and $\bsd{p}^{v}$. This overcomes the two major drawbacks of the LM method, namely, the increased system size and its saddle-point structure. Depending on the penalty parameter $\epsilon>0$, the constraint $\bsd{g}=\bsd{0}$ is relaxed and the exact solution is only recovered for $\epsilon\rightarrow\infty$. Additionally, the nodal LM in~\eqref{eq:lambda} has been scaled with the inverse of the diagonal matrix
\begin{equation}
\label{eq:kappa}
\bsd{\kappa}\left[j,j\right]=\int_{\Lambda_{{\mathrm{L}},h}}\hat{\Phi}_j \,ds.
\end{equation}
As proposed by Yang et al.\cite{Yang2005} this removes the dependency of the nodal LM on its "gap", i.e., in our case it makes its entries independent of the element lengths associated with its corresponding node. This can now be used to replace the LM vector such that the matrix-vector form of our hybrid model emerges as
\begin{equation}
\label{eq:mat_vec_homo}
\begin{bmatrix}
\bsd{K}^{\hat{v}\hat{v}}+\epsilon\bsd{D}^T\bsd{\kappa}^{-1}\bsd{D} & \bsd{G}^{\hat{v}l} & -\epsilon\bsd{D}^T\bsd{\kappa}^{-1}\bsd{M} \\
\bsd{H}^{l\hat{v}} & \bsd{K}^{l l} & \bsd{J}^{lv} \\
-\epsilon\bsd{M}^T\bsd{\kappa}^{-1}\bsd{D} & \bsd{L}^{vl} & \bsd{K}^{vv}+\epsilon\bsd{M}^T\bsd{\kappa}^{-1}\bsd{M}
\end{bmatrix}
\begin{bmatrix}
\bsd{p}^{\hat{v}} \\
\bsd{p}^{l} \\
\bsd{p}^{v}
\end{bmatrix}
=
\begin{bmatrix}
\bsd{F}^{\hat{v}} \\
\bsd{F}^{l} \\
\bsd{F}^{v}
\end{bmatrix}.
\end{equation}
As in the fully-resolved model~\eqref{eq:mat_vec_full}, main diagonal blocks are denoted as $\bsd{K}^{ii}$ and the coupling blocks $\bsd{G}^{\hat{v}l}$ and $\bsd{H}^{l\hat{v}}$ stem again from the transvascular 1D-3D exchange term. Additionally, the coupling blocks $\bsd{J}^{lv}$ and $\bsd{L}^{vl}$ account for exchange between homogenized vasculature and IF. The terms involving the mortar matrices $\bsd{D}$, $\bsd{M}$ and $\bsd{\kappa}$ couple blood flow in the larger vessels with the homogenized vasculature using our mortar penalty approach. Obviously, the LMs are no longer part of the system which is, consequently, not of saddle-point type anymore. The drawback, however, is that the choice of the penalty parameter influences the accuracy with which the constraint is fulfilled. Large penalty parameters yield better accuracy in terms of constraint fulfillment but can lead to an ill-conditioning of the system matrix. We will comment on the choice of the penalty parameter in Remark~\ref{rem:choice_penalty}.
\begin{remark}
\label{rem:order}
The concrete implementation of the hybrid model is slightly different than described here for illustrative purposes. The equations for IF flow and blood flow are evaluated simultaneously on the 3D domain and not assembled into two separate block matrices as written in~\eqref{eq:mat_vec_homo}. This means that the degrees of freedom are actually re-ordered in a node-wise manner compared to~\eqref{eq:mat_vec_homo} such that one row corresponding to the nodal IF pressure at a node $j$ is followed by a row corresponding to the homogenized blood pressure at this node $j$. Therefore, we actually solve a system which is blocked with $2\times 2$ submatrices, where the upper part corresponds to the resolved part of the vasculature and the lower part to the IF and the homogenized vasculature. For this system, we again employ the AMG(BGS) preconditioner\cite{Verdugo2016} with the GMRES iterative solver.
\end{remark}
\section{Setup of computational models}
\label{sec:setup}
\begin{figure}
     \centering
     \subfloat[][SW1222]{
     \begin{minipage}{0.99\textwidth}{
     \includegraphics[trim={12.0cm 0.0cm 7.5cm 0.0cm},clip,width=0.32\linewidth]{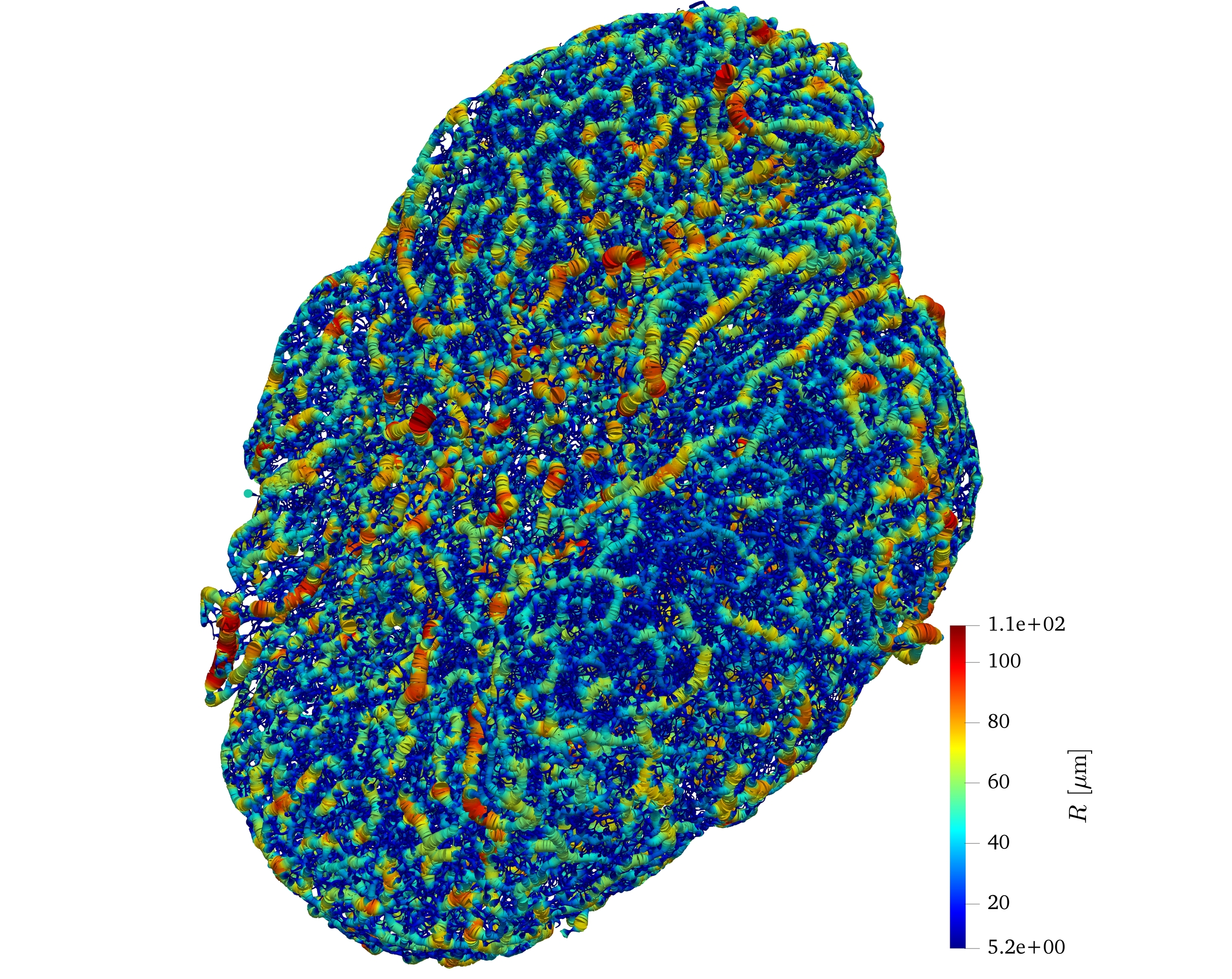}
     \includegraphics[trim={12.0cm 0.0cm 7.5cm 0.0cm},clip,width=0.33\linewidth]{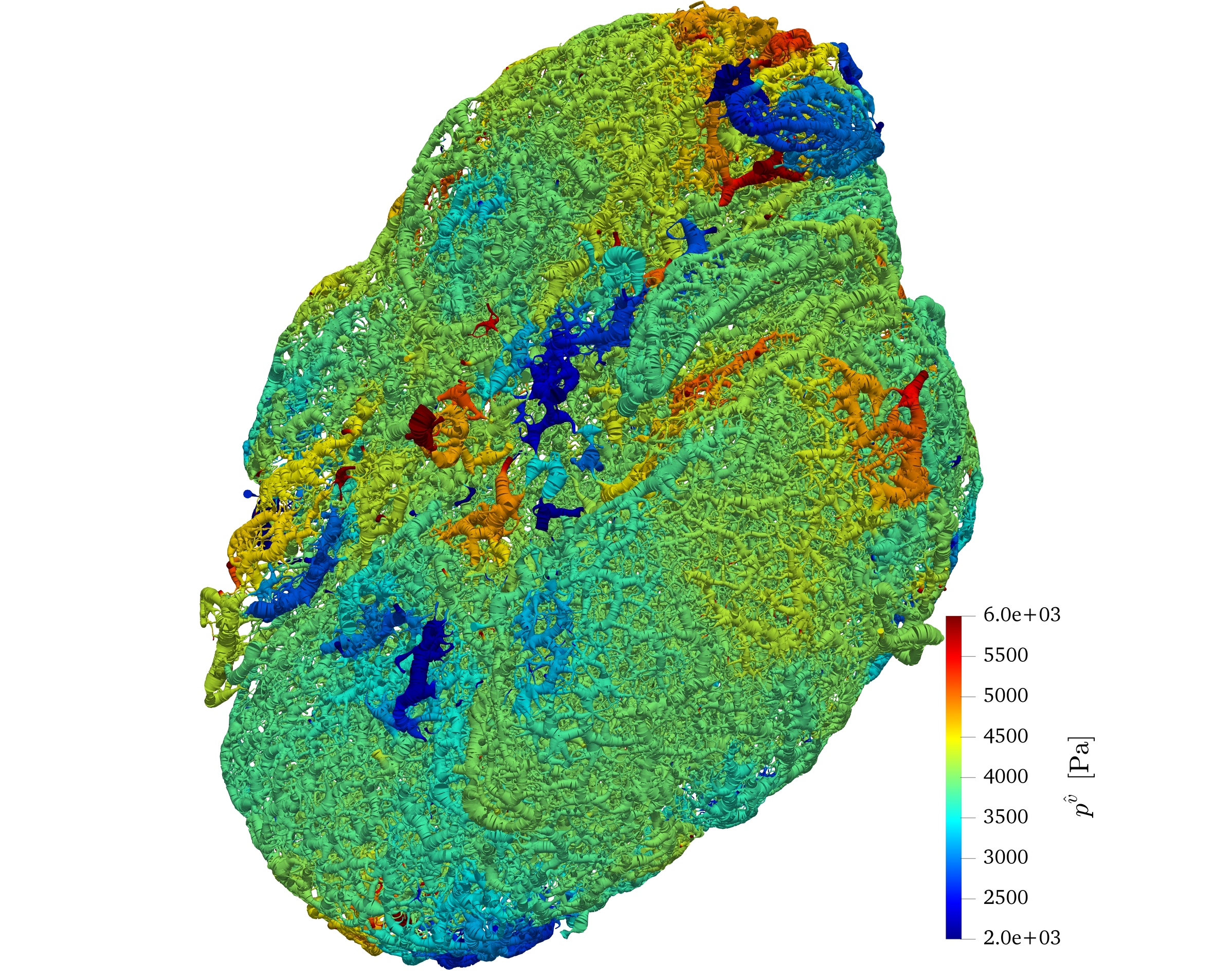}
     \includegraphics[trim={12.0cm 0.0cm 7.5cm 0.0cm},clip,width=0.33\linewidth]{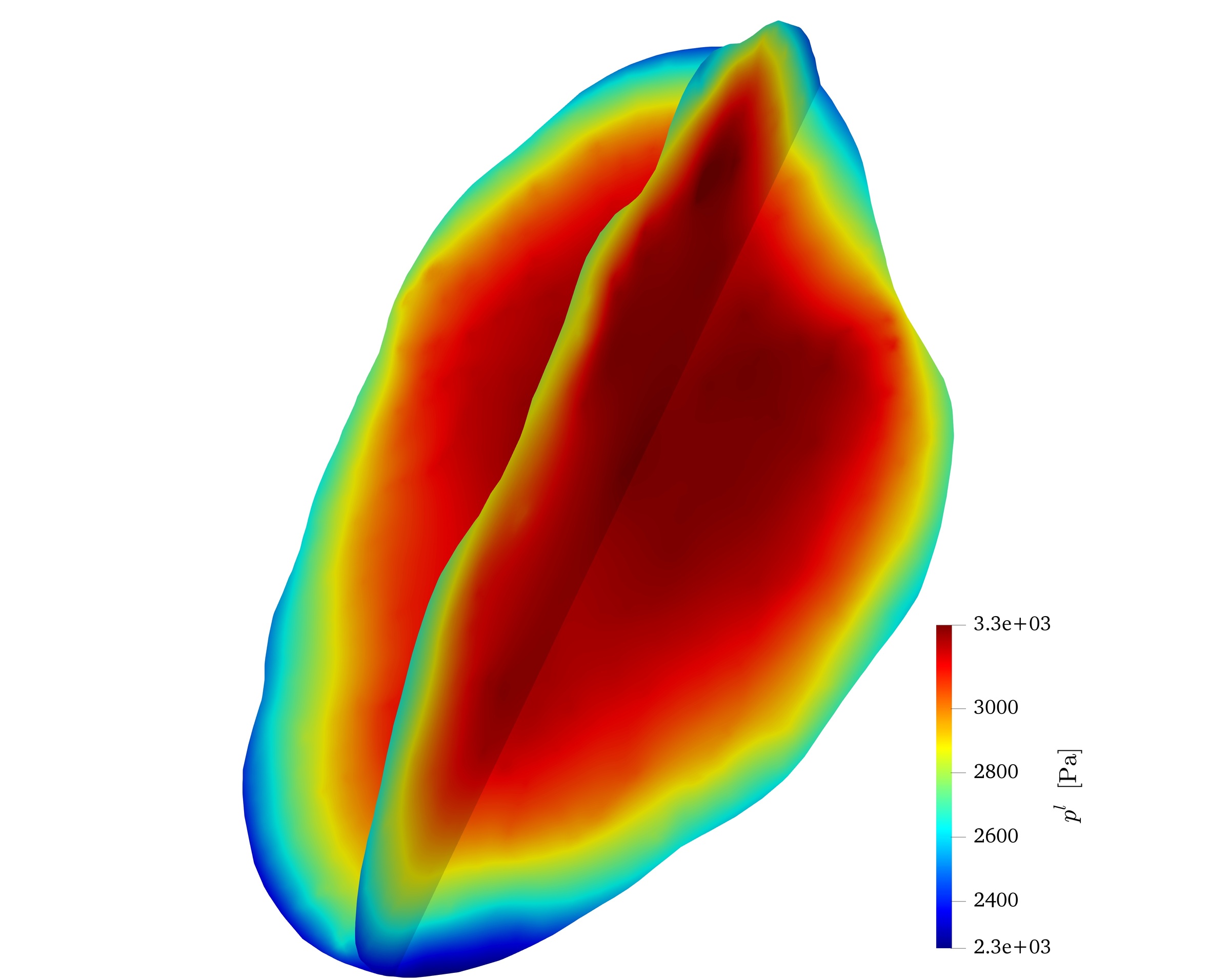}}
     \end{minipage}
     } \label{fig:SW1222}
     \subfloat[][LS174T]{
     \begin{minipage}{0.99\textwidth}{
     \includegraphics[trim={12.0cm 0.0cm 7.5cm 0.0cm},clip,width=0.33\linewidth]{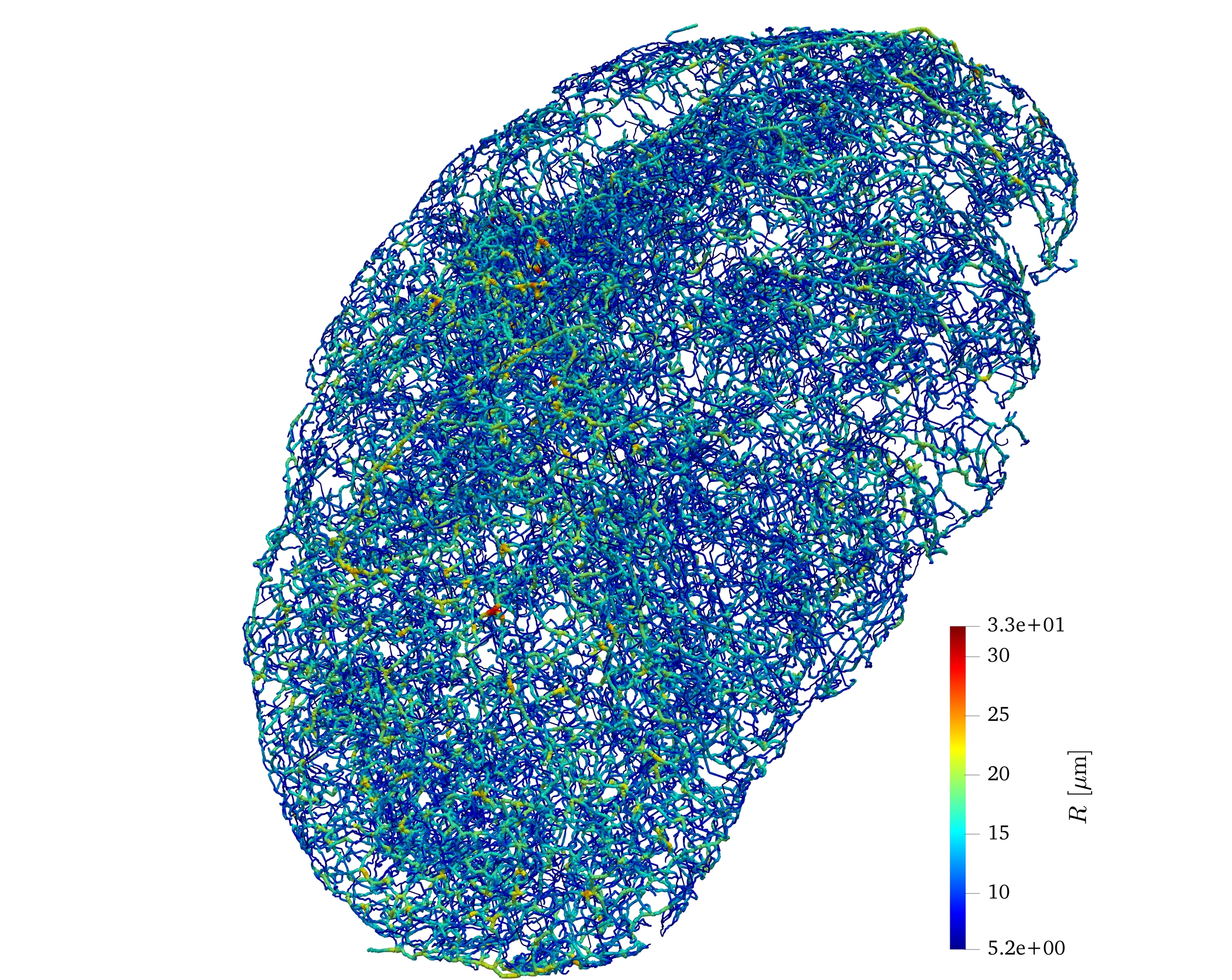}
     \includegraphics[trim={12.0cm 0.0cm 7.5cm 0.0cm},clip,width=0.33\linewidth]{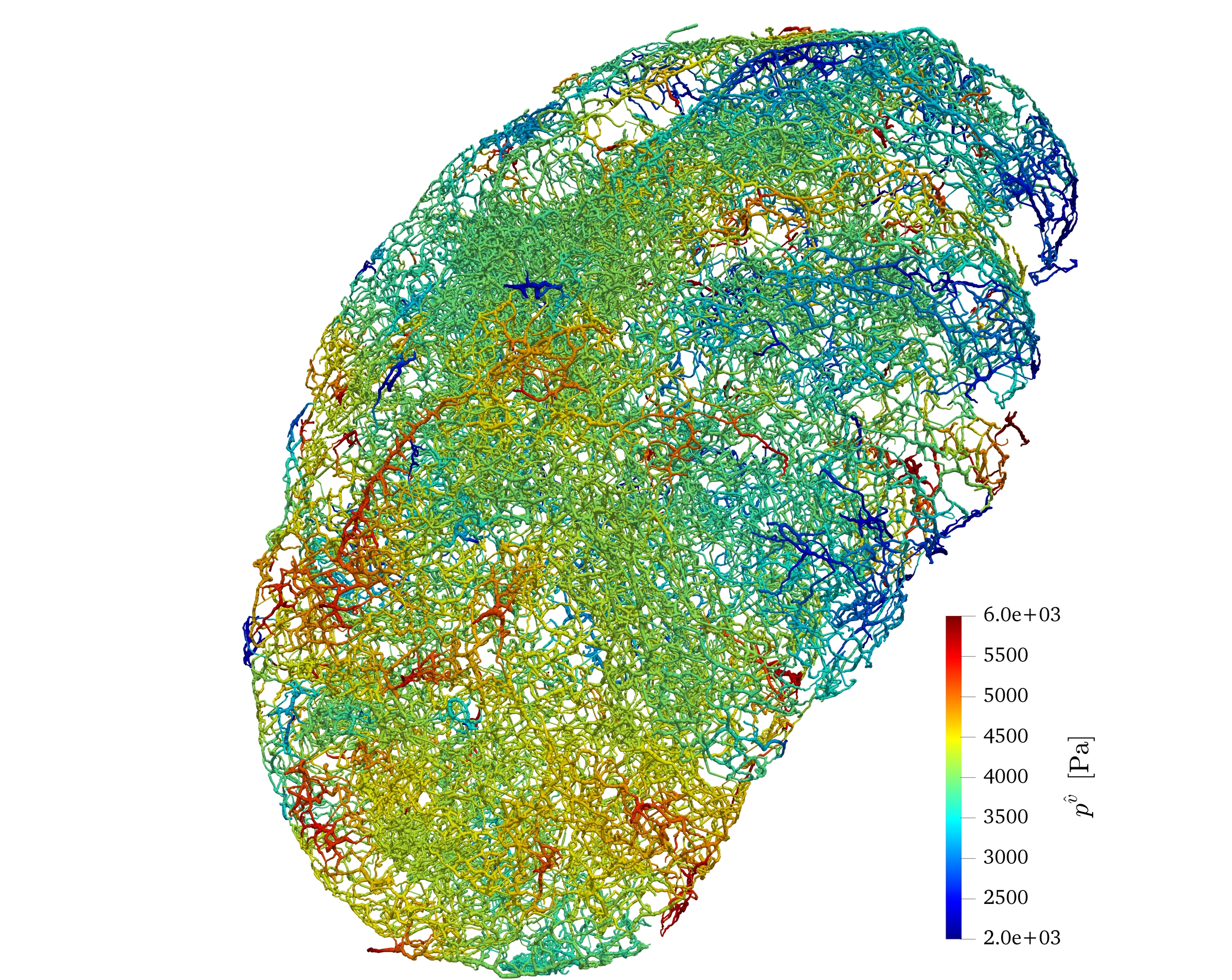}
     \includegraphics[trim={12.0cm 0.0cm 7.5cm 0.0cm},clip,width=0.33\linewidth]{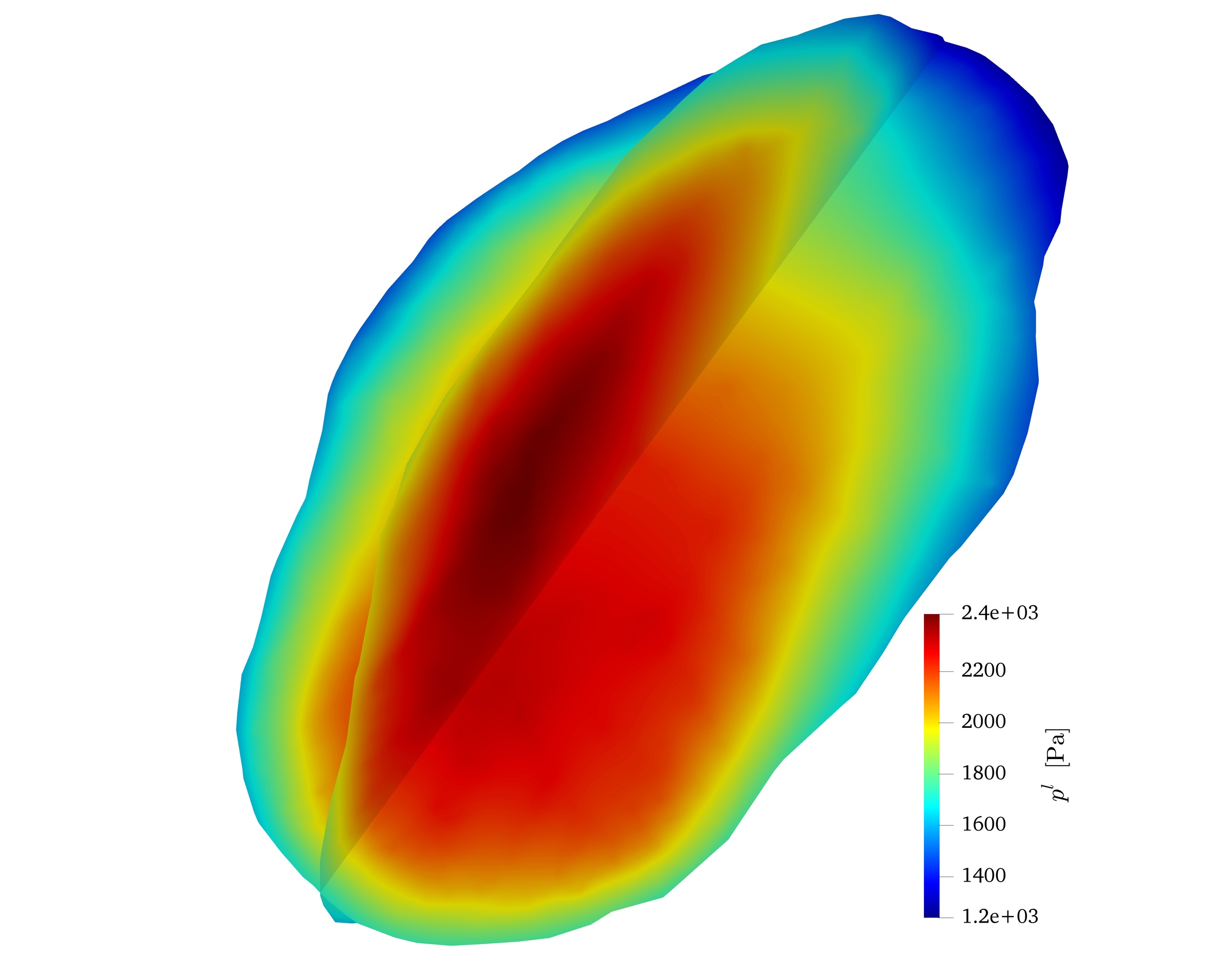}}
     \end{minipage}
     }\label{fig:LS174T} \hfill
     \subfloat[][GL261]{
     \begin{minipage}{0.99\textwidth}{
     \includegraphics[trim={12.0cm 0.0cm 7.5cm 30.0cm},clip,width=0.33\linewidth]{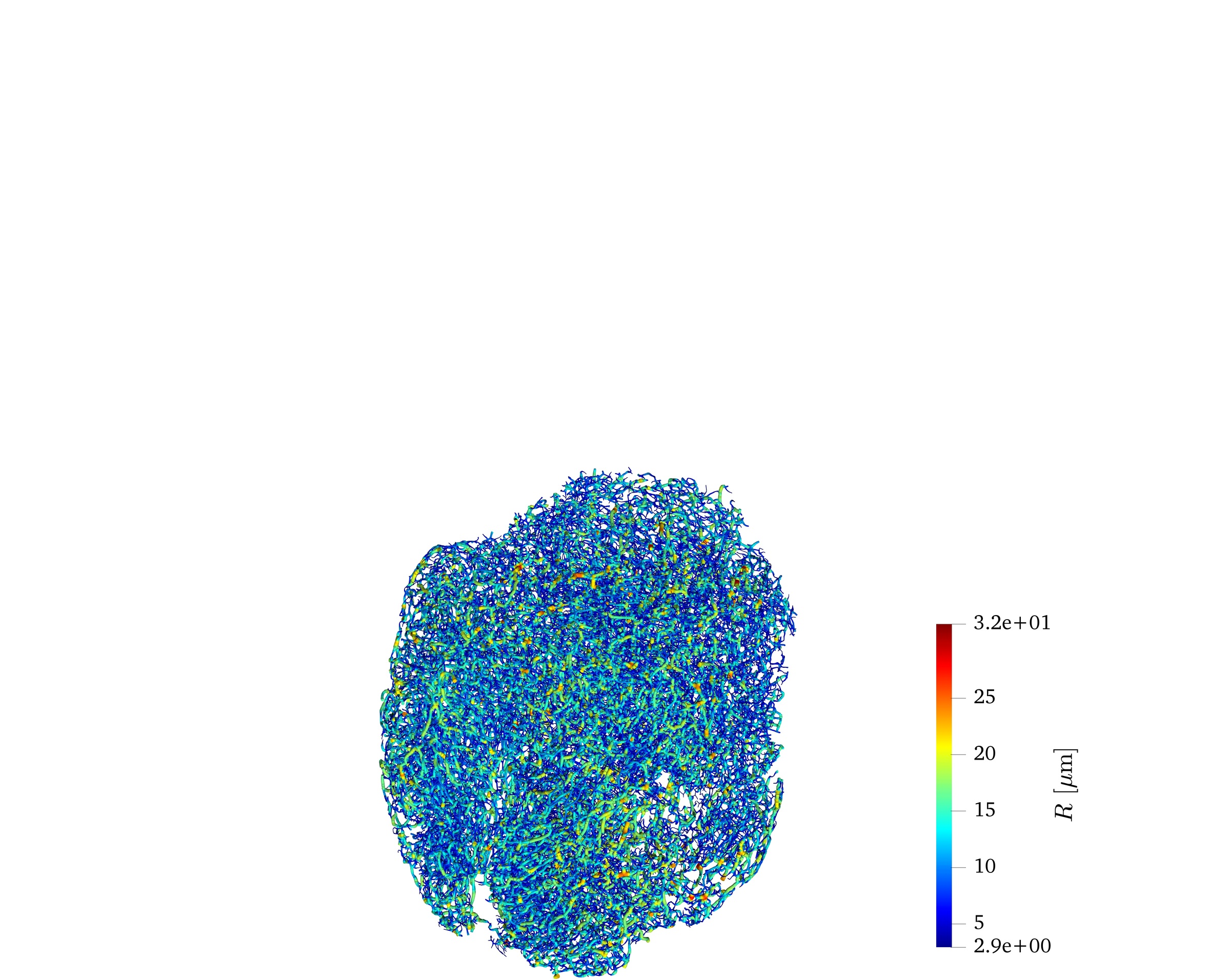}
     \includegraphics[trim={12.0cm 0.0cm 7.5cm 30.0cm},clip,width=0.33\linewidth]{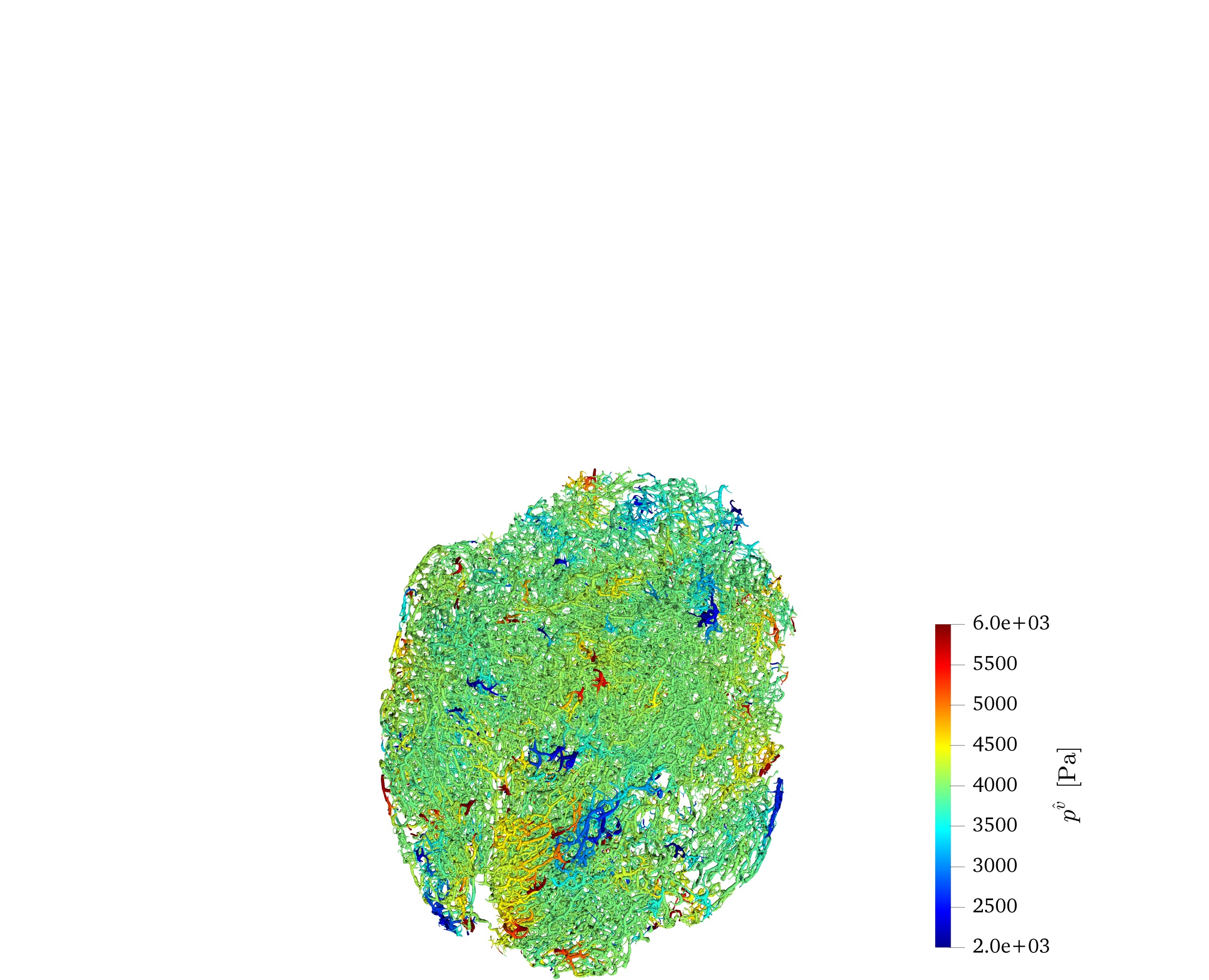}
     \includegraphics[trim={12.0cm 0.0cm 7.5cm 30.0cm},clip,width=0.33\linewidth]{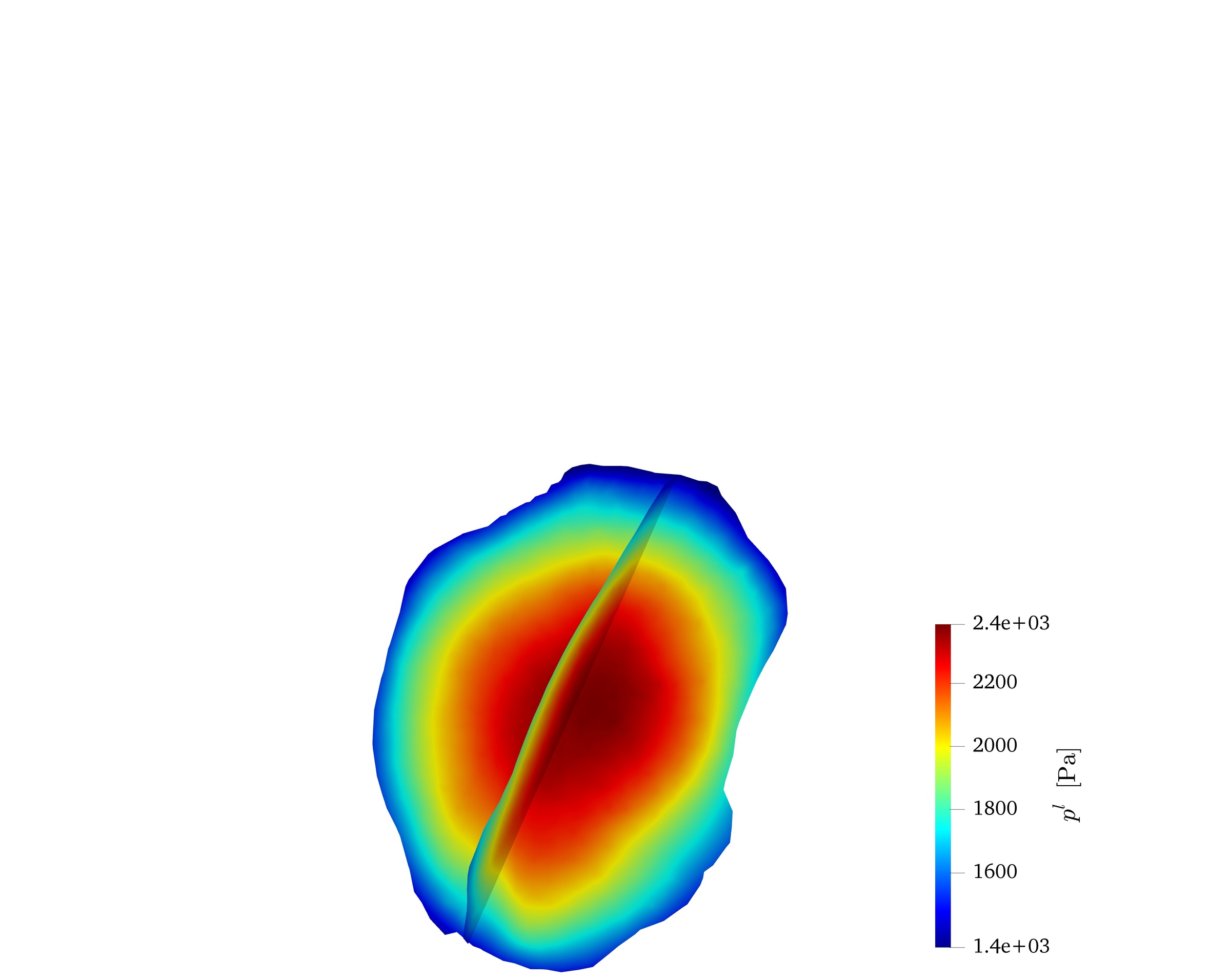}}
     \end{minipage}
     } \label{fig:GL261}
     \caption{Full topology and structure of the vascular networks (left, colour-coded by the respective radii), representative results for simulated blood pressures (middle) and IF pressures (right) in the fully-resolved model (Same spatial scale is used for all three cases)}
     \label{fig:datasets}
\end{figure}
This section describes the setup of our fully-resolved and of our hybrid model. We first analyse the real-world tumor data sets which we will employ for all our numerical tests. Subsequently, the assignment of boundary conditions in both models is described. Then, we illustrate how we create the hybrid model with homogenized vasculature starting from the full topology of the vascular networks. Finally, the definition of representative elementary volumes for homogenization is introduced.
\subsection{Analysis of real-world tumor data sets}
\label{sec:analysis_datasets}
\begin{table}
	\centering
	\resizebox{\textwidth}{!}{%
	\begin{tabular}{l l l l l} 
		\hline
		 & LS174T & GL261 & SW1222 & Unit \\
		\hline\hline
		No.\ of segments (elements) of 1D network & $\num{186092}$ & $\num{120340}$ & $\num{419198}$ & $-$\\
		No.\ of nodes of 1D network & $\num{178592}$ & $\num{110062}$ & $\num{385218}$ & $-$\\
		Tumor volume & $\num{190.5}$ & $\num{24.6}$ & $\num{235.5}$ & $\si{\cubic\milli\metre}$ \\
		Tumor dimensions & $4.46\times 7.59\times 10.88$  & $2.35\times 4.57\times 5.32$ & $6.44\times 8.07\times 11.08$ & $\si{\milli\metre}\times\si{\milli\metre}\times\si{\milli\metre}$ \\
		Blood vessel volume fraction & $\num{1.13}$ & $\num{4.01}$ & $\num{14.90}$ & $\si{\percent}$ \\
		Blood vessel surface-to-volume ratio & $\num{1.85e-3}$  & $\num{6.93e-3}$ & $\num{7.43e-3}$ & $\si{\per\micro\metre}$ \\
		Mean blood vessel diameter $\pm\;\text{std. dev.}$ & $\num{22.0 \pm 7.2}$  & $\num{17.6 \pm 10.0}$ & $\num{44.6 \pm 39.2}$ & $\si{\micro\metre}$ \\
		Mean blood vessel segment length $\pm\;\text{std. dev.}$ & $\num{27.2 \pm 6.8}$ & $\num{25.4 \pm 7.7}$ & $\num{28.7 \pm 9.2}$ &	$\si{\micro\metre}$	\\
		No.\ of boundary nodes of 1D network on tumor hull & $\num{1559}$ & $\num{2419}$ & $\num{1933}$ & $-$\\
		No.\ of boundary nodes of 1D network inside domain & $\num{1855}$ & $\num{6599}$ & $\num{13772}$ & $-$\\
		No.\ of elements of 3D domain & $\num{15955142}$ & $\num{15141173}$ & $\num{13231813}$ & $-$\\
		No.\ of nodes of 3D domain & $\num{2660273}$ & $\num{2524666}$ & $\num{2207655}$ & $-$\\
		Mean element size in $\Omega_v$ & $\num{76.4}$ & $\num{39.7}$ & $\num{78.1}$ & $\si{\micro\metre}$\\
		Edge length of REV & $\num{1250}$ & $\num{750}$ & $\num{1500}$ & $\si{\micro\metre}$	
	\end{tabular}}
		\caption{Details on tumor vasculature data sets and discretization}
\label{tab:analysis_datasets}
\end{table}
We have obtained three different vasculature data sets from REANIMATE\cite{Desposito2018,Sweeney2019}, which is a framework combining mathematical modeling with high-resolution imaging data to predict transport through tumors. We only briefly describe the experimental procedure here. More details are given in the two aforementioned papers. Two different colorectal cell lines, namely SW1222, LS174T, and one glioma cell line, GL261, were grown subcutaneously for 10 to 14 days in mice, resected and optically cleared, and finally imaged using optical projection tomography. The data was then segmented to obtain the complete blood vessel networks inside the tumors in the graph format as discussed before.

\begin{figure}
     \centering
     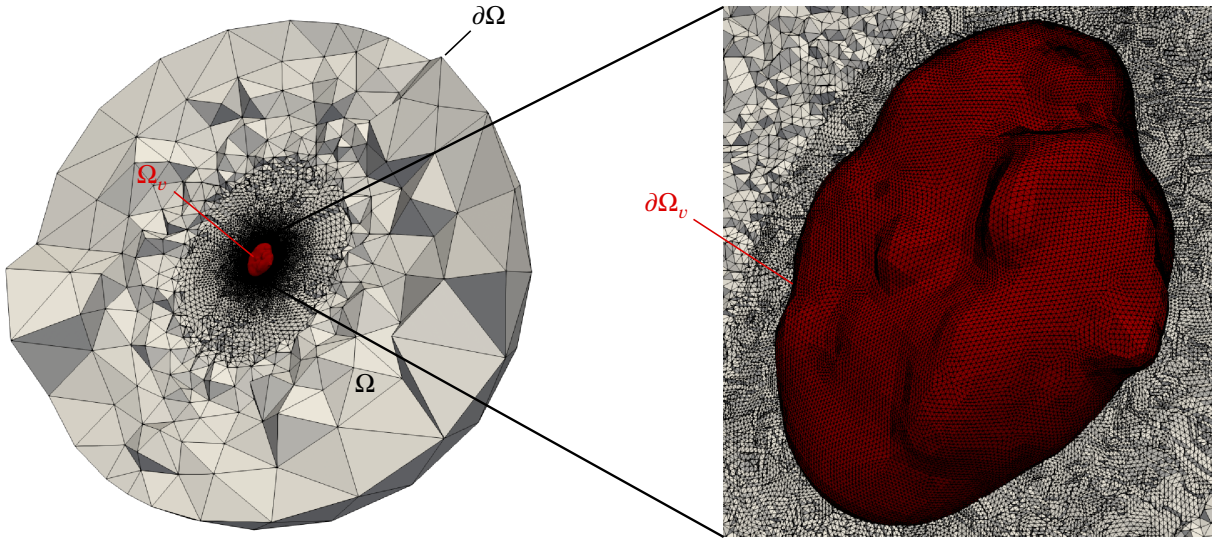
     \caption{Mesh of three-dimensional domain for SW1222 tumor. Tumor domain (equivalent to the domain $\Omega_v$ on which the additional porous network of smaller vessels is present in the hybrid model) is depicted in red and has been obtained as the alpha shape of nodes of the vascular network.}
     \label{fig:mesh}
\end{figure}

The topologies and blood vessel radii of the three distinct cases are illustrated in Figure~\ref{fig:datasets} together with representative results of blood vessel and IF pressure of the fully-resolved model. Further network data has been collected in Table~\ref{tab:analysis_datasets}: All three networks contain more than $\num{100000}$ blood vessel segments and nodes. The SW1222 case is the biggest tumor both in network size and tissue volume. The latter has been calculated by approximating the hull of the tumor using the \texttt{alphaShape} function of Matlab (MathWorks Inc., Natick, MA)\cite{alphaShape}. The hull is then smoothed, remeshed using Gmsh (version 4.4.1)\cite{Gmsh} and slightly enlarged to encompass all vasculature nodes. Its enclosed region is integrated to give the tumor volume, see also Figure~\ref{fig:mesh} for the SW1222 tumor. Note that this tumor domain corresponds to the domain $\Omega_v$ on which the additional porous network of smaller vessels is present in the hybrid model and where its additional governing equation~\eqref{eq:mass_homo_vasc} is defined and solved. Furthermore, all topologies are rotated such that their principal axes align with the coordinate axes. The three different cases show distinct vascular architectures, for instance, the SW1222 network is much denser with a higher blood vessel volume fraction and blood vessel surface-to-volume ratio than the two other types. In addition, its blood vessel diameters are generally larger and have a much higher variability. Finally, we have analyzed the boundary nodes, i.e., the tips which are only connected to one other node. All topologies have a comparable number of boundary nodes lying on the aforementioned enclosing alpha shape whereas the GL261 and SW1222 tumors have a much higher number of tips inside the domain than the LS174T tumor.
\subsection{Assignment of boundary conditions}
\label{sec:bc}
\def\svgwidth{0.9\textwidth}
The assignment of physiologically reasonable boundary conditions on large vascular networks is quite challenging since flows or pressures cannot be measured on the level of individual micro-vessels. Sweeney et al.\cite{Sweeney2019} developed an algorithm\cite{Sweeney2018code} to apply boundary conditions which match \textit{in-vivo} measurements of perfusion for the present data set. We will re-use this framework here to generate the boundary conditions for the fully-resolved case and briefly describe it in Section~\ref{sec:bc_full}. Boundary conditions for the hybrid model are detailed in Section~\ref{sec:bc_hybrid}.
\subsubsection{Fully-resolved model}
\label{sec:bc_full}
For the fully-resolved model, boundary conditions for the blood pressure $p^{\hat{v}}$ in the 1D network and the IF pressure $p^l$ need to be assigned. For the blood vessel pressure, we re-use the approach of Sweeney et al., which has been made publicly available\cite{Sweeney2018code} and is based on earlier work by Fry et al.\cite{Fry2012} Thereby, boundary conditions are assigned on the tips of the network, i.e., on the boundary nodes of the 1D representation of the vasculature both on the tumor hull and inside the tumor as given in Table~\ref{tab:analysis_datasets}. The following algorithm is applied: First, a high or low pressure of $\SI{5999.4}{\Pa}$ or $\SI{1999.8}{\Pa}$ (corresponding to $\SI{45}{\mmHg}$ or $\SI{15}{\mmHg}$) is randomly applied to the boundary points on the tumor surface until $\SI{5}{\percent}$ of \textit{all} end points of the 1D network have been assigned such a high/low pressure. Additionally, the method prevents that points which are in close proximity to each other are assigned the far apart pressure values which would produce unphysiologically large flows. Second, a no-flux boundary condition is randomly assigned to the interior boundary nodes until $\SI{33}{\percent}$ of \textit{all} boundary nodes have this condition. This value is consistent with the fraction of dead ends in tumor vasculature estimated from experimental studies.\cite{Morikawa2002} Third, the remaining $\SI{62}{\percent}$ of boundary nodes are given as unknowns to the flow optimization scheme of Fry et al.\cite{Fry2012} This scheme aims at solving a constrained optimization problem for incomplete pressure boundary data by minimizing the error of pressures and wall shear stress w.r.t.\ pre-defined target values. D'Esposito et al.\cite{Desposito2018} and Sweeney et al.\cite{Sweeney2019} have shown that this procedure for assignment of boundary conditions ensures that tumor perfusion is in good agreement with experimental data. Note that the entire algorithm is not deterministic due to the random selection of nodes for high/low boundary conditions on the external surface of the tumor and of nodes for no-flux boundary condition in its interior. Therefore, the analyses in the following sections will be performed on five different sets of pressure boundary conditions on the 1D network per tumor case.

Concerning the IF pressure, we want to prescribe the far-field pressure for the IF as $p^l_\infty=\SI{0}{\Pa}$ following Sweeney et al.\cite{Sweeney2019} In order to achieve this within our finite element approach, we enlarge the domain $\Omega$ radially to a sphere of radius $\SI{80}{\mm}$ as shown in Figure~\ref{fig:mesh} for the SW1222 case. This allows us to set a Dirichlet boundary condition of $p^l=\SI{0}{\Pa}$ on its boundary $\partial\Omega$ and, thereby, to mock the far-field pressure. We validated this approach in the following way for all three vascular networks: We solved the fully-resolved model and compared the IF pressure solution (for one specific set of pressure boundary conditions on the 1D network) with a case where the domain was only enlarged to a sphere with radius $\SI{40}{\mm}$ (with corresponding zero IF pressure Dirichlet boundary condition on its outer surface). No visible differences in the IF pressure distribution in our domain of interest inside and around the tumor domain could be detected. This indicates that the enlargement is big enough insofar as the solution in the domain of interest is not influenced by the size of the enlargement any more. We can also gradually coarsen the mesh when moving away further from the vascular domain as depicted in Figure~\ref{fig:mesh} since the IF pressure gradient flattens and tends to zero further away from the center of the domain. This enables the use of a sufficiently fine mesh for the region surrounding the embedded vascular network while the computational cost for extending the domain is not too high.
\subsubsection{Hybrid model}
\label{sec:bc_hybrid}
\begin{figure}
\centering
     \subfloat[][SW1222]{
     \includegraphics[trim={12.0cm 0.0cm 7.5cm 0.0cm},clip,width=0.32\linewidth]{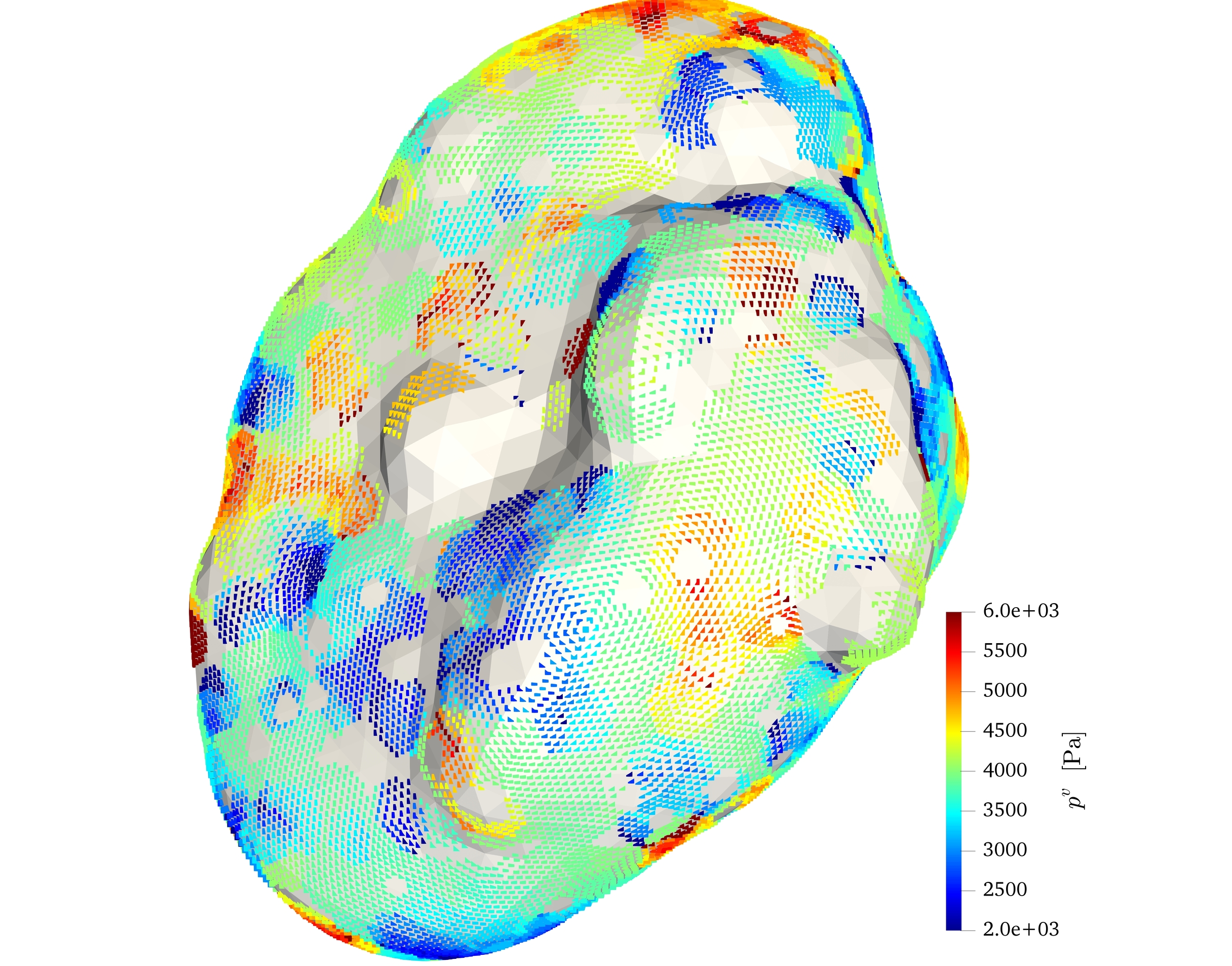}
     } \label{fig:SW1222_bc} \hfill
     \subfloat[][LS174T]{
     \includegraphics[trim={12.0cm 0.0cm 7.5cm 0.0cm},clip,width=0.32\linewidth]{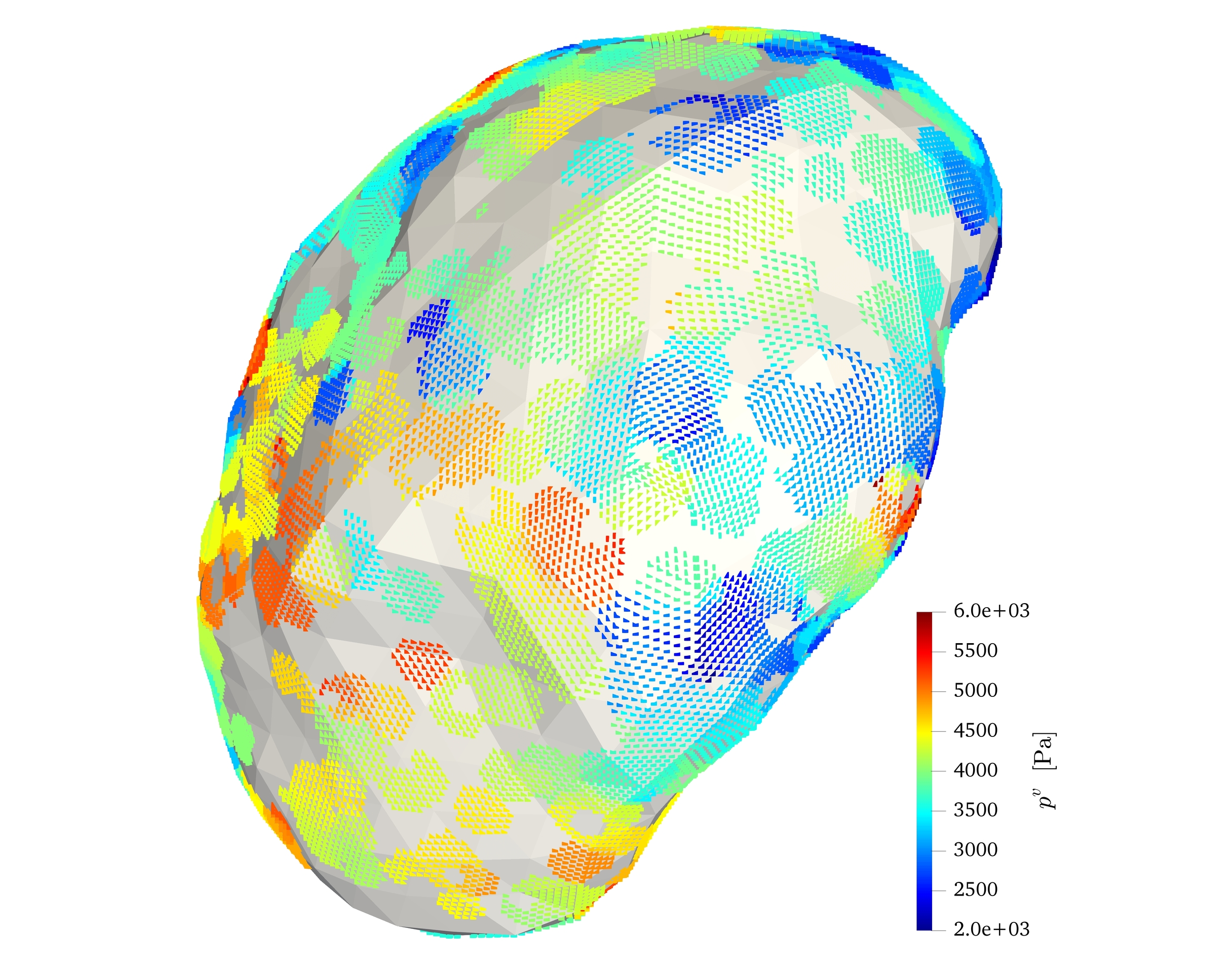}
     } \label{fig:LS174T_bc} \hfill
     \subfloat[][GL261]{
     \includegraphics[trim={12.0cm 0.0cm 7.5cm 0.0cm},clip,width=0.32\linewidth]{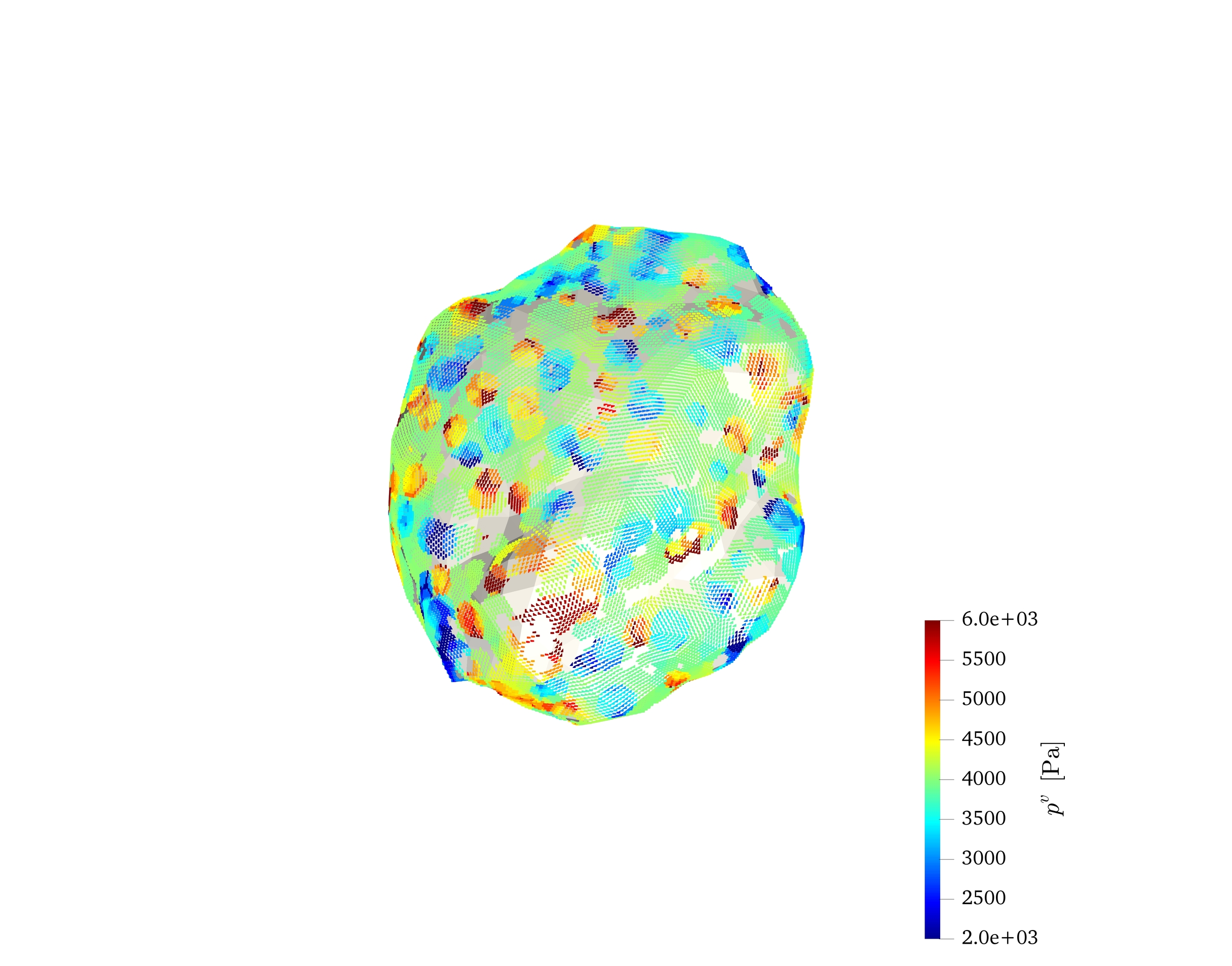}
     } \label{fig:GL261_bc}
     \caption{Exemplary distributions of boundary conditions for homogenized pressure $p^v$ on boundary of vascular domain $\partial\Omega_v$ in the hybrid model variant for all three cases}
     \label{fig:bc_homo}
\end{figure}
In addition to boundary conditions for the IF pressure $p^l$ and the blood pressure $p^{\hat{v}}$, the hybrid model requires boundary conditions for the pressure in the homogenized vasculature $p^v$. The IF pressure is treated as in the fully-resolved model and we set it to zero at the boundary of the domain $\partial\Omega$. In the following, we will always compare the accuracy of the hybrid variant w.r.t.\ the fully-resolved one for one specific set of pressure boundary conditions on the 1D network obtained with the procedure described in the previous sectoin. Thus, to perform this comparison the pressure boundary conditions on the 1D network are transferred from the fully-resolved model to the hybrid model in the following manner: The boundary conditions of blood pressure $p^{\hat{v}}$ on the larger vessels $\lbig$ can directly be taken from the boundary conditions of the fully-resolved model. If a node with a Dirichlet boundary condition in the fully-resolved vasculature $\Lambda$ is part of the larger vessels $\lbig$ we simply keep this boundary condition on the 1D discretization also in the hybrid model. Dirichlet boundary conditions on the smaller vessels $\lsmall$ cannot be assigned on the 1D discretization since smaller vessels are homogenized. However, we can employ them to assign boundary conditions for $p^v$ on the boundary of the domain of homogenized vessels $\partial\Omega_v$ as depicted in Figures~\ref{fig:notation_boundary_red} and~\ref{fig:mesh}. Similar to Vidotto et al.\cite{Vidotto2018}, we smooth these values to account for the homogenization of the smaller vessels: Each condition belonging to a node of the smaller vessels $\lsmall$ at the tumor surface is assigned to all 3D nodes lying on the surface $\partial\Omega_v$ within a distance of less than $\SI{400}{\micro\m}$ for the SW1222 and the LS174T tumor and less than $\SI{200}{\micro\m}$ for the GL261 tumor. Nodes of the 3D mesh which lie within this distance of multiple boundary nodes on $\lsmall$ are assigned the mean pressure value of all these boundary nodes. On the rest of the surface $\partial\Omega_v$ we set a no-flux boundary condition. We also do not set a boundary condition for $p^v$ on nodes of the 3D mesh in close proximity to end nodes of the 1D network since this would mean setting different boundary conditions on nodes whose pressures should be coupled due to the constraint on pressures $p^{\hat{v}}$ and $p^v$ and, thus, would lead to an overconstrainment of the system. The resulting distribution of boundary conditions $p^v$ over $\partial\Omega_v$ is illustrated in Figure~\ref{fig:bc_homo} for three exemplary cases.

\subsection{Distinction between fully-resolved and hybrid model}
\label{sec:diff_fully_hyb}
\begin{figure}
     \centering
     \subfloat[][full topology]{
     \includegraphics[trim={12.0cm 0.0cm 7.5cm 0.0cm},clip,width=0.32\linewidth]{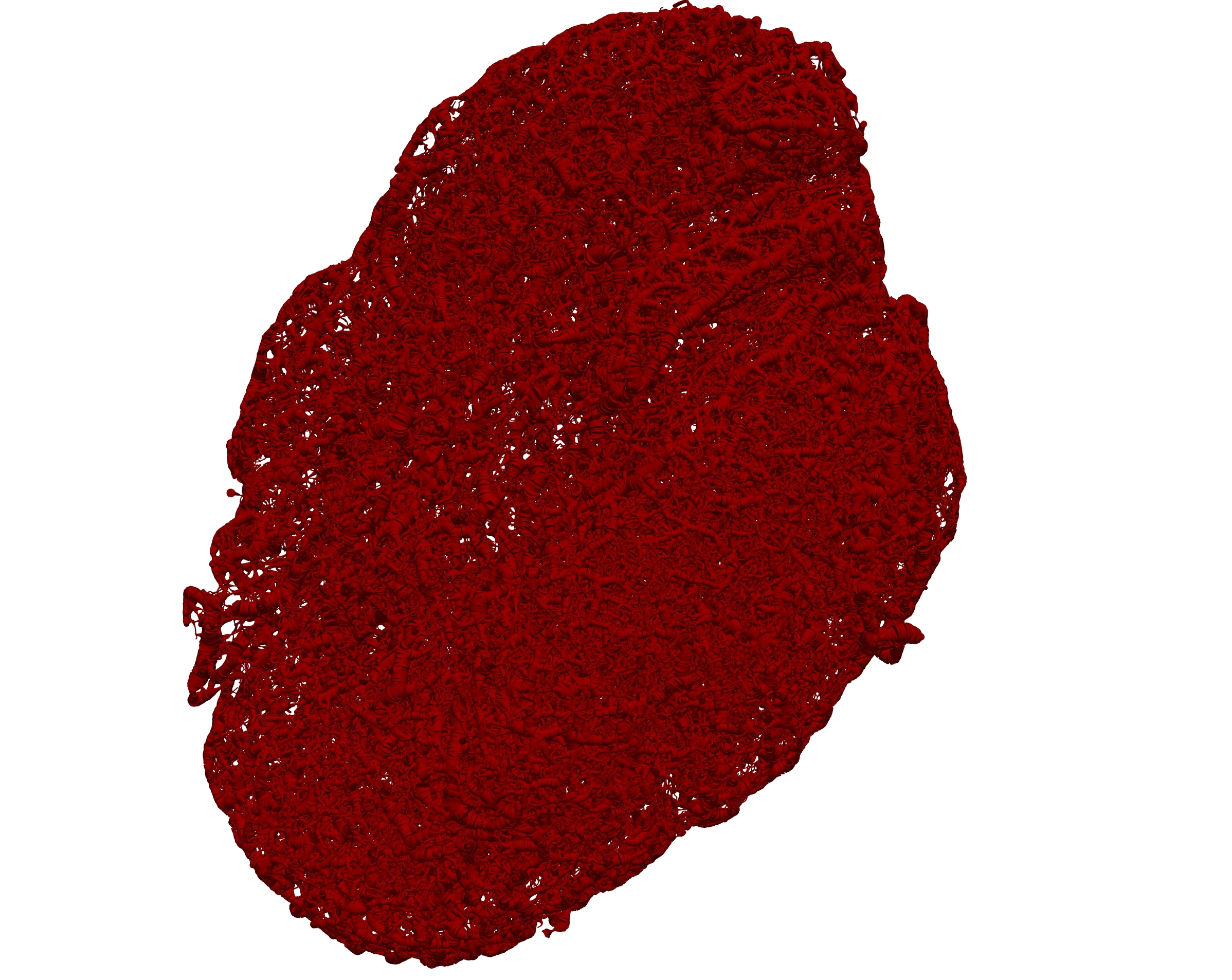} \label{fig:SW1222_full}
     }  \hfill
     \subfloat[][$\SI{10}{\percent}$ with highest flow]{
     \includegraphics[trim={12.0cm 0.0cm 7.5cm 0.0cm},clip,width=0.32\linewidth]{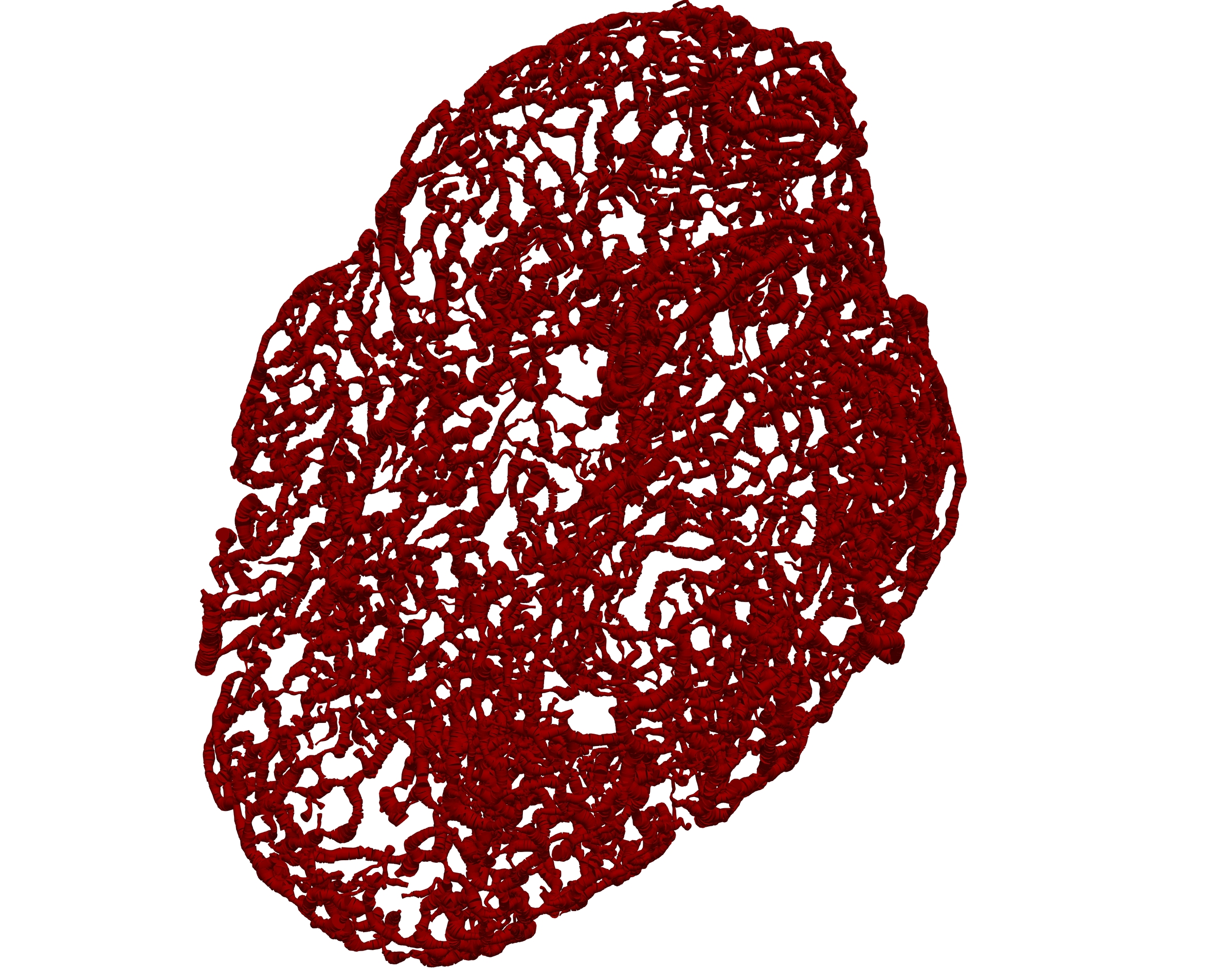} \label{fig:SW1222_largest_flow}
     }  \hfill
     \subfloat[][$\SI{10}{\percent}$ with biggest radius]{
     \includegraphics[trim={12.0cm 0.0cm 7.5cm 0.0cm},clip,width=0.32\linewidth]{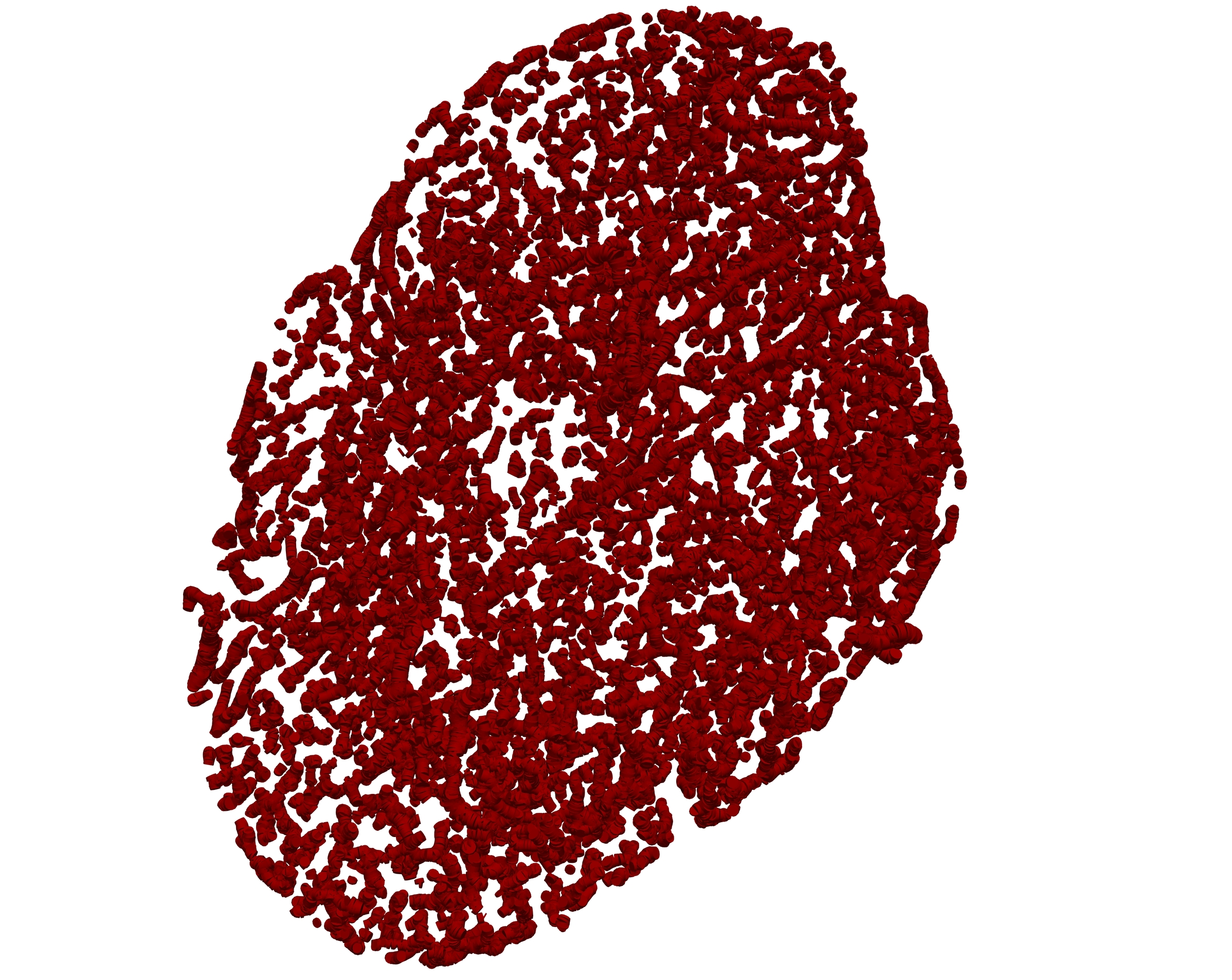} \label{fig:SW1222_largest_rad}
     } 
     \caption{Extraction of large vessels $\lbig$ from the entire network $\Lambda$ -- Comparison between flow-based criterion (and sorting out of small connected components) and radius-based criterion}
     \label{fig:SW1222_comp_flow_rad}
\end{figure}
\begin{table}
	\centering
	\begin{tabular}{l c c | c c | c c | c c} 
		\hline
		 & \multicolumn{2}{c}{case $\SI{5}{\percent}$} & \multicolumn{2}{c}{case $\SI{10}{\percent}$} & \multicolumn{2}{c}{case $\SI{15}{\percent}$} & \multicolumn{2}{c}{case $\SI{20}{\percent}$} \\
		 mean diameter $[\si{\micro\m}]$ & $D_{\lbig}$ & $D_{\lsmall}$  & $D_{\lbig}$ & $D_{\lsmall}$ & $D_{\lbig}$ & $D_{\lsmall}$  & $D_{\lbig}$ & $D_{\lsmall}$  \\
		 \hline\hline
		 SW1222 & $104.7$ & $41.4$ & $95.7$ &  $38.9$ & $88.4$ &  $36.9$ & $81.5$ &  $35.4$ \\
		 LS174T & $28.7$ & $21.7$ & $27.6$ &  $21.5$ & $27.0$ &  $21.2$ & $26.5$ &  $21.0$ \\
		 GL261 & $28.6$ & $17.1$ & $27.4$ &  $16.6$ & $26.5$ &  $16.1$ & $25.7$ &  $15.7$
	\end{tabular}
		\caption{Comparison of mean blood vessel radius in larger vessels $\lbig$ and smaller vessels $\lsmall$ (all values indicate the mean taken over five different sets of pressure boundary conditions on the 1D network produced by the methodology described in Section~\ref{sec:bc_full}, "case ${\mathrm{X}}\,\si{\percent}$" denotes the case where ${\mathrm{X}}\,\si{\percent}$ of the 1D blood vessels are retained in the hybrid approach)}
		\label{tab:analysis_radius}
\end{table}
As previously stated, we envision that our hybrid model could be applied in cases where the full structure of the vascular network is unknown such that only the topology of the larger vessels can be acquired via non-invasive imaging. However, in our data sets we actually have the full structure available. In line with the main goals of this paper, namely, to validate the hybrid approach, to quantify the error with respect to the fully-resolved case and to determine its optimal parameters for perfusion through solid tumors, we artificially create the hybrid model from the fully-resolved one. In the hybrid approaches of \cite{Shipley2019,Vidotto2018} this was realized by a radius-based criterion. Their employed data sets had a clear hierarchy typical for the microcirculation with larger arterioles branching into smaller capillaries which in turn connect and form larger venules. Thus, it was possible to exploit the hierarchical structure of the vasculature by keeping only the larger vessels in the set $I_{\mathrm{L}}$. 

For our tumor vasculature data sets this is not as straightforward. While there are some thicker vascular branches, especially in the SW1222 case, no clear hierarchical vascular architecture can be extracted from the topologies in Figure~\ref{fig:datasets} with a radius-based criterion. To illustrate this fact, we compare the full architecture of the SW1222 network with a network where only the top $\SI{10}{\percent}$ of vessels with the largest radius are kept in Figures~\ref{fig:SW1222_full} and~\ref{fig:SW1222_largest_rad}. Many small unconnected clusters of several blood vessel segments appear due to the heterogeneous, extremely variable distribution of the radius and lack of vascular hierarchy. Branches connecting these clusters which have a smaller radius are removed. Applying our or any hybrid model on this topology would not be possible as hybrid approaches also rely on a "sensible" topology for $\lbig$ which preserves the structure of the entire network via one or several connected subgraphs of larger vessels which feed respectively drain the smaller, homogenized vessels. Only then, the  1D blood flow model and corresponding boundary conditions can reasonably be applied on $\lbig$ together with suitable exchange terms into the smaller vessels. Thus, we instead distinguish between smaller and larger vessels based on the flow within the vessels. This yields a better preservation of the network architecture for the hybrid case, see Figures~\ref{fig:SW1222_full} and~\ref{fig:SW1222_largest_flow}. Now, connected subgraphs of larger vessels $\lbig$ emerge which connect in- and outlets of the main flow-carrying vessels with the smaller vessels. 

Hence, our strategy to obtain $\lbig$ is as follows: We first solve the fully-resolved model (using the boundary conditions described in Section~\ref{sec:bc_full}). Then, all elements except the ones with the highest flow are deleted from the vascular graph, e.g., the top $\SI{10}{\percent}$ with the highest flow are kept. However, there are still some very small clusters consisting of only a few segments present in the graph. We additionally delete connected components from the graph whose overall length is smaller than $\SI{250}{\micro\m}$, i.e., sub-components which are smaller than ten segments with the average segment lengths given in Table~\ref{tab:analysis_datasets}. By that, we delete an additional $\SIrange[range-phrase = -,range-units=single]{0.1}{0.8}{\percent}$ of segments which are part of these smaller sub-components. This methodology gives the set $I_{\mathrm{L}}$ of larger vascular branches which are kept in the hybrid approach as exemplarily shown in Figure~\ref{fig:SW1222_largest_flow}. Here, we show only the SW1222 case but equivalent results hold for the other two network topologies. In the following, we will denote cases where the top ${\mathrm{X}}\,\si{\percent}$ of elements with highest flow are kept and the small connected components are removed according to the procedure described above as "case ${\mathrm{X}}\,\si{\percent}$". 

Recall that the assignment of pressure boundary conditions on the fully-resolved vascular network is not deterministic. Moreover, different boundary conditions will produce distinct flow patterns in the vasculature and, hence, also different sets of large and small vessels in our procedure and a different topology for $\lbig$. Therefore, the following analysis will always be performed for five sets of pressure boundary conditions on the 1D network with corresponding distinct sets of large and small vessels $I_{\mathrm{L}}$ and $I_{\mathrm{S}}$. 

In Table~\ref{tab:analysis_radius} the mean diameters $D_{\lbig}$ and $D_{\lsmall}$ of larger and smaller vessels are compared. It is obvious that the diameters in the set of small vessels $I_{\mathrm{S}}$ which are removed from the hybrid model are considerably smaller than the diameters of the large vessels. This behaviour is most pronounced for the SW1222 topology where for the case $\SI{5}{\percent}$ the mean diameters in $\Lambda_{\mathrm{L}}$ are $\num{2.5}$ times bigger than in $\Lambda_{\mathrm{S}}$. Naturally, this ratio drops for all topologies when a higher percentage of segments is kept in the large vessel set. For the LS174T and GL261 data sets the difference in blood vessel diameters is not as large but this can be attributed to the fact that the diameters are less dispersed than in the SW1222 topology, see also the mean and standard deviation of the diameters in Table~\ref{tab:analysis_datasets}. Also in these cases, the diameters in $\Lambda_{\mathrm{L}}$ are larger by approximately one standard deviation of the diameter of the entire vasculature (as in the SW1222 case). In summary, our approach incorporates mainly the vessels with larger radii in the set $I_{\mathrm{L}}$ whereas also some segments with smaller radii are kept to preserve the main topology of the networks in the hybrid model. Therefore, there is also a significant congruence of the sets of large vessels $I_{\mathrm{L}}$ between different pressure boundary condition cases. For instance, in the case where $\SI{10}{\percent}$ of the blood vessels are kept in the hybrid model, the average percentage of identical retained segments between two different pressure boundary condition cases is \SI{45}{\percent} for the LS174T tumor, \SI{51}{\percent} for the GL261 tumor and \SI{78}{\percent} for the SW1222 tumor. In Remark~\ref{rem:real_life} we further comment on how the obtained topologies of larger vessels $\lbig$ relate to real \textit{in-vivo} tumor imaging data.

Next, we justify our line-based coupling approach between the large vessels $\lbig$ and the homogenized vasculature. For that, we have analyzed the connectivity between larger and smaller vessels for the fully-resolved topologies in Table~\ref{tab:analysis_conn}. Here, we denote by $\varphi=n_{{\mathrm{nodes}},\lbig\cap\lsmall}/n_{{\mathrm{nodes}},\lbig}$ the proportion of nodes of the larger vessels $\lbig$ which have a direct connection to a node of the smaller vessels $\lsmall$. The presented data illustrates that for the GL261 and the SW1222 tumor almost every third to every fourth node of the main branches $\lbig$ is directly connected to a node of the smaller blood vessel segments $\lsmall$, i.e., at every third to fourth node a smaller vessel branches away from $\lbig$. For the LS174T network, the connectivity is slightly smaller. Here, only $\SIrange[range-phrase = -,range-units=single]{13}{18}{\percent}$ of nodes in larger vessels are connected to smaller vessels. In all cases, these numbers obviously again drop when keeping a larger portion of the entire network in the set $I_{\mathrm{L}}$.

In the hybrid approach, information about these smaller branching vessels is lost since they are removed from the 1D representation of the vasculature. As stated above, we want to enforce equal pressures between larger and smaller vessels as this equality also holds in the fully-resolved model at branching points. The high connectivity between the two network parts supports our line-based mortar penalty coupling between the resolved and homogenized part of the vasculature in which we actually couple the entire network of big vessels $\lbig$ with the homogenized vasculature. Of course, we know the connecting nodes between larger and smaller vessels here as we know the full topology of all networks, so we could also enforce the coupling between resolved and homogenized part in a point-based manner at these locations. However, in the more realistic case when only the architecture of larger vessels is known without the exact locations where smaller vessels branch away, this is not the case. Therefore, we adopt our line-based coupling within the hybrid model hereafter to compare the results with the fully-resolved reference solution. Note that the network tips of $\lbig$ (both in the interior of the domain and on the tumor hull) are actually also coupled with the homogenized vasculature since the discrete constraint of a vanishing weighted pressure gap is enforced along the entire 1D discretization and, thus, also at the end nodes.

Finally, we analyze also the elements connecting larger and smaller vessels, i.e., those 1D elements of the smaller vessels where one node is part of $\lbig$ and the other part of $\lsmall$. We gather all these elements and compute their mean diameter and mean absolute flow value. Then, we calculate the coefficient of variation of these quantities, $CV_D$ and $CV_{\left\lvert Q\right\rvert}$ as the ratio of standard deviation of the diameter, resp. flow to its mean in these connectivity elements. The results are collected in Table~\ref{tab:analysis_conn}. Obviously, the SW1222 case shows the highest variability in both flow and diameter followed by the GL261 and the LS174T case. For all cases, the variability of the flow is larger than for the diameter since the volumetric flow in a segment depends on the fourth power of the diameter due to the employed Hagen-Poiseuille relationship. These results are consistent with the topology of the entire network where the variability of the blood vessel diameter is also larger for the SW1222 tumor than the GL261 and the LS174T tumor, see Table~\ref{tab:analysis_datasets}. In Section~\ref{sec:additional_comps} we will show that this higher variability makes it harder to match the flow from large into small vessels between the two models.
\begin{table}
	\centering
	\begin{tabular}{l c c c | c c c | c c c | c c c} 
		\hline
		 & \multicolumn{3}{c}{case $\SI{5}{\percent}$} & \multicolumn{3}{c}{case $\SI{10}{\percent}$} & \multicolumn{3}{c}{case $\SI{15}{\percent}$} & \multicolumn{3}{c}{case $\SI{20}{\percent}$} \\
		  & ${\varphi}$ & $CV_{D}$ & $CV_{\left\lvert Q\right\rvert}$  & ${\varphi}$ & $CV_{D}$ & $CV_{\left\lvert Q\right\rvert}$ & ${\varphi}$ & $CV_{D}$ & $CV_{\left\lvert Q\right\rvert}$  & ${\varphi}$ & $CV_{D}$ & $CV_{\left\lvert Q\right\rvert}$ \\
		 \hline\hline
		 SW1222 & $\num{0.29}$ & $\num{0.39}$ & $\num{2.40}$ & $\num{0.28}$ & $\num{0.45}$ & $\num{2.14}$ & $\num{0.26}$ & $\num{0.50}$ & $\num{1.77}$  & $\num{0.24}$ & $\num{0.55}$ & $\num{1.53}$   \\
		 LS174T & $\num{0.18}$ & $\num{0.23}$ & $\num{0.88}$ & $\num{0.16}$ & $\num{0.24}$ & $\num{0.85}$ & $\num{0.15}$ & $\num{0.24}$ & $\num{0.83}$ & $\num{0.13}$ &  $\num{0.24}$ & $\num{0.82}$  \\
		 GL261 & $\num{0.30}$ & $\num{0.35}$ & $\num{1.21}$ & $\num{0.29}$ & $\num{0.37}$ & $\num{1.16}$ & $\num{0.28}$ &  $\num{0.38}$ & $\num{1.13}$ & $\num{0.27}$ & $\num{0.40}$ & $\num{1.09}$
	\end{tabular}
		\caption{Analysis of connectivity between fully-resolved and homogenized part of vasculature: $\varphi$ is the fraction of nodes of larger vessels with a direct connection to smaller vessels, $CV_{D}$ and $CV_{\left\lvert Q\right\rvert}$ are measures of the variability of the diameter and flow, respectively, in the segments connecting larger and smaller vessels (data includes the mean taken over five different sets of pressure boundary conditions on the 1D network produced by the methodology described in Section~\ref{sec:bc_full}, "case ${\mathrm{X}}\,\si{\percent}$" denotes the case where ${\mathrm{X}}\,\si{\percent}$ of the 1D blood vessels are retained in the hybrid approach)}
\label{tab:analysis_conn}
\end{table}
\begin{remark}
\label{rem:dataset_vidotto}
We believe that our hybrid approach is also applicable to more organized, hierarchical networks as, for example, the topology used by Vidotto et al.\cite{Vidotto2018} In this publication the network was partitioned by a radius-based threshold, see Figure 1 therein. The larger vascular structures contain very short branches going away from the main vessels. At the tips of these short branches, the node-based coupling is performed. If one instead removed these very short branches and left only the major, flow-carrying vessels in $\lbig$, a line-based coupling along these vessels could again be implemented.
\end{remark}
\subsection{Determination of representative elementary volume size}
\label{sec:rev}
\setlength\figureheight{0.15\textwidth}
\setlength\figurewidth{0.35\textwidth}
\begin{figure}
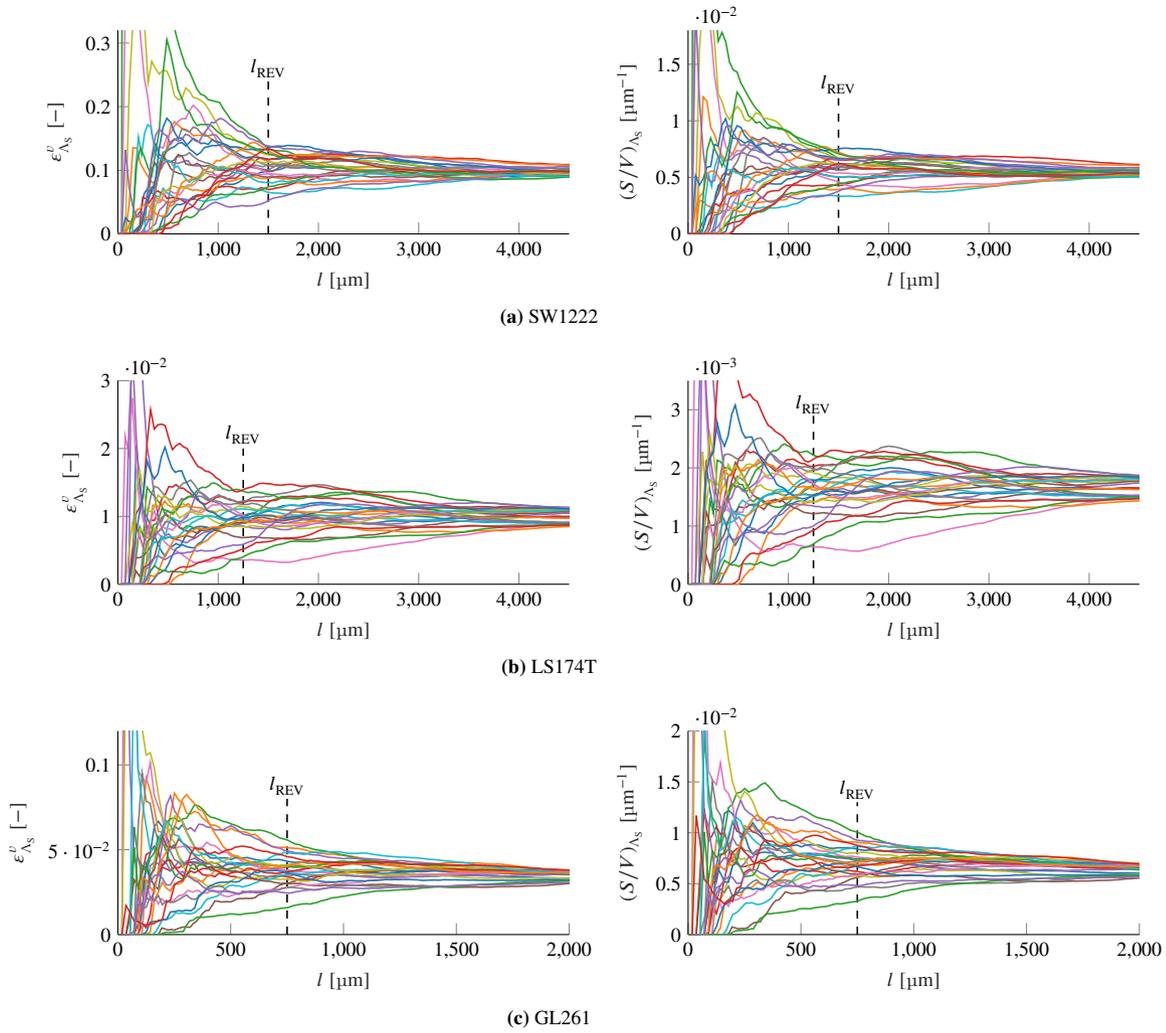

\subfloat[SW1222\label{fig:SW1222_rev}]{
\footnotesize
     \hspace{2cm}\input{./figures/SW1222_rev_vf.tikz}\hspace{1.5cm}
     \input{./figures/SW1222_rev_sv.tikz}}\\
\subfloat[LS174T\label{fig:LS174T_rev}]{
\footnotesize
     \hspace{2cm}\input{./figures/LS174T_rev_vf.tikz}\hspace{1.5cm}
     \input{./figures/LS174T_rev_sv.tikz}}\\
\subfloat[GL261\label{fig:GL261_rev}]{
\footnotesize
     \hspace{2cm}\input{./figures/GL261_rev_vf.tikz}\hspace{1.5cm}
     \input{./figures/GL261_rev_sv.tikz}}
\caption{Determination of representative elementary volume (REV) size -- evolution of blood vessel volume fraction $\porosity^v_{\lsmall}$ and surface-to-volume ratio $(S/V)_{\lsmall}$ of smaller blood vessels $\lsmall$ is shown for increasing possible REV sizes}
\label{fig:rev}
\end{figure}
The existence of a resentative elementary volume (REV) is an important concept for different homogenization procedures.\cite{Davit2013}  In general, such a volume should be big enough to smooth out fluctuations of spatial heterogeneities yet small enough to resolve the physical effects of interest. In this section, we investigate the choice of REVs in the context of our model and the employed data sets. Naturally, we will investigate the properties of the smaller vessels $\lsmall$ in the following since this is the part of the vasculature which is homogenized and treated as a porous continuum in the hybrid approach. Furthermore, five different sets of pressure boundary conditions on the 1D network are studied. This is again due to the fact that different pressure boundary conditions on the 1D network will lead to different flow patterns in the vascular network and, therefore, also different sets $I_{\mathrm{L}}$ and $I_{\mathrm{S}}$ of large and small vessels (potentially with a different distribution throughout the domain) with the employed flow-based criterion.  

For this purpose, we have devised the following procedure: 
\begin{enumerate}
\item For each network topology we create five different cases with a different set of pressure boundary conditions on the 1D network for the fully-resolved model as described in Section~\ref{sec:bc_full}. 
\item We partition all cases into the distinct sets of large and small vessels as described in Section~\ref{sec:diff_fully_hyb}. We here investigate the case $\SI{10}{\percent}$ for all different topologies but equivalent results have been obtained for leaving the top $\SI{5}{\percent}$, $\SI{15}{\percent}$ or $\SI{20}{\percent}$ of vessels with the largest flow in the system. 
\item We select random positions in the vasculature domain $\Omega_v$ in the range $\left[x_{min}+0.15\cdot l_x,x_{max}-0.15\cdot l_x\right]$, $\left[y_{min}+0.15\cdot l_y,y_{max}-0.15\cdot l_y\right]$ and $\left[z_{min}+0.15\cdot l_z,z_{max}-0.15\cdot l_z\right]$, where $l_i$ denotes the domain lengths in the respective coordinate directions and $(\cdot)_{min}$ and $(\cdot)_{max}$ the minimum and maximum coordinate value in each direction in $\Omega_v$. In this way, the random positions are chosen such that they do not lie too close to the boundaries of the domain.
\item For each of the random positions within the domain we define a cube with edge length $l_{edge}=1/300\cdot max(l_x,max(l_y,l_z))$. The random position is chosen as the center of that cube.
\item The size of the cubes is successively increased in all coordinate directions by $l_{edge}$ while keeping their centers fixed. The blood vessel volume fraction $\porosity^v_{\lsmall}$ and the surface-to-volume ratio $(S/V)_{\lsmall}$ of smaller blood vessels $\lsmall$ is computed for each cube at each size. If a cube protrudes from the domain during this enlargement, these quantities are calculated on the intersection of the cube with the domain $\Omega_v$.
\end{enumerate}
Per case with different boundary conditions this is performed for ten randomly generated cube centers. The results are shown in Figure~\ref{fig:rev} for only three REVs per boundary condition case, that is, in total 15 cases to not clutter the plots. The evolution of the blood vessel volume fraction $\porosity^v_{\lsmall}$ and of the surface-to-volume ratio $(S/V)_{\lsmall}$ of the smaller blood vessels $\lsmall$ for increasing the edge length of the cubes is illustrated. Therein, we denote the length scale as $l=\sqrt[3]{V_{cube\,\cap\,\Omega_v}}$ to account for cases when a larger cube protrudes from the domain $\Omega_v$. All three topologies exhibit similar features: $\porosity^v_{\lsmall}$ and $(S/V)_{\lsmall}$ fluctuate strongly for smaller lengths. Then, most curves stabilize and remain almost stationary while increasing the size of the averaging volume. Finally, for even larger volumes the curves slowly converge to the values of these quantities across the entire domain. This behaviour can be expected in porous media\cite{Davit2013} and we consequently define the length of the REVs $l_{\mathrm{REV}}$ at the point where the initial oscillations of too small control volumes fade out and the values stabilize.

\begin{figure}
     \centering
     \subfloat[][SW1222, $n_{\mathrm{REV}}=78$]{
     \includegraphics[trim={12.0cm 0.0cm 7.5cm 0.0cm},clip,width=0.32\linewidth]{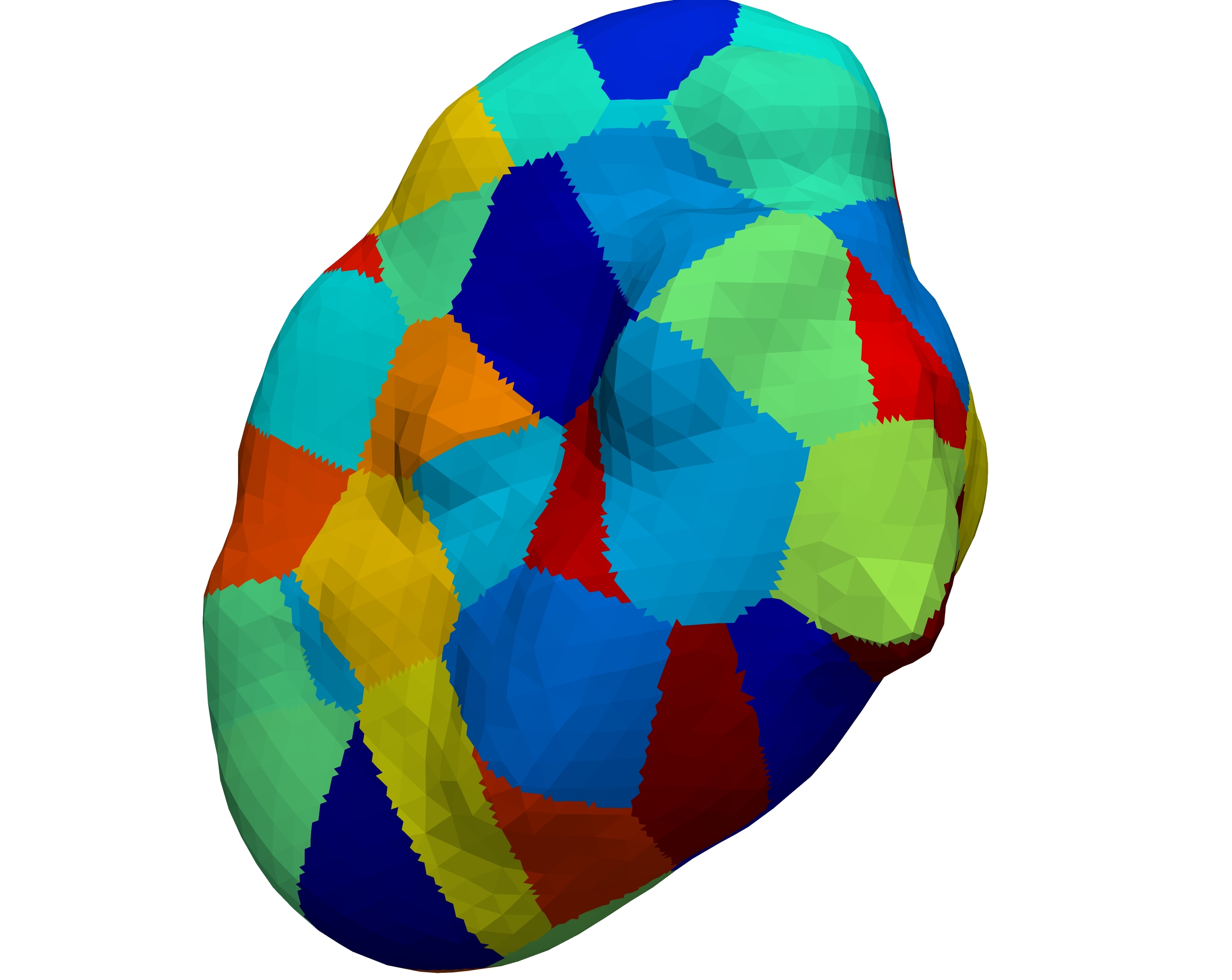}
     } \label{fig:networks_SW1222_rev} \hfill
     \subfloat[][LS174T, $n_{\mathrm{REV}}=114$]{
     \includegraphics[trim={12.0cm 0.0cm 7.5cm 0.0cm},clip,width=0.32\linewidth]{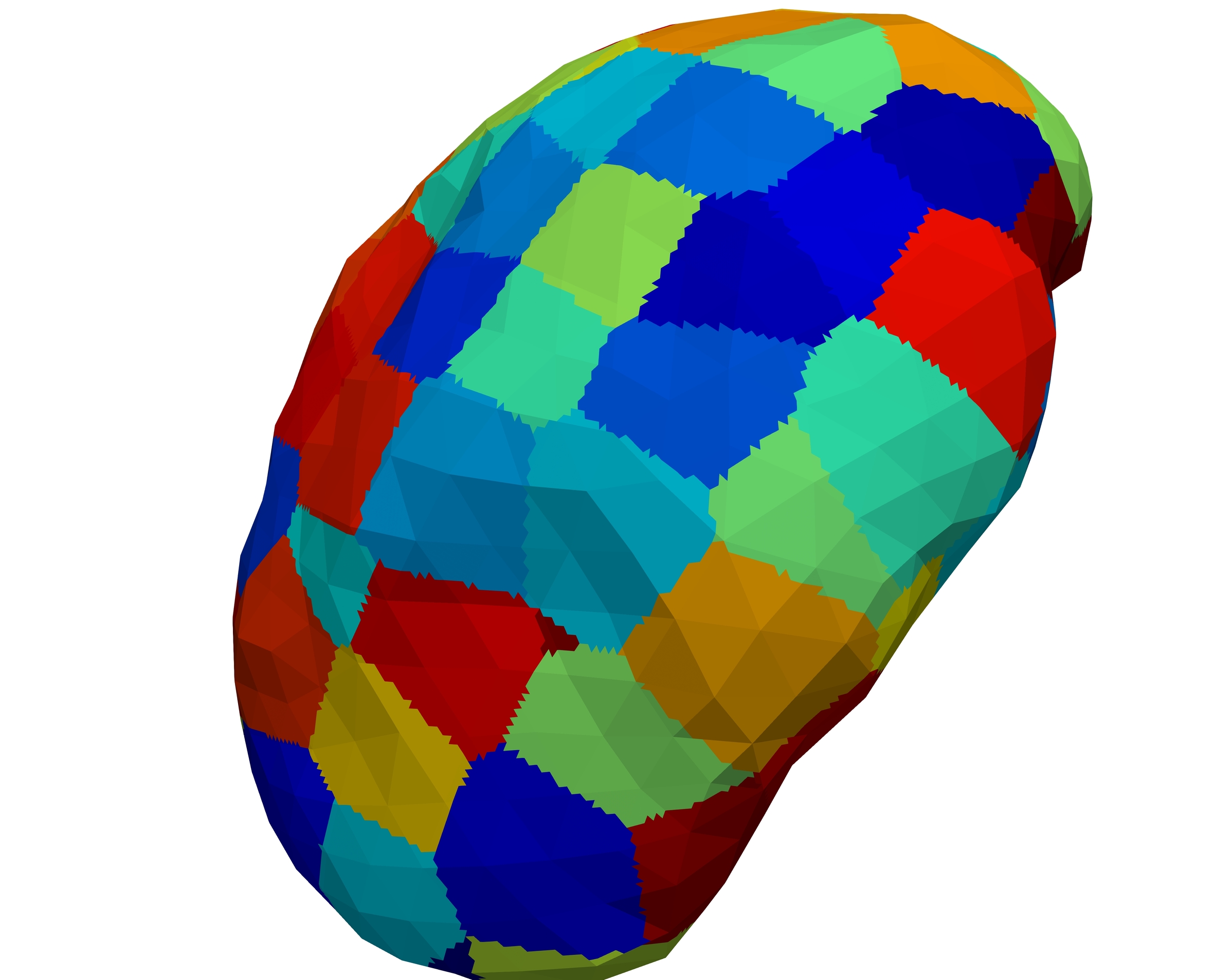}
     } \label{fig:networks_LS174T_rev} \hfill
     \subfloat[][GL261, $n_{\mathrm{REV}}=55$]{
     \includegraphics[trim={12.0cm 0.0cm 7.5cm 0.0cm},clip,width=0.32\linewidth]{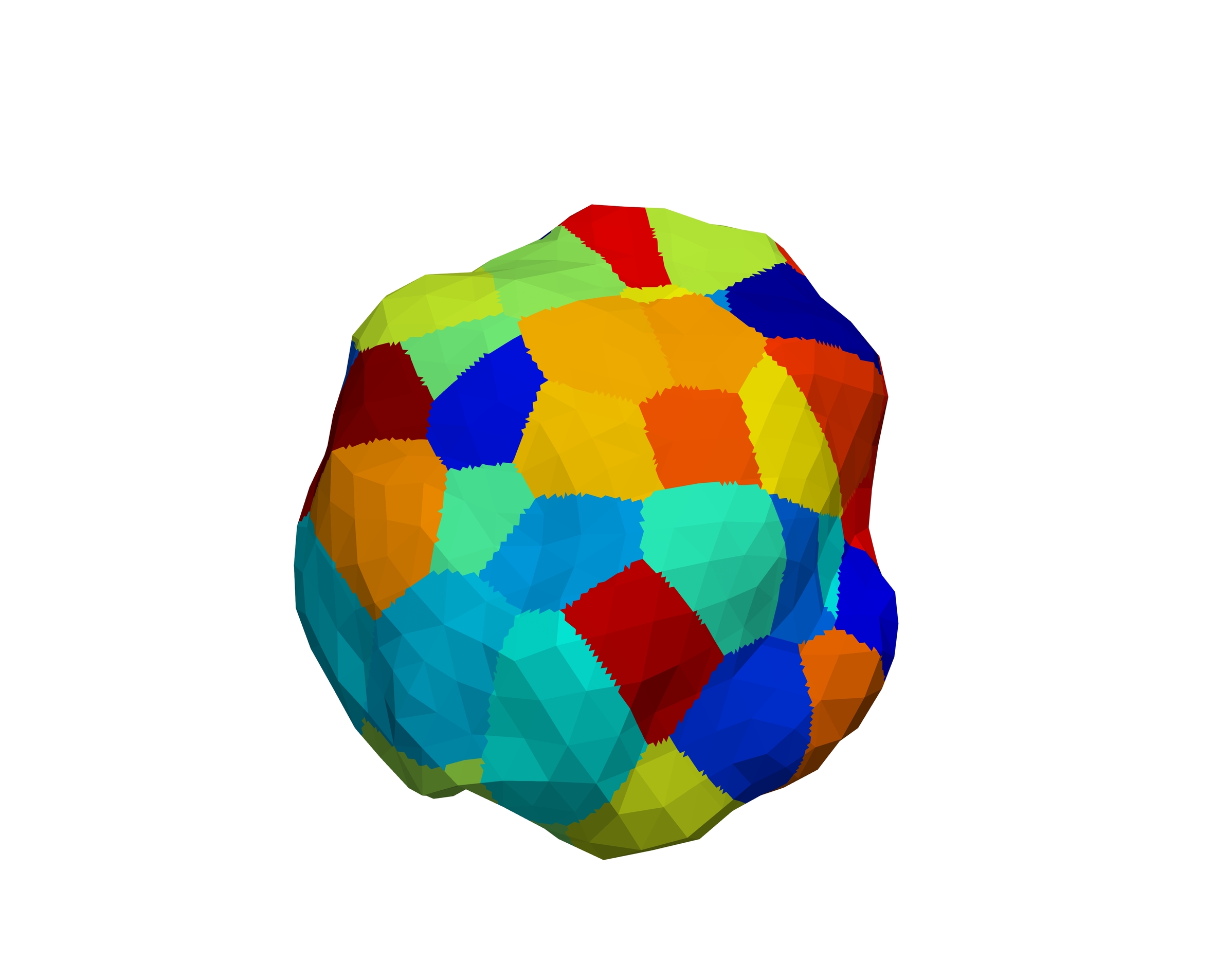}
     } \label{fig:networks_GL261_rev}
     \caption{Representative elementary volumes of all three tumor domains}
     \label{fig:networks_rev}
\end{figure}

Splitting the domains into these REVs of equal size is not an easy task due to their irregular, elliptic shape. We first defined a regular grid of REV centers and performed an initial Voronoi tesselation based on this grid. Due to the shape of the domain this resulted in too small or too large REVs. Therefore, we performed an optimization of the Voronoi tesselation where the objective function had the goal to define REVs of equal volume and equal dimensions. The resulting REVs are visualized in Figure~\ref{fig:networks_rev}. The mean deviation of the REVs from the previously determined volume and lengths from Figure~\ref{fig:rev} in all three coordinate directions is less than $\SI{5}{\percent}$ for the domains of all three tumor types.

\begin{figure}
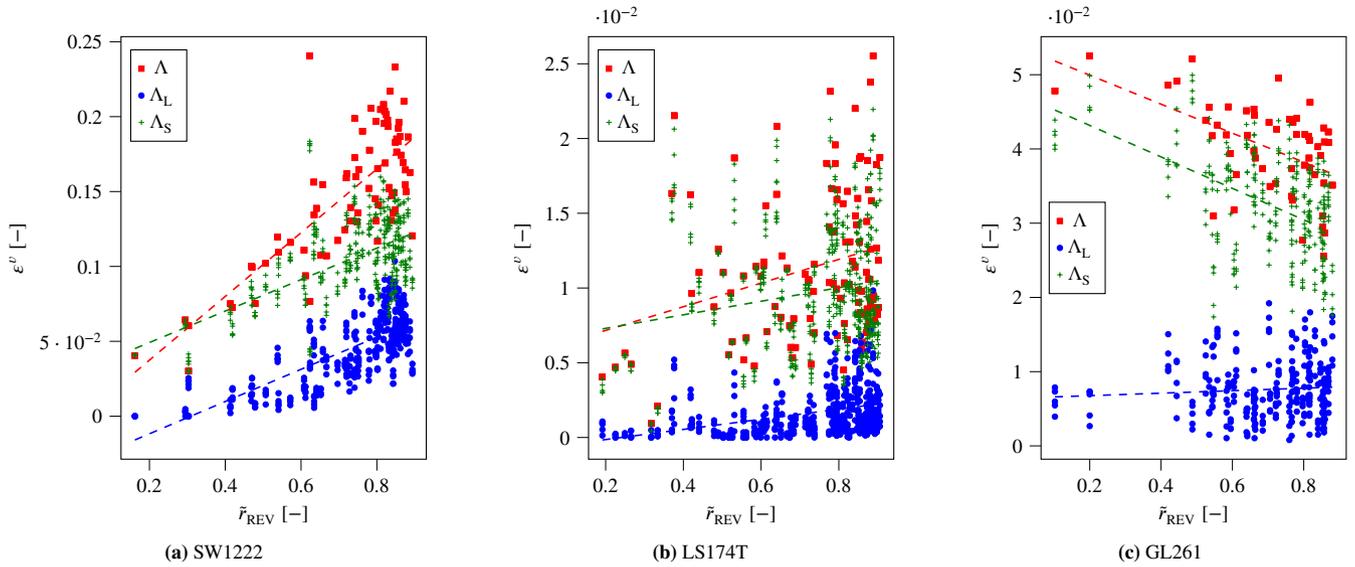

\subfloat[SW1222\label{fig:SW1222_vf_over_rad}]{
\footnotesize
     \input{./figures/SW1222_vf_over_rad.tikz}}\hfill
\subfloat[LS174T\label{fig:LS174T_vf_over_rad}]{
\footnotesize
     \input{./figures/LS174T_vf_over_rad.tikz}}\hfill
\subfloat[GL261\label{fig:GL261_vf_over_rad}]{
\footnotesize
     \input{./figures/GL261_vf_over_rad.tikz}}
\caption{Dependency of volume fraction of big vessels $\lbig$, small vessels $\lsmall$  and entire vasculature $\Lambda$ over non-dimensionalized radial distance from center of domain (data is taken from five different sets of 1D blood pressure boundary conditions if $\SI{10}{\percent}$ of 1D blood vessels are retained in hybrid model for each network structure, dashed lines indicate linear least squares fits)}
\label{fig:vf_over_rad}
\end{figure}

Finally, we employ these REVs to study the distribution of blood vessels inside the domain. For that, we define the non-dimensionalized radial distance of each REV $\tilde{r}_{\mathrm{REV}}$ as the distance of the center of the REV to the center of the domain divided by the distance of the center of the domain to the tumor hull in direction of the center of the REV. Again, this analysis is performed for all three tumor types for five different sets of pressure boundary conditions on the 1D network since those influence the flow in the 1D vasculature and, consequently, also the composition of $\lbig$ and $\lsmall$ as previously mentioned. The results for the volume fraction of big vessels $\porosity^v_{\lbig}$, small vessels $\porosity^v_{\lsmall}$ and the entire vasculature $\porosity^v_{\Lambda}$ are shown in Figure~\ref{fig:vf_over_rad}. The clearest structure is evident for the the SW1222 case: towards the tumor hull, $\porosity^v_{\lsmall}$ and $\porosity^v_{\lbig}$ and, thus, also the sum of the former two, $\porosity^v_{\Lambda}$, gradually increase. Close to the center of the domain, there is still a significant amount of smaller blood vessels while almost no larger blood vessels are present. This is consistent with experimental data showing higher blood vessel density and perfusion in the tumor periphery\cite{Desposito2018,Forster2017} with only a few major vessels penetrating into the center of the tumor.\cite{Holash1999a} These trends are also present in the LS174T tumor, albeit, far less pronounced than for the SW1222 tumor. By contrast, the GL261 vascular network shows a completely different behaviour. While the vascular density of large vessels remains almost constant over the tumor radius, the one of the smaller blood vessels $\lsmall$ drops and, thus, also the overall volume fraction $\porosity^v_{\Lambda}$.
\setlength\figureheight{0.4\textwidth}
\setlength\figurewidth{0.33\textwidth}
\begin{remark}
\label{rem:real_life}
The validity of the obtained topologies and distributions for $\lbig$ and $\lsmall$ and the applicability of the proposed hybrid approach is supported by state-of-the art optoacoustic \textit{in-vivo} imaging techniques.\cite{Li2020a} The currently attainable spatial resolution is less than \SI{50}{\micro\m} throughout the tumor domain which is in the range of the diameter of larger vessels from Table~\ref{tab:analysis_radius}. Furthermore, the larger vessels which are retained in the hybrid model are more concentrated at the tumor periphery (at least for the SW1222 and LS174T case) and are, thus, more accessible to imaging. Qualitatively, the topology of the larger vessels from Figure~\ref{fig:SW1222_largest_flow} is in good agreement with corresponding optoacoustic imaging data from tumors\cite{Li2020a} where larger feeding vessels are visible at the tumor rim. From these experiments, one can extract a similar topology of larger vessels $\lbig$ to apply our hybrid model. A connected subgraph of larger peripheral feeder blood vessels as in Figure~\ref{fig:SW1222_largest_flow} rather than single clusters as in Figure~\ref{fig:SW1222_largest_rad} is attainable. Hence, we conclude that the employed methodology of splitting into larger and smaller vessels yields a valid scenario resembling real experimental data and can, therefore, be used to investigate our hybrid embedded/homogenized approach for solid tumor perfusion.
\end{remark}
\section{Numerical experiments}
\label{sec:num_ex}
In this section we perform several numerical experiments to evaluate the performance of the hybrid model in comparison to the fully-resolved one. We first define a comparison metric and optimize the parameters of the hybrid model such that the best possible correspondence between the models is achieved according to this metric. Subsequently, we study several other quantities to compare the two models and present a further improvement of the hybrid model via a vascular volume fraction dependent permeability for the homogenized vessels.

The tumor hull is smoothed and triangulated using Gmsh (version 4.4.1)\cite{Gmsh} and its enclosed volume is meshed with linear tetrahedral elements using Trelis 17.0.\cite{Trelis} An exemplary mesh for the SW1222 topology is shown in Figure~\ref{fig:mesh} and parameters of the 3D mesh are given in Table~\ref{tab:analysis_datasets}. Note that the 3D mesh is completely independent of the discretization of the 1D networks, that is, the nodes of the two meshes do not match which is an advantageous feature provided by our recently introduced hybrid approach.\cite{Kremheller2019} Both the fully-resolved and the hybrid FEM model have been implemented in the in-house parallel multi-physics research code BACI.\cite{baci} Parameters for both models are listed in Table~\ref{tab:params}.
\begin{remark}
\label{rem:choice_penalty}
In preliminary simulations, we determined the proper range for the penalty parameter $\epsilon$. As a compromise between accuracy and a well-conditioned system matrix, we defined the following criterion:
\begin{equation}
\label{eq:crit_penalty}
\delta=\frac{1}{n_{{\mathrm{nodes}},\lbig}}\sum_{i=1}^{n_{{\mathrm{nodes}},\lbig}}\frac{\left|\bsd{\kappa}^{-1}[i,i]\bsd{g}[i]\right|}{\bsd{p}^{\hat{v}}[i]}< \SI{1}{\percent}.
\end{equation}
This rule states that the mean relative pressure error $\delta$ in terms of the length-independent nodal pressure difference vector $\bsd{\kappa}^{-1}\bsd{g}$ and the nodal pressure $\bsd{p}^{\hat{v}}$ in the 1D network is less than $\SI{1}{\percent}$. The values for all cases are given in Table~\ref{tab:params}. Note that the penalty parameter has units of $[\mathrm{length}]^2/[\mathrm{time}\cdot\mathrm{pressure}]$ such that the LM field represents a 1D-3D mass transfer term, or volumetric flow per length. This allows interpreting the penalty parameter as very large permeability governing the mass transfer between resolved and homogenized vasculature in the hybrid model.
\end{remark}
\begin{remark}
\label{rem:comp_performance}
In our opinion, the main advantage of the hybrid model is not a reduction of computational cost compared to the full model, but the fact that it relies only on data available through non-invasive imaging. Nevertheless, we also did a first preliminary evaluation comparing the computational costs of the two models and found that the hybrid model was not significantly faster than the fully-resolved one and in some cases even slower. The effort for finding 1D-3D elements interacting with each other, building the integration segments and evaluating the coupling terms along the 1D vasculature is obviously smaller for the hybrid model since less 1D vessels are present. However, this is balanced or even outweighed by its increased effort in several other aspects: The evaluation of the 3D elements is more costly since two equations per node (in $\Omega_v$) have to be evaluated, the system size, which is dominated by the number of 3D nodes, and, thus, the linear solver time is increased and the condition of the system is worse compared to the fully-resolved case due to the penalty approach, which in turn raises the linear solver time. However, for all our studies we used the same 3D meshes for both hybrid and full model. The cost for the hybrid model could be greatly reduced by employing a coarser 3D mesh. Vidotto et al.\cite{Vidotto2018} showed that this still gave acceptable results in terms of REV pressures for their approach.
\end{remark}
\begin{threeparttable}
	\centering
	\begin{tabular}{l l l l l l} 
		\hline
		Quantity & Symbol & Value & Unit & Reference & Eqns.  \\
		\hline\hline
		Density of blood & $\rho^{\hat{v}}$, $\rho^v$ & $\num{1060}$ & $\si{\kg\per\cubic\m}$ & \cite{Formaggia2009} & \eqref{eq:1D_blood_flow},\eqref{eq:1D_blood_flow_large},\eqref{eq:mass_homo_vasc} \\
		Viscosity of blood & $\mu^{\hat{v}}$ & \tnote{[a]} & $\si{\Pa\s}$ & \cite{Pries2005} & \eqref{eq:1D_blood_flow},\eqref{eq:1D_blood_flow_large} \\
		Density of interstitial fluid and blood plasma & $\rho^{l}$ & $\num{1000}$ & $\si{\kg\per\cubic\m}$ & known & \eqref{eq:mass_IF},\eqref{eq:starling},\eqref{eq:M_leak},\eqref{eq:mass_IF_homo} \\
		Hydraulic conductivity of interstitial fluid & $k^{l}/\mu^l$ & $\num{1.2782e-1}$ & $\si{\square\micro\m\per\pascal\per\second}$ & \cite{Boucher1998} & \eqref{eq:mass_IF},\eqref{eq:mass_IF_homo} \\
		Hydraulic conductivity for transvascular flow & $L_{p,\hat{v}}$,$L_{p,v}$ & $\num{2.1e-5}$ & $\si{\micro\m\per\Pa\per\s}$ & \cite{Baxter1989} & \eqref{eq:starling},\eqref{eq:M_leak} \\
		Oncotic reflection coefficient & $\sigma$ & $0.82$ & - & \cite{Sweeney2019} & \eqref{eq:starling},\eqref{eq:M_leak} \\
		Oncotic pressure of blood & $\pi^b$ & $2666.4$ & $\si{\Pa}$ & \cite{Sweeney2019} & \eqref{eq:starling},\eqref{eq:M_leak} \\
		Oncotic pressure of interstitial fluid & $\pi^l$ & $1999.8$ & $\si{\Pa}$ & \cite{Sweeney2019} & \eqref{eq:starling},\eqref{eq:M_leak} \\
		Hydraulic conductivity of vasculature & $k^{v}/\mu^v$ & see Table~\ref{tab:results} & $\si{\square\micro\m\per\pascal\per\second}$ & - & \eqref{eq:M_leak} \\
		Surface-to-volume ratio for transvascular flow & $(S/V)_{\lsmall}$ & see Table~\ref{tab:results} & $\si{\per\micro\m}$ & - & \eqref{eq:M_leak} \\
		\hline
		Penalty parameter & $\epsilon$ &  & $\si{\square\micro\m\per\pascal\per\second}$  & Remark~\ref{rem:choice_penalty} & \eqref{eq:lambda} \\
		\quad\quad SW1222: case $\SI{5}{\percent}$ & & $\num{400}$ & & & \\
		\quad\quad SW1222: case $\SI{10}{\percent}$, $\SI{15}{\percent}$, $\SI{20}{\percent}$ & & $\num{100}$ & & & \\
		\quad\quad LS174T: case $\SI{5}{\percent}$ & & $\num{100}$ & & & \\
		\quad\quad LS174T: case $\SI{10}{\percent}$, $\SI{15}{\percent}$, $\SI{20}{\percent}$ & & $\num{50}$ & & & \\
		\quad\quad GL261: case $\SI{5}{\percent}$ & & $\num{100}$ & & & \\
		\quad\quad GL261: case $\SI{10}{\percent}$, $\SI{15}{\percent}$, $\SI{20}{\percent}$ & & $\num{50}$ & & &
	\end{tabular}
	\begin{tablenotes}
	\item[{[a]}] The value for blood viscosity is calculated separately in each 1D element using the algebraic relationship of Pries and Secomb~\cite{Pries2005} with hematocrit value fixed to $0.45$.
	\end{tablenotes}
\caption{Parameters and values}
\label{tab:params}
\end{threeparttable}
\subsection{Definition of metric for comparison of the two models}
\label{sec:comp_models}
To assess the performance of the hybrid model in predicting microvascular flow and IF pressure inside solid tumors in comparison with the fully-resolved model, a suitable metric is warranted. Ideally, the hybrid model should match the fully-resolved one in terms of blood and IF pressure as well as blood and IF flow to obtain an accurate representation of the perfusion through the tumor. Therefore, we define our metric as a combination of these quantities. The first contribution is the correspondence of blood pressures in the large vessels $\lbig$ which are present both in the fully-resolved and the hybrid model. We define the \textit{coefficient of determination} $R^2$ in terms of nodal blood pressures in the large vessels between the two models as 
\begin{equation}
\label{eq:r_squared_large}
R^2_{\mathrm{L}}=1-\frac{\sum_{i=1}^{n_{{\mathrm{nodes}},\lbig}}\left(\left.\bsd{p}^{\hat{v}}\left[i\right]\big|\right._{\mathrm{full}}-\left.\bsd{p}^{\hat{v}}\left[i\right]\big|\right._{\mathrm{hyb}}\right)^2}{\sum_{i=1}^{n_{{\mathrm{nodes}},\lbig}}\left(\left.\bsd{p}^{\hat{v}}\left[i\right]\big|\right._{\mathrm{full}}-\mu_{I_{\mathrm{L}}}\left(\left.\bsd{p}^{\hat{v}}\big|\right._{\mathrm{full}}\right)\right)^2},
\end{equation}
where $\mu_{I_{\mathrm{L}}}\left(\left.\bsd{p}^{\hat{v}}\big|\right._{\mathrm{full}}\right)$ is the mean blood pressure in the large vessels of the fully-resolved model. A value of $R^2=1$ would mean a perfect correspondence of both models while smaller values suggest larger deviations. A negative $R^2$ indicates that the hybrid model performs worse than simply taking the mean value of the fully-resolved model. The second contribution to our metric is the correspondence of blood pressures in the small vessels $\lsmall$ between the fully-resolved and the hybrid model, which we calculate as
\begin{equation}
\label{eq:r_squared_small}
R^2_{\mathrm{S}}=1-\frac{\sum_{i=1}^{n_{{\mathrm{nodes}},\lsmall}}\left(\left.\bsd{p}^{\hat{v}}\left[i\right]\big|\right._{\mathrm{full}}-\left.\bsd{p}^{v}\left(\bsd{X}\left[i\right]\right)\big|\right._{\mathrm{hyb}}\right)^2}{\sum_{i=1}^{n_{{\mathrm{nodes}},\lsmall}}\left(\left.\bsd{p}^{\hat{v}}\left[i\right]\big|\right._{\mathrm{full}}-\mu_{I_{\mathrm{S}}}\left(\left.\bsd{p}^{\hat{v}}\big|\right._{\mathrm{full}}\right)\right)^2}.
\end{equation}
Since the smaller vessels $\lsmall$ are not retained in the hybrid model, we compare nodal blood pressures in the smaller vessels of the fully-resolved model with the homogenized blood pressure field $\bsd{p}^{v}$ of the hybrid model evaluated at the nodal positions $\bsd{X}\left[i\right]$ of the smaller vessels. Again, this is formulated in terms of a coefficient of determination, now involving all nodes in the small vessels and $\mu_{I_{\mathrm{S}}}\left(\left.\bsd{p}^{\hat{v}}\big|\right._{\mathrm{full}}\right)$ is the mean blood pressure in the small vessels of the fully-resolved model.

Equivalently, the coefficient of determination of the IF pressure is given by
\begin{equation}
\label{eq:r_squared_IF}
R^2_{\mathrm{IF}}=1-\frac{\sum_{i=1}^{n_{{\mathrm{nodes}},\Omega}}\left(\left.\bsd{p}^{l}\left[i\right]\big|\right._{\mathrm{full}}-\left.\bsd{p}^{l}\left[i\right]\big|\right._{\mathrm{hyb}}\right)^2}{\sum_{i=1}^{n_{{\mathrm{nodes}},\Omega}}\left(\left.\bsd{p}^{l}\left[i\right]\big|\right._{\mathrm{full}}-\mu\left(\left.\bsd{p}^{l}\big|\right._{\mathrm{full}}\right)\right)^2}
\end{equation}
with the mean IF pressure $\mu\left(\left.\bsd{p}^{l}\big|\right._{\mathrm{full}}\right)$ of the full model in the tissue domain $\Omega$. Instead of the point-wise comparison of pressures in~\eqref{eq:r_squared_small} and~\eqref{eq:r_squared_IF}, one could also compare mean REV pressures of the two models. We will additionally calculate and compare mean (blood and IF) pressures inside the REVs in Section~\ref{sec:additional_comps}. With the previous three equations, the metrics for blood and IF pressure have been defined. Also the flow in the larger vessels $\lbig$ is covered since larger vessels present in both models have the same diameter, length and blood viscosity. Therefore, if the nodal pressures match, also the flow between the nodes, i.e., inside the elements is identical. The same applies for flow in the IF if the same 3D mesh and hydraulic conductivity $k^l/\mu^l$ is employed in both models which we will assume hereafter. What is still missing, is a metric for comparison of blood flow inside the smaller blood vessels $\lsmall$ which are homogenized in the hybrid model. We define this measure as follows
\begin{equation}
\label{eq:r_squared_flow}
R^2_{{\mathrm{flow}},\lsmall}=1-\frac{\sum_{i=1}^{n_{{\mathrm{REV}}}}\sum_{j=1}^{3}\left(\left.Q_j^{\hat{v}}\big|\right._{\lsmall,{\mathrm{full}}}-\left.Q_j^{v}\big|\right._{\lsmall,{\mathrm{hyb}}}\right)^2}{\sum_{i=1}^{n_{{\mathrm{REV}}}}\sum_{j=1}^{3}\left(\left.Q_j^{\hat{v}}\big|\right._{\lsmall,{\mathrm{full}}}-\mu\left(\left.Q^{\hat{v}}\big|\right._{\lsmall,{\mathrm{full}}}\right)\right)^2},
\end{equation}
i.e., for each of the $n_{{\mathrm{REV}}}$ REVs we compare the volumetric flow $Q_j$ of the fully-resolved and the hybrid model in all three coordinate directions and compare it with each other via the coefficient of determination of flow in the smaller vessels $R^2_{{\mathrm{flow}},\lsmall}$. 

Next, we will detail how we calculate the flows in the REVs in both models. In the center of each REV we define a square $\Box_j$ with dimensions $l_{\mathrm{REV}}\times l_{\mathrm{REV}}$ such that its normal $\bs{n}_j$ is aligned with coordinate direction $j$. The volumetric flow in the homogenized part of the vasculature in coordinate direction $j$ is then given by
\begin{equation}
\label{eq:volflow_homo}
\left.Q_j^{v}\big|\right._{\lsmall,{\mathrm{hyb}}}=\int_{\Box_j}-\frac{k^v}{\mu^v}\bs{n}_j\cdot \grad p^v\,\mathrm{dA}
\end{equation}
as the surface integral of the flux through the square. For the fully-resolved model, we define it as
\begin{equation}
\label{eq:volflow_full}
\left.Q_j^{\hat{v}}\big|\right._{\lsmall,{\mathrm{full}}}=\sum_{\Box_j\cap \lsmall}-\frac{\pi R^4}{8\mu^{\hat{v}}}\frac{\partial p^{\hat{v}}}{\partial s}\cdot \sgn\left(\bs{t}\cdot \bs{n}_j\right),
\end{equation}
which is the sum of the volumetric flow of all segments which are part of the smaller vessels and cut by the square $\Box_j$. Therein, $\bs{t}$ is the tangential vector of a segment pointing from its first to its second node and $\sgn\left(\cdot\right)$ denotes the sign function.

Finally, we define the total coefficient of determination between the two models as the sum of the contributions from blood pressure in large vessels~\eqref{eq:r_squared_large}, blood pressure in small vessels~\eqref{eq:r_squared_small}, IF pressure~\eqref{eq:r_squared_IF} and flow in small vessels~\eqref{eq:r_squared_flow} as
\begin{equation}
\label{eq:r_squared_tot}
R^2_{\mathrm{tot}}=\frac{1}{4}\left(R^2_{\mathrm{L}}+R^2_{\mathrm{S}}+R^2_{\mathrm{IF}}+R^2_{{\mathrm{flow}},\lsmall}\right).
\end{equation}
This metric, where all four contributions are weighted equally, will be employed to study the accuracy of the hybrid model w.r.t.\ the full model and to find the optimal parameters of the hybrid model.
\subsection{Optimization of parameters of the hybrid model}
\label{sec:optimization}
Compared to the fully-resolved model, the hybrid one has two additional parameters, which are the hydraulic conductivity of the homogenized vessels $k^{v}/\mu^v$ in~\eqref{eq:mass_homo_vasc} governing blood flow and the surface-to-volume ratio $(S/V)_{\lsmall}$ accounting for transvascular flow from the homogenized vessels into the IF in~\eqref{eq:M_leak}. Our goal in this section is to determine these parameters such that the agreement in terms of blood flow and blood and IF pressures of the hybrid model with the fully-resolved model is maximized. For that we employ the total coefficient of variation~\eqref{eq:r_squared_tot} between the two models deduced in the previous section and formulate the following optimization problem in terms of the parameters of the hybrid model:
\begin{equation}
\label{eq:optimization}
\argmax\limits_{k^{v}/\mu^v,\,(S/V)_{\lsmall}}R^2_{\mathrm{tot}},
\end{equation}
that is, we aim to find the parameters of the hybrid model, for which the correspondence of the two models is optimized. With these optimal parameters we can then evaluate the accuracy of the hybrid model w.r.t.\ the fully-resolved one. For the optimization procedure, we parallelized the least-squares method of the SciPy package (version 1.5.2)\cite{scipy} and interfaced it to the software framework QUEENS.\cite{queens} Internally, SciPy employs the Levenberg-Marquardt algorithm to solve the nonlinear least-squares problem~\eqref{eq:optimization}. Derivatives of the metric~\eqref{eq:r_squared_tot} w.r.t.\ the parameters are approximated using forward finite differences. This implies that the hybrid model has to be solved three times per iteration step. In preliminary simulations, we confirmed that different initial conditions (from a sensible parameter range) converged to the same optimum.

Since the full topology of the vasculature is available, we could also obtain these parameters by a suitable homogenization procedure for the permeability as previously done for other hybrid or continuum models.\cite{Penta2015,Penta2015a,Mascheroni2017b,Shipley2019,Vidotto2018} We did not follow this approach here for the following reasons: First, the chaotic structure of the blood vessel network implying also a very chaotic blood flow pattern typical for the solid tumors would make this very challenging. Second, we want to create a best-case scenario by fitting the parameters of the hybrid model such that possible errors introduced by a homogenization scheme are minimal.

The general algorithm can be described as follows: 
\begin{enumerate}
\item Obtain a set of boundary conditions for the full model as described in Section~\ref{sec:bc_full} and solve the full model to generate a reference solution.
\item Extract the topology of larger vessels for the hybrid model from the full model, conf.~Section~\ref{sec:diff_fully_hyb}, and apply boundary conditions on the hybrid model, conf.~Section~\ref{sec:bc_hybrid}.
\item Find the optimal parameters of the hybrid model by maximizing the total coefficient of variation~\eqref{eq:optimization}. During the optimization procedure repeated evaluations of the hybrid model with different parameters are required.
\end{enumerate}
\setlength\figureheight{0.25\textwidth}
\setlength\figurewidth{0.25\textwidth}
\begin{figure}
\centering
\subfloat[][$R^2_{\mathrm{L}}=0.983$\label{fig:r_squared_art_press}]{\footnotesize
%
%
\begin{tikzpicture}[trim axis left,trim axis right]

\begin{axis}[%
width=\figurewidth,
height=\figureheight,
at={(0\figurewidth,0\figureheight)},
scale only axis,
xmin=2000,
xmax=6000,
ymin=2000,
ymax=6000,
axis on top,
axis background/.style={fill=white},
axis x line*=bottom,
axis y line*=left,
xlabel={$\left.\bsd{p}^{\hat{v}}\big|\right._{\rm full}\,[\si{Pa}]$},
ylabel={$\left.\bsd{p}^{\hat{v}}\big|\right._{\rm hyb}\,[\si{Pa}]$},
legend style={legend cell align=left, align=left, draw=white!15!black}
]
\addplot [plot graphics/node/.append ]
graphics [xmin=2000, xmax=6000, ymin=2000, ymax=6000] {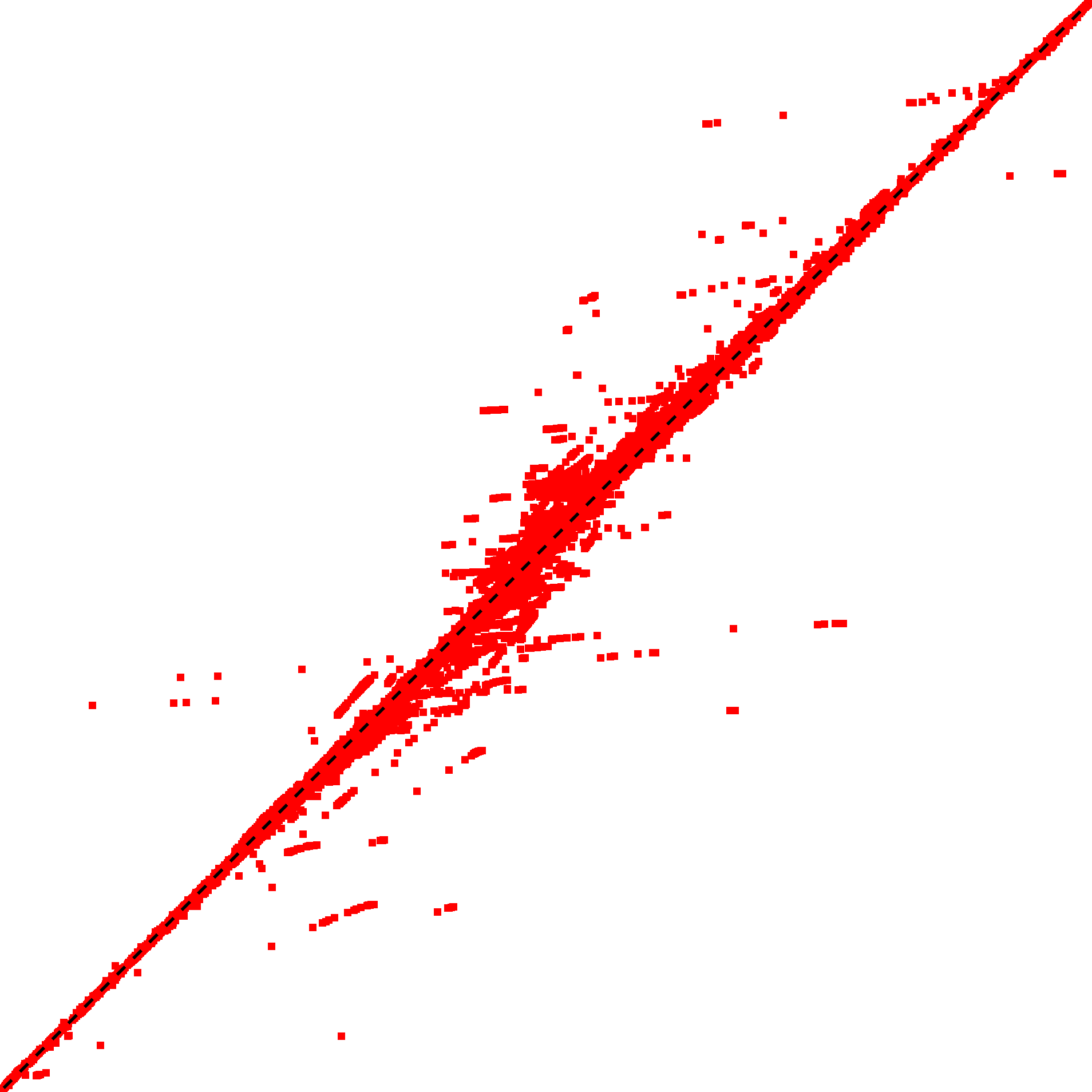};
\end{axis}
\end{tikzpicture}%
\hspace{1.5cm}}
\subfloat[][$R^2_{\mathrm{S}}=0.670$\label{fig:r_squared_homo_press}]{\footnotesize
%
%
\begin{tikzpicture}[trim axis left,trim axis right]

\begin{axis}[%
width=\figurewidth,
height=\figureheight,
at={(0\figurewidth,0\figureheight)},
scale only axis,
xmin=2000,
xmax=6000,
ymin=2000,
ymax=6000,
axis on top,
axis background/.style={fill=white},
axis x line*=bottom,
axis y line*=left,
xlabel={$\left.\bsd{p}^{\hat{v}}\big|\right._{\rm full}\,[\si{Pa}]$},
ylabel={$\left.\bsd{p}^{v}\left(\bsd{X}\right)\big|\right._{\rm hyb}\,[\si{Pa}]$},
legend style={legend cell align=left, align=left, draw=white!15!black}
]
\addplot [plot graphics/node/.append ]
graphics [xmin=2000, xmax=6000, ymin=2000, ymax=6000] {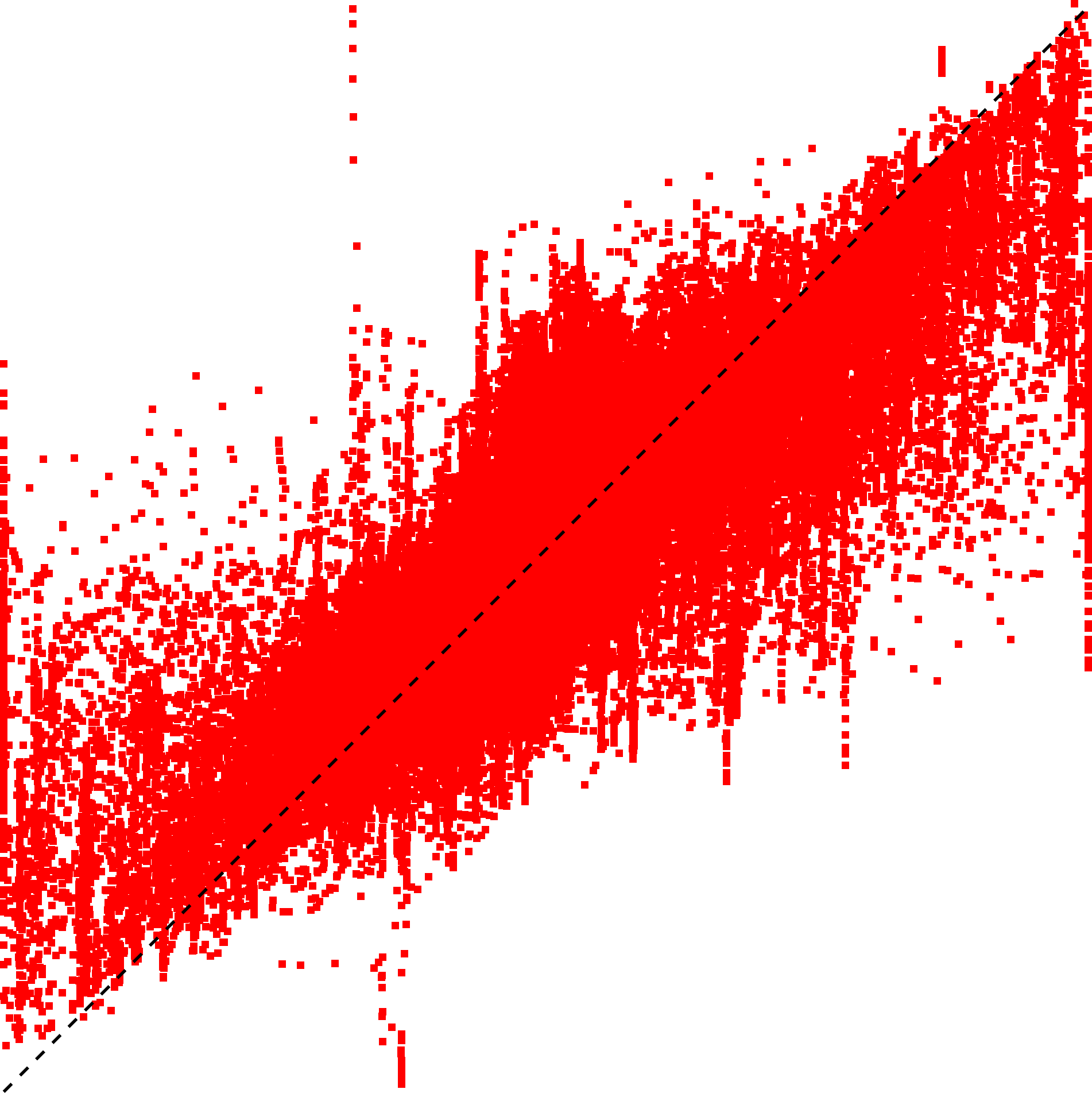};
\end{axis}
\end{tikzpicture}%
} \\
\subfloat[][$R^2_{\mathrm{IF}}=0.997$\label{fig:r_squared_if_press}]{\footnotesize
%
%
\begin{tikzpicture}[trim axis left,trim axis right]

\begin{axis}[%
width=\figurewidth,
height=\figureheight,
at={(0\figurewidth,0\figureheight)},
scale only axis,
xmin=0,
xmax=3547.82763671875,
ymin=0,
ymax=3601.23168945312,
axis on top,
xtick={0, 1000.0,2000.0,3000.0},
ytick={0, 1000.0,2000.0,3000.0},
axis background/.style={fill=white},
axis x line*=bottom,
axis y line*=left,
xlabel={$\left.\bsd{p}^{l}\big|\right._{\rm full}\,[\si{Pa}]$},
ylabel={$\left.\bsd{p}^{l}\big|\right._{\rm hyb}\,[\si{Pa}]$},
legend style={legend cell align=left, align=left, draw=white!15!black}
]
\addplot [plot graphics/node/.append ]
graphics [xmin=0, xmax=3547.8276367187, ymin=0, ymax=3601.23168945312] {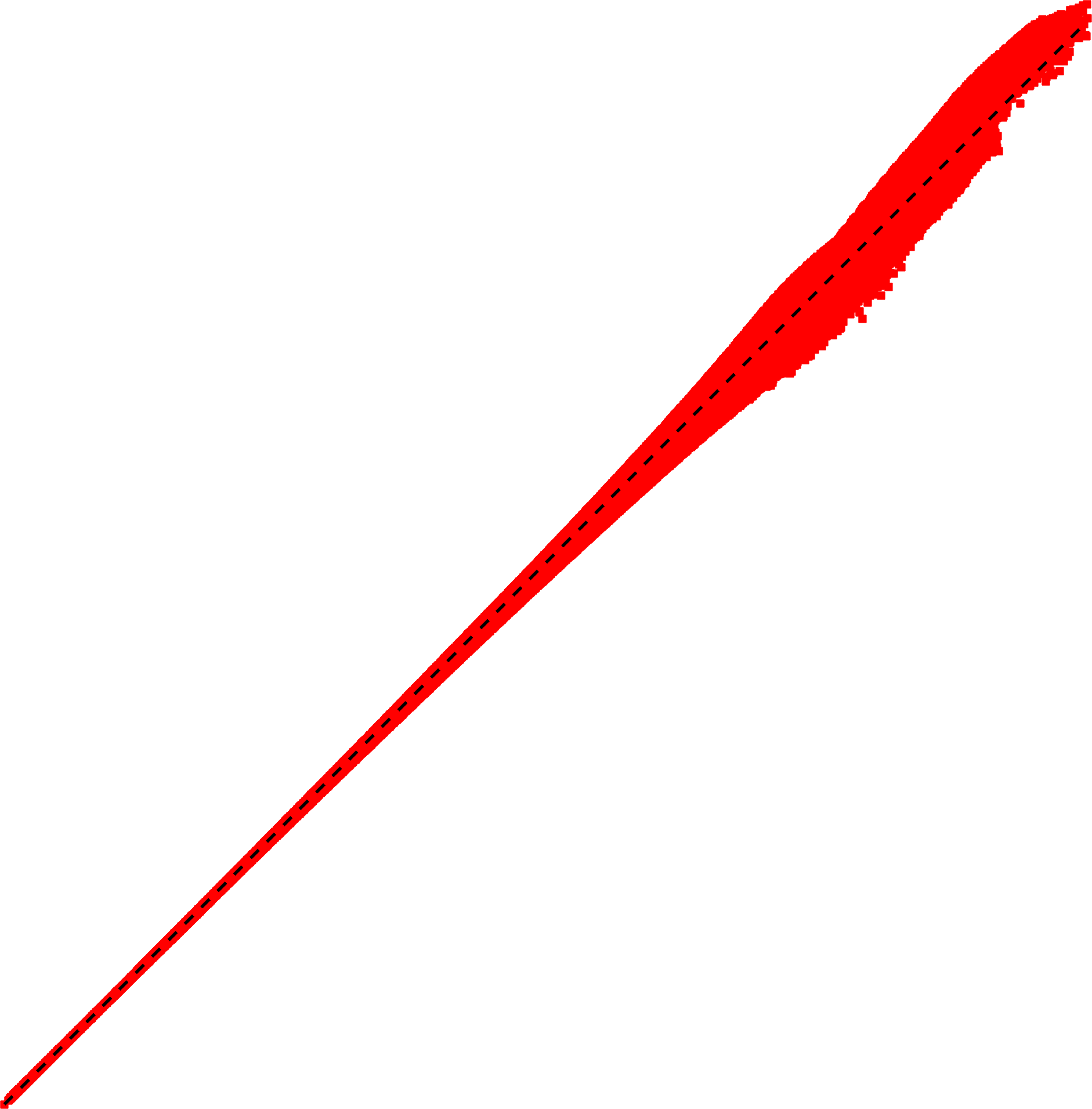};
\end{axis}
\end{tikzpicture}%
\hspace{1.5cm}}
\subfloat[][$R^2_{{\mathrm{flow}},\lsmall}=0.214$\label{fig:r_squared_flux}]{\footnotesize
%
%
\begin{tikzpicture}[trim axis left,trim axis right]

\begin{axis}[%
width=\figurewidth,
height=\figureheight,
at={(0\figurewidth,0\figureheight)},
scale only axis,
xmin=-24147574,
xmax=22590508,
ymin=-24147574,
ymax=22590508,
axis on top,
xtick={-2e7, -1e7,0,1e7,2e7},
ytick={-2e7, -1e7,0,1e7,2e7},
axis background/.style={fill=white},
axis x line*=bottom,
axis y line*=left,
xlabel={$\left.Q^{\hat{v}}\big|\right._{\lsmall,{\rm full}}\,[\si{\cubic\micro\m\per\s}]$},
ylabel={$\left.Q^{v}\big|\right._{\lsmall,{\rm hyb}}\,[\si{\cubic\micro\m\per\s}]$},
legend style={legend cell align=left, align=left, draw=white!15!black}
]
\addplot [plot graphics/node/.append ]
graphics [xmin=-24147574, xmax=22590508, ymin=-24147574, ymax=22590508] {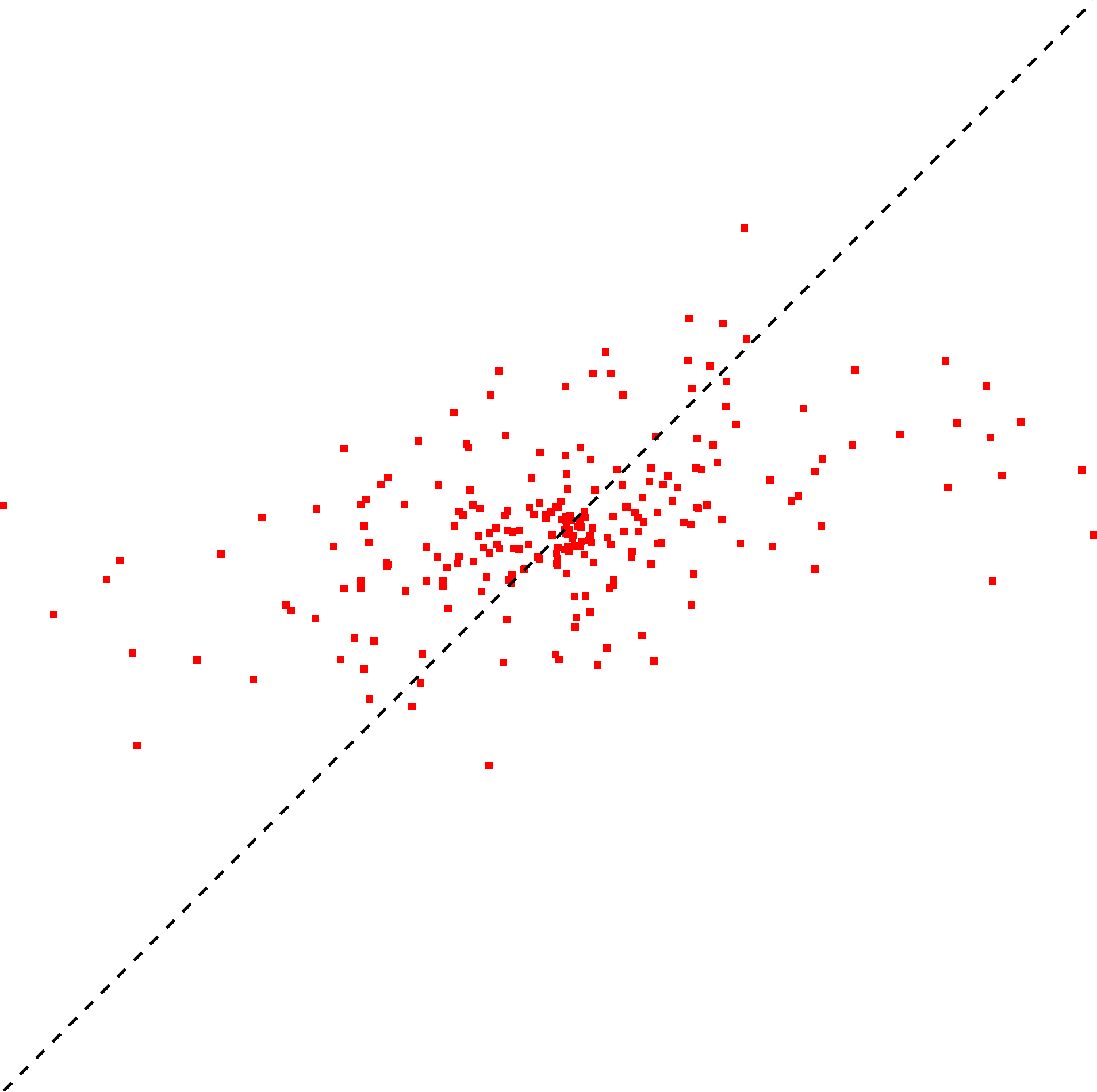};
\end{axis}
\end{tikzpicture}%
}
\caption{Exemplary comparison of hybrid model (with optimized parameters) with fully-resolved model for one specific network topology (SW1222, $\SI{10}{\percent}$ of 1D blood vessels have been retained in hybrid model). In each subfigure, solution of hybrid model is plotted over solution of fully-resolved model and the dashed line indicates perfect agreement between the models with 1:1-correspondence. Comparison of blood pressure in large vessels is depicted in subfigure a), blood pressure in small vessels in b), IF pressure in c) and flow in small vessels in d). Coefficient of determination for agreement between both variants is given for each quantity and overall coefficient of determination calculated according to~\eqref{eq:r_squared_tot} is $R^2_{\mathrm{tot}}=0.716$ for this case.}
\label{fig:r_squared}
\end{figure}

Representative results of the optimization scheme are depicted in Figure~\ref{fig:r_squared} for all four contributions to the total coefficient of determination. Very good agreement between the two models in terms of nodal pressures in the larger vessels $\bsd{p}^{\hat{v}}$ can be observed in Figure~\ref{fig:r_squared_art_press}. This can be expected because the same boundary conditions on the large vessels are applied in both cases. Thus, large and small pressure values show very good agreement, further away from these boundary conditions in the medium pressure range, deviations become larger. The clusters with the largest errors are separate branches which are not directly connected to nodes of the 1D network carrying boundary conditions. The correspondence for the nodal IF pressures $\bsd{p}^{l}$ in Figure~\ref{fig:r_squared_if_press} is also very good. For low IF pressures this is again due to the zero pressure boundary condition assigned on $\partial\Omega$ for both cases, but also for higher IF pressures inside the tumor, which is the actual domain of interest, the pressure differences are very small, in this case, the maximum absolute error is $\SI{237.1}{\Pa}$ corresponding to a maximum relative error of $\SI{8.4}{\percent}$. The pressure in the smaller vessels, resp. the homogenized vasculature in the hybrid model, exhibits larger errors, see Figure~\ref{fig:r_squared_homo_press}. Overall, the agreement is still reasonable. We found that the error is largest for branches ending in tips with boundary conditions on the 1D vasculature either inside the domain or on the tumor hull. For instance, this is the case for the larger errors around $\left.\bsd{p}^{\hat{v}}\middle|\right._{\mathrm{full}}\approx \SI{3300}{\Pa}$. The boundary conditions on these tips inside the domain are not retained in the hybrid model and for the tips on the tumor hull, they are smeared over several 3D nodes as described in Section~\ref{sec:bc_hybrid}. Hence, while the error in the medium pressure range is distributed symmetrically, larger deviations at both ends of the pressure spectrum towards the smeared values are present. This error due to point-wise non-matching boundary conditions can also not be improved by the optimization of the parameters. However, in Section~\ref{sec:additional_comps}, we will show that averaged REV pressures of both models are in very good accordance. We believe that this is a more interpretable and fairer comparison metric as the hybrid model cannot be expected to exactly match the pressure distribution of the fully-resolved one (in particular on the boundary) since the information about the exact topology of the smaller vessels is not represented. Finally, the results for the flow in the smaller vessels are shown in Figure~\ref{fig:r_squared_flux}. Here, the poorest agreement of the two models is present, especially, larger flows are not met properly.

\begin{table}
	\centering
	\resizebox{\textwidth}{!}{%
	\begin{tabular}{l l l l l l l l l l} 
		\hline
		Network & Case & $k^v/\mu^v\,[\si{\square\micro\m}]$ & $(S/V)_{\lsmall}\,[\si{\per\micro\m}]$ & $E_{S/V}\,[\si{\percent}]$ & $R^2_{\mathrm{L}}$ & $R^2_{\mathrm{S}}$ & $R^2_{\mathrm{IF}}$ & $R^2_{{\mathrm{flow}},\lsmall}$ & $R^2_{\mathrm{tot}}$  \\
		\hline\hline
		SW1222 & case $\SI{5}{\percent}$ & $\num{16.059 \pm 2.840}$ & $\num{6.423 \pm 0.101e-3}$ & $\num{2.17}$ & 0.944 & 0.488 & 0.994 & 0.163 & 0.647 \\
		& case $\SI{10}{\percent}$ & $\num{3.846 \pm 0.506}$ & $\num{5.722 \pm 0.069e-3}$ & $\num{1.93}$ & 0.988 & 0.654 & 0.998 & 0.176 & 0.704 \\
		& case $\SI{15}{\percent}$ & $\num{1.612 \pm 0.247}$ & $\num{5.077 \pm 0.051e-3}$ & $\num{2.66}$ & 0.993 & 0.739 & 0.999 & 0.262 & 0.748 \\
		& case $\SI{20}{\percent}$ & $\num{0.655 \pm 0.122}$ & $\num{4.532 \pm 0.048e-3}$ & $\num{3.84}$ & 0.989 & 0.792 & 0.999 & 0.193 & 0.743 \\
		 \hline
		LS174T & case $\SI{5}{\percent}$ & $\num{1.799 \pm 0.105}$ & $\num{1.721 \pm 0.073e-3}$ & $\num{3.21}$ & 0.905 & 0.643 & 0.990 & 0.282 & 0.705 \\
		& case $\SI{10}{\percent}$ & $\num{1.117 \pm 0.096}$ & $\num{1.581 \pm 0.038e-3}$ & $\num{3.18}$ & 0.916 & 0.683 & 0.991 & 0.260 & 0.713 \\
		& case $\SI{15}{\percent}$ & $\num{0.745 \pm 0.032}$ & $\num{1.478 \pm 0.020e-3}$ & $\num{2.70}$ & 0.930 & 0.695 & 0.992 & 0.244 & 0.715 \\
		& case $\SI{20}{\percent}$ & $\num{0.522 \pm 0.064}$ & $\num{1.382 \pm 0.021e-3}$ & $\num{2.30}$ & 0.944 & 0.718 & 0.993 & 0.207 & 0.715 \\
	    \hline
		GL261 & case $\SI{5}{\percent}$ & $\num{1.754 \pm 0.288}$ & $\num{6.184 \pm 0.099e-3}$ & $\num{3.35}$ & 0.917 & 0.195 & 0.997 & 0.199 & 0.577 \\
		& case $\SI{10}{\percent}$ & $\num{0.802 \pm 0.093}$ & $\num{5.688 \pm 0.097e-3}$ & $\num{3.63}$ & 0.927 & 0.233 & 0.996 & 0.107 & 0.566 \\
		& case $\SI{15}{\percent}$ & $\num{0.479 \pm 0.073}$ & $\num{5.200 \pm 0.076e-3}$ & $\num{4.12}$ & 0.941 & 0.295 & 0.996 & 0.113 & 0.586 \\
		& case $\SI{20}{\percent}$ & $\num{0.321 \pm 0.061}$ & $\num{4.756 \pm 0.048e-3}$ & $\num{4.21}$ & 0.950 & 0.346 & 0.996 & 0.134 & 0.607
	\end{tabular}}
	\caption{Results of the optimization procedure for hydraulic conductivity and surface-to-volume ratio of homogenized vasculature in the hybrid model. Relative error w.r.t.\ calculated surface-to-volume ratio and $R^2$-values for agreement between both variants in terms of blood pressure in large vessels, blood pressure in small vessels, IF pressure and flow in small vessels is additionally provided. Overall coefficient of determination $R^2_{\mathrm{tot}}$ between fully-resolved and hybrid model is calculated according to~\eqref{eq:r_squared_tot}. (All data includes the mean taken over five different sets of pressure boundary conditions on the 1D network produced by the methodology described in Section~\ref{sec:bc_full}, "case ${\mathrm{X}}\,\si{\percent}$" denotes the case where ${\mathrm{X}}\,\si{\percent}$ of the 1D blood vessels are retained in the hybrid approach)}
\label{tab:results}
\end{table}

Further results for all cases have been collected in Table~\ref{tab:results}. For each tumor network, we generate five different sets of pressure boundary conditions on the 1D network from which different flow patterns and, therefore, also different sets of larger and smaller blood vessels emerge as discussed in Section~\ref{sec:diff_fully_hyb}. Then, we investigate different cases, where $\SIrange[range-phrase = -,range-units=single]{5}{20}{\percent}$ of the larger vessels are kept in the hybrid model. We found that five sets of pressure boundary conditions on the 1D network were enough to study our hybrid model since randomly picking only four out of the five boundary condition cases changed the mean result by at most $\SI{8}{\percent}$. Moreover, taking the mean parameter of a case ${\mathrm{X}}\,\si{\percent}$ over all different boundary condition cases instead of the optimal value for each specific case only changed the total coefficient of determination by less than $\SI{2}{\percent}$. Furthermore, we compare the result of the optimization procedure for $(S/V)_{\lsmall}$ with the calculated surface-to-volume ratio of the smaller vessels for each case. The relative error $E_{S/V}$ is smaller than $\SI{5}{\percent}$ for all cases validating that the optimization procedure converges to a physically reasonable result. The permeability is largest for the SW1222 topology which can be expected considering the much denser network of this case. For all tumors, it decreases if a larger proportion of the 1D vessels is kept in the model, which is also sensible since the smaller the proportion of homogenized vessels, the less permeable these vessels.

As already described above, all cases exhibit a very good correspondence in terms of blood pressures in larger vessels and IF pressures, proven by the values for $R^2_{\mathrm{L}}$ and $R^2_{\mathrm{IF}}$ in Table~\ref{tab:results}. If the fidelity of the hybrid model is increased by resolving a larger proportion of the network structure, the agreement between the two models grows likewise. This is also the case for the coefficient of determination of blood pressure in smaller vessels $R^2_{\mathrm{S}}$. Here, the SW1222 and the LS174T case exhibit comparable accuracy whereas the GL261 case experiences a larger discrepancy. We can attribute this to the fact that this topology has the largest number of tips at the tumor hull and also the largest number of dead ends considering that it is the smallest data set, see Table~\ref{tab:analysis_datasets}. Hence, the pressure error is largest due to non-matching boundary conditions between fully-resolved and hybrid model as mentioned above. However, we will show in Section~\ref{sec:additional_comps} that in terms of REV blood pressures its conformity with the hybrid model is as good as the other cases. The difference in flow in the small vessels is the largest source of error in all cases. Also taking more 1D vessels into account for the hybrid model does not necessarily improve the behaviour. We believe that this is due to the chaotic flow patterns in the smaller vessels and to the fact that we define the permeability tensor as isotropic and constant over the entire domain $\Omega_v$. Apparently, this is insufficient to resolve the flow in the homogenized vasculature in comparison to the full model. We tried to increase the agreement by giving a higher weight to the coefficient of determination of flow in the smaller vessels $R^2_{{\mathrm{flow}},\lsmall}$ in the definition of the total coefficient of determination~\eqref{eq:r_squared_tot} but could not achieve any significant improvements. However, the agreement in terms of flow in the entire (resolved and homogenized) vasculature, which will be investigated in the next section, is much better. 
\subsection{Additional comparisons of results of both models}
\label{sec:additional_comps}
\begin{table}
	\centering
	\begin{tabular}{l l l l l l l l} 
		\hline
		Network & Case & $\overline{E^v_{\mathrm{abs}}}\,[\si{\Pa}]$ & $\overline{E^v_{\mathrm{rel}}}\,[\si{\percent}]$ & $\overline{E^l_{\mathrm{abs}}}\,[\si{\Pa}]$ & $\overline{E^l_{\mathrm{rel}}}\,[\si{\percent}]$ & $R^2_{{\mathrm{flow}},\lbig\rightarrow\lsmall}$ & $R^2_{{\mathrm{flow}},\Lambda}$ \\
		\hline\hline
		SW1222 & case $\SI{5}{\percent}$ & $\num{90.2}$ & $\num{2.25}$ & $\num{49.3}$ & $\num{1.66}$ & $\num{0.192}$ & $\num{0.992}$ \\
		& case $\SI{10}{\percent}$ & $\num{57.2}$ & $\num{1.43}$ & $\num{32.2}$ & $\num{1.08}$ & $\num{0.091}$ & $\num{0.999}$ \\
		& case $\SI{15}{\percent}$ & $\num{44.1}$ & $\num{1.10}$ & $\num{24.9}$ & $\num{0.83}$ & $\num{0.142}$ & $\num{1.000}$ \\
		& case $\SI{20}{\percent}$ & $\num{42.1}$ & $\num{1.05}$ & $\num{21.5}$ & $\num{0.72}$ & $\num{0.100}$ & $\num{1.000}$ \\
		 \hline
		LS174T & case $\SI{5}{\percent}$ & $\num{113.7}$ & $\num{2.88}$ & $\num{73.0}$ & $\num{3.83}$ & $\num{0.640}$ & $\num{0.813}$ \\
		& case $\SI{10}{\percent}$ & $\num{99.4}$ & $\num{2.53}$ & $\num{68.3}$ & $\num{3.58}$ & $\num{0.567}$ & $\num{0.892}$ \\
		& case $\SI{15}{\percent}$ & $\num{97.6}$ & $\num{2.48}$ & $\num{63.0}$ & $\num{3.30}$ & $\num{0.513}$ & $\num{0.931}$ \\
		& case $\SI{20}{\percent}$ & $\num{91.4}$ & $\num{2.33}$ & $\num{59.2}$ & $\num{3.10}$ & $\num{0.444}$ & $\num{0.956}$ \\
	    \hline
		GL261 & case $\SI{5}{\percent}$ & $\num{104.2}$ & $\num{2.64}$ & $\num{41.6}$ & $\num{2.16}$ & $\num{0.472}$ & $\num{0.962}$ \\
		& case $\SI{10}{\percent}$ & $\num{99.5}$ & $\num{2.52}$ & $\num{44.2}$ & $\num{2.29}$ & $\num{0.489}$ & $\num{0.984}$ \\
		& case $\SI{15}{\percent}$ & $\num{87.8}$ & $\num{2.22}$ & $\num{44.9}$ & $\num{2.33}$ & $\num{0.435}$ & $\num{0.993}$ \\
		& case $\SI{20}{\percent}$ & $\num{79.7}$ & $\num{2.02}$ & $\num{45.0}$ & $\num{2.33}$ & $\num{0.407}$ & $\num{0.996}$
	\end{tabular}
	\caption{Additional error measures for the agreement of both models. Shown are the absolute and relative error of the hybrid approach in terms of mean REV blood pressure in smaller vessels and mean REV interstitial fluid pressure and the $R^2$-values for agreement between both variants in terms of flow from large to small vessels and flow in the entire vasculature. (All data includes the mean taken over five different sets of pressure boundary conditions on the 1D network produced by the methodology described in Section~\ref{sec:bc_full}, "case ${\mathrm{X}}\,\si{\percent}$" denotes the case where ${\mathrm{X}}\,\si{\percent}$ of the 1D blood vessels are retained in the hybrid approach)}
\label{tab:results_contd}
\end{table}
We further study the agreement of the hybrid model with the optimized parameters from the previous section in terms of several other quantities. For that, we define the mean REV IF pressure in the $j$-th REV as
\begin{equation}
\label{eq:if_press_int}
\overline{p^l}(j)=\frac{1}{V_{{\mathrm{REV}}_j}}\int_{{\mathrm{REV}}_j }p^l\,dV.
\end{equation}
This is employed to study the absolute and relative mean IF pressure error between the two models in each REV as
\begin{equation}
\label{eq:if_press_err_rev}
E^l_{\mathrm{abs}}(j)={\mathrm{abs}}\left(\left.\overline{p^l}(j)\big|\right._{\mathrm{full}}-\left.\overline{p^l}(j)\big|\right._{\mathrm{hyb}}\right)\;\;{\mathrm{and}}\;\;E^l_{\mathrm{rel}}(j)=\frac{{\mathrm{abs}}\left(\left.\overline{p^l}(j)\big|\right._{\mathrm{full}}-\left.\overline{p^l}(j)\big|\right._{\mathrm{hyb}}\right)}{\left.\overline{p^l}(j)\big|\right._{\mathrm{full}}}.
\end{equation}
Equivalently, we define the mean blood pressure in the homogenized vasculature of the hybrid model in the $j$-th REV as 
\begin{equation}
\label{eq:blood_press_int_hyb}
\left.\overline{p^v}(j)\big|\right._{\mathrm{hyb}}=\frac{1}{V_{{\mathrm{REV}}_j}}\int_{{\mathrm{REV}}_j }p^v\,dV
\end{equation}
and as
\begin{equation}
\label{eq:blood_press_int_full}
\left.\overline{p^v}(j)\big|\right._{\mathrm{full}}=\frac{1}{{n_{{\mathrm{nodes}},\lsmall\cap {\mathrm{REV}}_j }}}\sum_{i=1}^{{n_{{\mathrm{nodes}},\lsmall\cap {\mathrm{REV}}_j }}}\bsd{p}^{\hat{v}}\left[i\right]
\end{equation}
for the smaller vessels of the fully-resolved model. The latter is simply the mean blood pressure of all ${n_{{\mathrm{nodes}},\lsmall\cap {\mathrm{REV}}_j }}$ nodes of the smaller blood vessels which lie inside the $j$-th REV. This allows us to define the absolute and relative mean blood pressure error (in the smaller vessels) between the two models in each REV as
\begin{equation}
\label{eq:blood_press_err_rev}
E^v_{\mathrm{abs}}(j)={\mathrm{abs}}\left(\left.\overline{p^v}(j)\big|\right._{\mathrm{full}}-\left.\overline{p^v}(j)\big|\right._{\mathrm{hyb}}\right)\;\;{\mathrm{and}}\;\;E^v_{\mathrm{rel}}(j)=\frac{{\mathrm{abs}}\left(\left.\overline{p^v}(j)\big|\right._{\mathrm{full}}-\left.\overline{p^v}(j)\big|\right._{\mathrm{hyb}}\right)}{\left.\overline{p^v}(j)\big|\right._{\mathrm{full}}}.
\end{equation}
Furthermore, we denote by $\overline{\left(\cdot\right)}$ the mean value of these error measures over all $n_{\mathrm{REV}}$ REVs. Note also that both the mean REV blood and IF pressure vary considerably between different REVs. The pressure difference between single REVs varies in a range of $\SIrange[range-phrase = -,range-units=single]{800}{1200}{\Pa}$ for the IF and a range of $\SIrange[range-phrase = -,range-units=single]{800}{2000}{\Pa}$ for blood. The data of this analysis is collected in Table~\ref{tab:results_contd}. Overall, a remarkable agreement of the mean REV pressures for both blood and IF can be observed in all cases. As in Table~\ref{tab:results}, the SW1222 tumor has the best agreement, but also the GL261 case which previously showed the biggest nodal blood pressure errors in the homogenized vessels is very accurate in terms of mean REV blood pressure. As described above, the error is located mainly on the tips of the smaller vessels, of which the GL261 has the most compared to its network size. Nevertheless, the average blood pressure in the REVs is still matched very well for this and all other cases even though locally the pressure error is larger. We can expect that these small-scale spatial fluctuations of blood pressures cannot be represented correctly in the homogenized vessels of the hybrid model while macroscopically the average REV pressures show good agreement. Anticipating a validation with experimental data, it is anyhow not possible to perform a point-wise comparison of (blood and IF) pressures such that the average REV pressure is the more relevant and meaningful metric.

Additionally, we investigate the volumetric flow between large and small vessels and compare the results of both models. In the hybrid model, the flow between large and small vessels is given by the LM field $\lambda=\iema M{\hat{v}}{v}$ interpreted as a mass transfer term, or volumetric flow per length, as detailed in Section~\ref{sec:hybrid}. Note that this can represent both a flow from large 1D vessels into the homogenized vasculature if locally the pressure in the resolved vasculature is bigger than the homogenized pressure or, vice versa, a flow from the homogenized vasculature into the larger vessels if the homogenized pressure is bigger than the blood pressure in the 1D vasculature. Consequently, for each REV the flow between the two compartments is given by the integral of the LM field along the part of the larger vessels $\lbig\cap{\mathrm{REV}}_j$ inside the specific REV $j$, or
\begin{equation}
\label{eq:mass_transf_1D_into_small_hyb}
\left.\iema M{\hat{v}}{v}\left(j\right)\big|\right._{\mathrm{hyb}}=\int_{\lbig\cap{\mathrm{REV}}_j }\lambda\,ds.
\end{equation}
In the full model we can directly evaluate the mass transfer between large and small vessels inside the connecting elements of both sets, which are those elements of the smaller vessels where one node is part of $\lbig$ and the other part of $\lsmall$. Assuming that the first node is part of the larger vessels and the second one part of the smaller vessels, the flow between large and small vessels in the $j$-th REV is given by
\begin{equation}
\label{eq:mass_transf_1D_into_small_full}
\left.\iema M{\hat{v}}{v}\left(j\right)\big|\right._{\mathrm{full}}=\sum_{i=1}^{n_{{\mathrm{ele}},\lbig\rightarrow\lsmall\cap {\mathrm{REV}}_j}}-\frac{\pi R^4}{8\mu^{\hat{v}}}\frac{\partial p^{\hat{v}}}{\partial s}
\end{equation}
as the sum of the volumetric flows in the elements connecting large and small vessels which lie inside the specific REV $j$. The number of these elements is denoted by $n_{{\mathrm{ele}},\lbig\rightarrow\lsmall\cap {\mathrm{REV}}_j}$ in the previous equation. To compare the mass transfer between large and small vessels in both models, we again define a coefficient of determination as
\begin{equation}
\label{eq:r_squared_flow_into_small}
R^2_{{\mathrm{flow}},\lbig\rightarrow\lsmall}=1-\frac{\sum_{j=1}^{n_{{\mathrm{REV}}}}\left(\left.\iema M{\hat{v}}{v}\left(j\right)\big|\right._{\mathrm{full}}-\left.\iema M{\hat{v}}{v}\left(j\right)\big|\right._{\mathrm{hyb}}\right)^2}{\sum_{j=1}^{n_{{\mathrm{REV}}}}\left(\left.\iema M{\hat{v}}{v}\left(j\right)\big|\right._{\mathrm{full}}-\mu\left(\left.\iema M{\hat{v}}{v}\left(j\right)\big|\right._{\mathrm{full}}\right)\right)^2},
\end{equation}
with the respective mass transfer terms for the hybrid and the full model for each REV. Again, $\mu\left(\cdot\right)$ denotes the mean of the mass transfer between large and small vessels of the full model over all $n_{\mathrm{REV}}$ REVs.

The reference solution of the fully resolved model for this volumetric flow per REV varies considerably between the different REVs and both positive values, representing an overall outflow from the larger vessels into the smaller vessels in a specific REV, and negative values, representing an overall inflow into the larger vessels from the smaller vessels in a specific REV, are present. This indicates that in- or outflow from larger to smaller vessels is indeed a meaningful quantity describing the spatially varying flow patterns inside the vascular network. To reproduce this behaviour in the hybrid model variant, a good agreement with the reference solution is desirable. The results for the coefficient of determination $R^2_{{\mathrm{flow}},\lbig\rightarrow\lsmall}$ are again assembled in Table~\ref{tab:results_contd}. The LS174T case shows the best agreement with the fully-resolved model while the SW1222 case delivers the worst results. We believe that this can be attributed to the much higher dispersion of the diameter and, thus, also the flow in the connectivity elements, which we have already studied by the coefficient of variability in Table~\ref{tab:analysis_conn}. The LS174T case, which has the least dispersed distribution of both values, performs best in matching the flow between larger and smaller vessels in the hybrid model. There is a small decline of the agreement for higher percentages of retained vessels in all cases. However, the flow between large and small vessels is not included in the parameter optimization procedure. Hence, we assume that the better performance in terms of the other quantities is at the expense of this metric.

Finally, we study the correspondence between the two models in terms of the blood flow in the entire vasculature $\Lambda$. Previously, in Table~\ref{tab:results} only flow in the smaller vessels $\lsmall$, respectively, the homogenized vasculature was investigated. For the full model, the total flow in $\Lambda$ in each REV in coordinate direction $j$ is calculated as in~\eqref{eq:volflow_full}, but now both large and small vessels are taken into account. For the hybrid model, the total flow can be obtained as the sum of the flow in the homogenized vessels as given by~\eqref{eq:volflow_homo} and the flow in the larger, resolved vessels, that is, eqn.~\eqref{eq:volflow_full} evaluated for the larger vessels of the hybrid model. The two quantities are compared in Table~\ref{tab:results_contd} defining a coefficient of determination for flow in the entire vasculature $R^2_{{\mathrm{flow}},\Lambda}$ as in~\eqref{eq:r_squared_flow}. Evidently, the agreement between the two models is very good and much better than the previously reported agreement of flow in the smaller vessels $R^2_{{\mathrm{flow}},\lsmall}$. This is due to the fact that, as expected, flow is dominated by the larger vessels, the values calculated for flow in the entire vasculature are one to two orders of magnitude larger than the in the small vessels depending on the investigated case. As we are able to match the pressure in the large vessels very well and, thus, also the flow therein, very good accordance can be achieved for the total flow in both small and big vessels. As flow in the big vessels is decisive for the overall perfusion of the domain and could also be more easily acquired with experiments for further validation this is an encouraging result for the applicability of the hybrid approach. Nevertheless, we demonstrate how to enhance the correspondence of the hybrid model also in terms of flow in the smaller vessels in the following section.
\section{Improvements for the hybrid model}
\label{sec:poss_improvements}
In this section, we discuss some possible improvements for the hybrid model and implement one of them. The most straightforward one would be to define the permeability of the homogenized vessels not as a constant over the entire domain $\Omega_v$ but per REV. Instead of an isotropic permeability tensor, one could easily integrate anisotropic effects based on the blood vessel structure inside each REV. Both has been done in the hybrid model of Vidotto et al.\cite{Vidotto2018}, where a diagonal permeability tensor with different permeabilities in all three coordinate directions was employed. This could potentially augment the agreement in terms of mass fluxes in the homogenized vasculature, which is the main source of error in the hybrid model. However, we did not integrate this into our optimization procedure since we believe that this would result in overfitting of the chaotic flow in the tumor such that we would meet every single boundary condition case very well but with largely different results for the permeability tensors between the cases with distinct flow patterns. With a single scalar permeability the results for the permeability between different boundary condition cases did not fluctuate greatly. Moreover, in a real-world case where only the architecture of the larger vessels is known, it seems unrealistic to deduce the entire permeability field from the limited amount of information.

\setlength\figureheight{0.28\textwidth}
\setlength\figurewidth{0.36\textwidth}
\begin{figure}
\subfloat[SW1222, $R^2=0.50$\label{fig:SW1222_flux_over_vf}]{
\footnotesize
\begin{tikzpicture}

\begin{axis}[
width=0.951\figurewidth,
height=\figureheight,
at={(0\figurewidth,0\figureheight)},
tick align=outside,
tick pos=left,
x grid style={white!69.0196078431373!black},
xmin=0.0224945146084105, xmax=0.189649244978678,
xtick style={color=black},
y grid style={white!69.0196078431373!black},
ymin=-2968535, ymax=62339235,
xlabel={$\porosity^v_{\lsmall}\;[-]$},
ylabel={$\left.|Q^{\hat{v}}|\big|\right._{\lsmall,{\rm full}}\,[\si{\cubic\micro\m\per\s}]$},
ytick style={color=black}
]
\addplot [only marks, mark size=1pt, mark=*, draw=red, fill=red, colormap/viridis]
table{%
x                      y
0.0994470484991292 18874558
0.100785309341806 47358244
0.113852757135704 16207194
0.121201773878491 35925760
0.0993596423170869 24312668
0.0855408081979865 22462128
0.0990262535930206 40528552
0.0895536639797236 11688727
0.128827155588096 36001988
0.0921026832157892 9929383
0.0673359014351949 12043097
0.0671512485283101 10512879
0.112276415827774 13652961
0.146631258473562 8305019.5
0.109114319372284 3623271.25
0.12254200506648 15310096
0.0680587644600702 319121.71875
0.081620890041237 6594304
0.0789538851383993 16006866
0.0861616263060857 7868630
0.0888185123847265 14740691
0.0769486958263945 21923254
0.0874693431302347 21478116
0.102648149471861 29574038
0.124681034797766 11975647
0.12892515482162 10891911
0.103623001652431 25499216
0.115697364546483 6788092.5
0.0405489209446827 715073.8125
0.0781721892895976 4899817.5
0.09143792968974 16616790
0.0908118800620469 17099618
0.036511253804381 0
0.061157309194769 2070730.75
0.141440875692905 11642301
0.0892362012542406 9511021
0.0742243251980862 8978894
0.10515086605486 15282636
0.14151581999966 20727736
0.102506876557989 9725466
0.147371538914539 34460760
0.12600638173889 30713384
0.0900637605597526 6598891
0.054969351510923 1909462.75
0.0972057763072615 25724932
0.0989336784793227 9539464
0.0966057510270006 4244290.5
0.0433741961383365 12288427
0.0300924568979681 2540289
0.125471673750838 21622966
0.111807773628237 4652550
0.0779097290046946 27692670
0.0691890255100204 3965516.25
0.108454876424555 16665975
0.133185206964482 40096532
0.109063310605069 15145908
0.119254713393152 18350840
0.0851725463249965 20288568
0.143271265437056 23192700
0.124250128726713 1960191.875
0.0787394640005441 12265690
0.0899227520697399 4207614
0.108601180029476 12544684
0.114781670105077 7142945.5
0.144727612471881 21769850
0.139197787276691 19803962
0.125567160977874 31518366
0.126026706103656 17575116
0.18205130268912 40726932
0.108052161052157 7974299
0.144073922796977 26886942
0.135731023443197 22482088
0.121353872296531 9673386
0.142095708943613 22768078
0.133301681784892 14299178
0.137811562478661 33521834
0.131120850283658 33927632
0.113022019590073 57929312
0.0994470484991292 14001042
0.100785309341806 16031011
0.113852757135704 9830645
0.121201773878491 39422336
0.0993596423170869 5043340
0.0855408081979865 11042357
0.0990262535930206 26102552
0.0895536639797236 6826096
0.128827155588096 12218526
0.0921026832157892 15445856
0.0673359014351949 18509468
0.0671512485283101 39798188
0.112276415827774 17235724
0.146631258473562 12703000
0.109114319372284 16532806
0.12254200506648 11562388
0.0680587644600702 2535846.25
0.081620890041237 16533095
0.0789538851383993 7254420.5
0.0861616263060857 5153414
0.0888185123847265 13217885
0.0769486958263945 13493175
0.0874693431302347 4896252.5
0.102648149471861 13112503
0.124681034797766 8779258
0.12892515482162 4585502.5
0.103623001652431 28115768
0.115697364546483 13425622
0.0405489209446827 1156124
0.0781721892895976 3577902
0.09143792968974 15838383
0.0908118800620469 13118768
0.036511253804381 2789732
0.061157309194769 3588164.5
0.141440875692905 39647684
0.0892362012542406 9450672
0.0742243251980862 6928158
0.10515086605486 1162892
0.14151581999966 31092232
0.102506876557989 17708872
0.147371538914539 36662772
0.12600638173889 14030326
0.0900637605597526 5842434
0.054969351510923 7170634.5
0.0972057763072615 12162074
0.0989336784793227 2955422.5
0.0966057510270006 7959463.5
0.0433741961383365 10593350
0.0300924568979681 2499841.75
0.125471673750838 10217499
0.111807773628237 26182744
0.0779097290046946 12710961
0.0691890255100204 913259.75
0.108454876424555 7349778.5
0.133185206964482 34239304
0.109063310605069 11801435
0.119254713393152 23929518
0.0851725463249965 1647548.25
0.143271265437056 10004708
0.124250128726713 16608669
0.0787394640005441 15679373
0.0899227520697399 8892243
0.108601180029476 6383390.5
0.114781670105077 11563596
0.144727612471881 9277566
0.139197787276691 23539540
0.125567160977874 24555962
0.126026706103656 25573806
0.18205130268912 19303946
0.108052161052157 18825766
0.144073922796977 43908588
0.135731023443197 11709382
0.121353872296531 8458655
0.142095708943613 23700344
0.133301681784892 22429286
0.137811562478661 12749064
0.131120850283658 29440788
0.113022019590073 24661330
0.0994470484991292 24105878
0.100785309341806 35783452
0.113852757135704 17198416
0.121201773878491 24256878
0.0993596423170869 15221422
0.0855408081979865 8045047
0.0990262535930206 10102123
0.0895536639797236 35979876
0.128827155588096 7491469.5
0.0921026832157892 17414534
0.0673359014351949 16791144
0.0671512485283101 37887836
0.112276415827774 27307332
0.146631258473562 40093056
0.109114319372284 28368186
0.12254200506648 9608452
0.0680587644600702 876761.375
0.081620890041237 20505912
0.0789538851383993 23859608
0.0861616263060857 6878058
0.0888185123847265 7272644.5
0.0769486958263945 7151082
0.0874693431302347 4385558
0.102648149471861 13929343
0.124681034797766 12578756
0.12892515482162 41001224
0.103623001652431 26235352
0.115697364546483 11525059
0.0405489209446827 1328220.375
0.0781721892895976 19576256
0.09143792968974 28244714
0.0908118800620469 22444210
0.036511253804381 313948.15625
0.061157309194769 7413928.5
0.141440875692905 30104728
0.0892362012542406 13495082
0.0742243251980862 749100.625
0.10515086605486 18778604
0.14151581999966 33175080
0.102506876557989 25336854
0.147371538914539 25688292
0.12600638173889 20913730
0.0900637605597526 17867196
0.054969351510923 5806149
0.0972057763072615 15019671
0.0989336784793227 12514571
0.0966057510270006 13682435
0.0433741961383365 4441821
0.0300924568979681 968221.4375
0.125471673750838 15719497
0.111807773628237 10763535
0.0779097290046946 4370879
0.0691890255100204 5501294.5
0.108454876424555 24521934
0.133185206964482 46591568
0.109063310605069 25115930
0.119254713393152 18067210
0.0851725463249965 23077982
0.143271265437056 18218692
0.124250128726713 16971546
0.0787394640005441 12047318
0.0899227520697399 16447646
0.108601180029476 9581752
0.114781670105077 6226869
0.144727612471881 22525800
0.139197787276691 25530460
0.125567160977874 34959880
0.126026706103656 20657052
0.18205130268912 34641900
0.108052161052157 13758662
0.144073922796977 17643698
0.135731023443197 59370700
0.121353872296531 22557476
0.142095708943613 24859234
0.133301681784892 18354742
0.137811562478661 15526859
0.131120850283658 34266000
0.113022019590073 22582554
};
\addplot [semithick, dashed, black]
table {%
0.0300924568979681 2318462.78889468
0.18205130268912 32241439.3267969
};
\end{axis}

\end{tikzpicture}}\hfill
\subfloat[LS174T, $R^2=0.66$\label{fig:LS174T_flux_over_vf}]{
\footnotesize
     \input{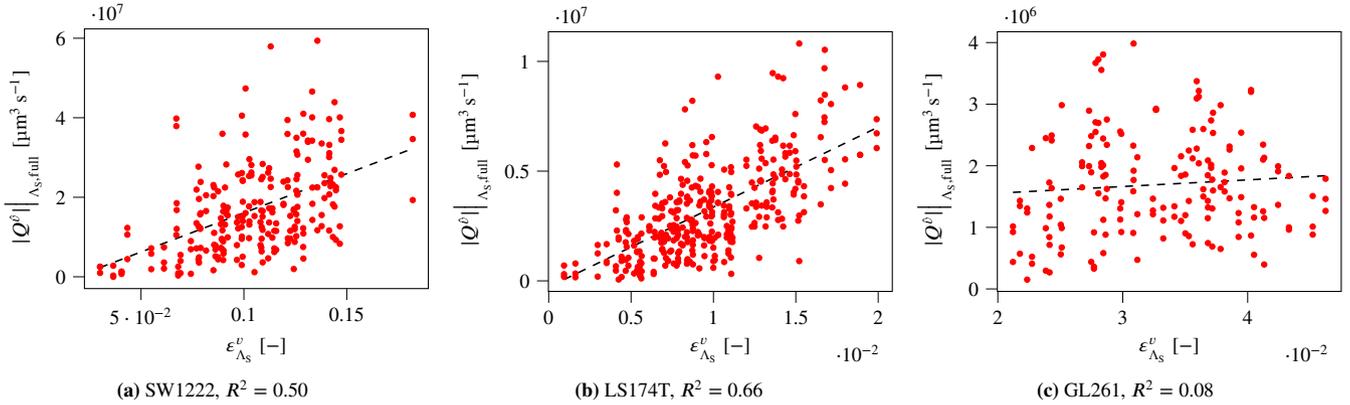}}\hfill
\subfloat[GL261, $R^2=0.08$\label{fig:GL261_flux_over_vf}]{
\footnotesize
\begin{tikzpicture}

\begin{axis}[
width=0.951\figurewidth,
height=\figureheight,
at={(0\figurewidth,0\figureheight)},
tick align=outside,
tick pos=left,
x grid style={white!69.0196078431373!black},
xmin=0.019962238431155, xmax=0.0474891072050218,
xtick style={color=black},
y grid style={white!69.0196078431373!black},
ymin=-40536.73515625, ymax=4175297.84453125,
xlabel={$\porosity^v_{\lsmall}\;[-]$},
ylabel={$\left.|Q^{\hat{v}}|\big|\right._{\lsmall,{\rm full}}\,[\si{\cubic\micro\m\per\s}]$},
ytick style={color=black}
]
\addplot [only marks, mark size=1pt, mark=*, draw=red, fill=red, colormap/viridis]
table{%
x                      y
0.0267458026881794 2055704
0.0394806071068686 1489803.75
0.028728546600394 1624299.75
0.0350072622188242 1857209.875
0.0243076428572129 2492192.75
0.0238392500756294 986266.125
0.0273020854183447 1993492.625
0.0326635585595695 2925065.5
0.0424336800354085 1908420.25
0.0332425872272534 1110759.75
0.0217666949129658 571435.9375
0.0212134597390581 1015142.625
0.0298850795619941 2518186.25
0.0278240914732981 2705904.25
0.025117223501796 2983490.25
0.0297995788285001 1126571.625
0.0370729902963626 2014530.25
0.0241218921039095 1727171.375
0.0452058046103372 1507764.625
0.0250383798770863 1463328.125
0.0227385815950048 522133.625
0.0392699195422144 1080282.25
0.0241396795109998 269092.21875
0.0280529708338105 3726543.5
0.0368060175488641 1845550.75
0.0284307872029342 3805989.25
0.0402676661719229 3229487
0.0385437370386167 1471292.5
0.0368199075366995 2085825.875
0.0412345566659551 2342991
0.0405090738058525 1159365.25
0.0324166833267889 1464375
0.0308706624294841 3983669
0.0311706938234881 1214161.75
0.0378300778516918 2983327
0.028289947343966 3555402
0.0308430129888187 1584897
0.0360863484145798 3215432
0.0223492914355186 1305843.75
0.0284483120365114 581711.9375
0.0462378858971187 1266296.125
0.0371642139909929 688278.6875
0.035288748053439 2246962.75
0.0286273046998917 1852212.5
0.0355919114427826 2583862.5
0.0276965499227147 918664.4375
0.0413222136728524 2129811
0.0381475803343246 1582022
0.0346682274868035 1103800.75
0.0364442000333885 1500558.375
0.0341327734325715 1962206.75
0.04330042676533 962277.375
0.0274793756290511 2480803.25
0.0359171092074486 3092706.25
0.0372267788338402 2861031.75
0.0267458026881794 1649077.5
0.0394806071068686 1956250.875
0.028728546600394 2750457.5
0.0350072622188242 902749.8125
0.0243076428572129 1636004.375
0.0238392500756294 296210.25
0.0273020854183447 2893482
0.0326635585595695 2909408
0.0424336800354085 1991194.25
0.0332425872272534 1317153
0.0217666949129658 1346592.875
0.0212134597390581 438098.875
0.0298850795619941 923703.0625
0.0278240914732981 2547986.75
0.025117223501796 988347.125
0.0297995788285001 2556006.75
0.0370729902963626 1966548.125
0.0241218921039095 841681.6875
0.0452058046103372 882012.3125
0.0250383798770863 562085.4375
0.0227385815950048 408737.75
0.0392699195422144 844399.125
0.0241396795109998 715368.25
0.0280529708338105 1550893.25
0.0368060175488641 1429703.5
0.0284307872029342 2442091.25
0.0402676661719229 866546.9375
0.0385437370386167 2312186
0.0368199075366995 1323050.25
0.0412345566659551 1260629.25
0.0405090738058525 554658
0.0324166833267889 1212265.375
0.0308706624294841 1768664.75
0.0311706938234881 472304.375
0.0378300778516918 1653619.375
0.028289947343966 1980364.875
0.0308430129888187 909056.125
0.0360863484145798 3121249.5
0.0223492914355186 1240474.625
0.0284483120365114 1826001.125
0.0462378858971187 1461847.375
0.0371642139909929 1305430.375
0.035288748053439 1427927.375
0.0286273046998917 1470010
0.0355919114427826 2118966.75
0.0276965499227147 329284.03125
0.0413222136728524 397283.125
0.0381475803343246 1733964.375
0.0346682274868035 2153419.25
0.0364442000333885 1618576.125
0.0341327734325715 1341289.5
0.04330042676533 999987
0.0274793756290511 438894.5
0.0359171092074486 2278143.5
0.0372267788338402 2532678.25
0.0267458026881794 1993588.5
0.0394806071068686 2537594.5
0.028728546600394 2339583
0.0350072622188242 1197660.25
0.0243076428572129 2410299.75
0.0238392500756294 2447643.75
0.0273020854183447 2319931.5
0.0326635585595695 2908804.25
0.0424336800354085 1364901.5
0.0332425872272534 819775.6875
0.0217666949129658 1431555.5
0.0212134597390581 924205.375
0.0298850795619941 1406903.5
0.0278240914732981 3668908.25
0.025117223501796 673677.9375
0.0297995788285001 1305853.5
0.0370729902963626 2736223.5
0.0241218921039095 1364555.75
0.0452058046103372 1011682.875
0.0250383798770863 1042506.75
0.0227385815950048 2288474
0.0392699195422144 1231853
0.0241396795109998 1175180.125
0.0280529708338105 2694314.25
0.0368060175488641 1149287.125
0.0284307872029342 2007985.5
0.0402676661719229 3199437.75
0.0385437370386167 2262063.75
0.0368199075366995 744958.8125
0.0412345566659551 2099246.25
0.0405090738058525 1327705.75
0.0324166833267889 1272833.125
0.0308706624294841 1919407.375
0.0311706938234881 2137026
0.0378300778516918 649279.5625
0.028289947343966 2042996
0.0308430129888187 2326304.5
0.0360863484145798 2038974.625
0.0223492914355186 151092.109375
0.0284483120365114 1122293.125
0.0462378858971187 1788995.125
0.0371642139909929 1775286.5
0.035288748053439 858147.1875
0.0286273046998917 881283.875
0.0355919114427826 725364.5
0.0276965499227147 360398.625
0.0413222136728524 1153308.5
0.0381475803343246 1889151.125
0.0346682274868035 1836024
0.0364442000333885 2595811.25
0.0341327734325715 1171700.25
0.04330042676533 1834952.5
0.0274793756290511 1893166.125
0.0359171092074486 3373078.5
0.0372267788338402 1598327
};
\addplot [semithick, dashed, black]
table {%
0.0212134597390581 1568111.97665769
0.0462378858971187 1838016.54172011
};
\end{axis}

\end{tikzpicture}}
\caption{Dependency of absolute flow in small vessels $\lsmall$ of the full model on volume fraction of small vessels in REVs for one representative case per tumor topology where $\SI{10}{\percent}$ of 1D blood vessels have been retained in hybrid model (dashed lines indicate linear least squares fits with corresponding $R^2$-values)}
\label{fig:flux_over_vf}
\end{figure}
Instead we tried to enhance the model by taking information of volume fractions of the smaller vessels into account. Our rationale behind this approach is that while the complete structure of the smaller vessels cannot be obtained non-invasively, regions with higher or smaller microvascular density of small vessels could still be identified. This information could then be employed to enrich the hybrid model. The overall trend we observed in Table~\ref{tab:results} is that the higher the volume fraction of the homogenized vessels, the larger their permeability. It also reasonable to assume that areas with a higher vascular volume fraction are more permeable to blood flow. Therefore, we investigated the relationship of the volume fraction of smaller vessels $\porosity^v_{\lsmall}$ in each REV on the perfusion of the smaller blood vessels in the full model. Results are shown in Figure~\ref{fig:flux_over_vf}. Here, the absolute volumetric flow in each coordinate direction (calculated as in~\eqref{eq:volflow_full} but not taking the flow direction into account) is plotted over the volume fraction of the smaller vessels $\porosity^v_{\lsmall}$. The clearest picture emerges for the LS174T topology with a good correlation of flow in smaller vessels with their volume fraction. A similar, yet less distinctive trend is present for the SW1222 case whereas no relationship can be observed for the GL261 tumor.  

Therefore, inside each REV $j$ we defined the isotropic permeability tensor as
\begin{equation}
\label{eq:perm_per_rev}
\frac{k^v}{\mu^v}\left(j\right)\cdot \tns{I}=\alpha\cdot\porosity^v_{\lsmall}\left(j\right)\cdot\tns{I},
\end{equation}
that is, a simple linear dependency of the permeability in the $j$-th REV on the volume fraction of smaller vessels in the $j$-th REV with proportionality constant $\alpha$. We also tested a nonlinear Kozeny-Karman law, but obtained slightly better results with the linear fit. Thus, we will exclusively study this linear dependency hereafter. Next, the optimization of the nonlinear least-squares problem~\eqref{eq:optimization} is performed for the proportionality constant $\alpha$. Results are shown in Table~\ref{tab:results_with_vf} for the case $\SI{10}{\percent}$, which can be compared with the cases with constant permeability over the entire domain from Table~\ref{tab:results}. We obtained a slightly better agreement in terms of flow in the smaller vessels $R^2_{{\mathrm{flow}},\lsmall}$ and, thus, also for the total coefficient of determination $R^2_{\mathrm{tot}}$ for the SW1222 and GL261 case. Compared to that, the correspondence of flow in the smaller vessels was markedly better than the constant permeability case for the LS174T topology. This is coherent with Figure~\ref{fig:flux_over_vf} where the latter network showed the most evident correlation of blood flow on volume fraction. Thus, one could expect that no significant improvement was possible for the GL261 case where volume fraction and flow seem to be decoupled. However, also for the SW1222 topology, which showed at least a moderate dependency of blood flow on volume fraction, the agreement could not be increased significantly. Therefore, at least for one of our cases it was beneficial to include blood vessel volume fraction information into the hybrid model while it was not detrimental for the other two. 

It also is conceivable that at least preferential directions of smaller vessels or their tortuosity can be detected non-invasively even though their complete structure cannot be resolved. A further enhancement of the model could be achieved when taking this information about the anisotropy of smaller vessels or their tortuosity into account during the homogenization procedure.\cite{Penta2015,Penta2015a,Mascheroni2017b} However, we want to emphasize that our whole study is based on numerical results. Experimental findings indicate no dependency between blood vessel diameter and flow in tumors\cite{Dewhirst2017,Dewhirst1989,Leunig1992} and a high vascular density does not automatically imply efficient perfusion, nutrient supply and drug delivery for solid tumors.\cite{Jain2005} These properties could make it impossible to deduce permeabilities of blood vessels inside tumors from macroscopic quantities such as blood vessel volume fractions or preferential directions. By contrast, non-invasive measurements of perfusion\cite{Desposito2018,Thomas2000} could prove helpful to enhance the hybrid model.

\begin{table}
	\centering
	\begin{tabular}{l l l l l l l l l l} 
		\hline
		Network & $\alpha\,[\si{\square\micro\m}]$ & $R^2_{\mathrm{L}}$ & $R^2_{\mathrm{S}}$ & $R^2_{\mathrm{IF}}$ & $R^2_{{\mathrm{flow}},\lsmall}$ & $R^2_{\mathrm{tot}}$  \\
		\hline\hline
		SW1222 & $\num{37.0 \pm 5.9}$ & 0.988 & 0.653 & 0.998 & 0.189 & 0.707 \\
		 \hline
		LS174T & $\num{129.7 \pm 10.6}$ & 0.928 & 0.691 & 0.991 & 0.362 & 0.743 \\
	    \hline
		GL261 & $\num{26.1 \pm 3.5}$ & 0.929 & 0.245 & 0.996 & 0.117 & 0.572
	\end{tabular}
	\caption{Results of the optimization procedure for non-constant permeability depending on volume fraction of smaller vessels. $\SI{10}{\percent}$ of 1D blood vessels have been retained in hybrid model for each tumor topology. Shown are the proportionality constant $\alpha$ relating permeability and blood vessel volume fraction of smaller vessels inside each REV according to~\eqref{eq:perm_per_rev}. $R^2$-values for agreement between both variants in terms of blood pressure in large vessels, blood pressure in small vessels, IF pressure and flow in small vessels is additionally provided. Overall coefficient of determination $R^2_{\mathrm{tot}}$ between fully-resolved and hybrid model is calculated according to~\eqref{eq:r_squared_tot}. (All data includes mean taken over five different sets of pressure boundary conditions on the 1D network per case)}
\label{tab:results_with_vf}
\end{table}
Similarly, improvements are possible for flow from the larger into the smaller vessels. In this contribution, we have assumed equal pressures in resolved and homogenized vasculature and, thereby, infinite (or at least a very large) permeability governing the flow between the two compartments such that a constraint of equal pressures holds along the resolved 1D vasculature. This has the major advantage that the coupling between resolved and homogenized vasculature is essentially parameter-free. Only the penalty parameter has to be chosen large enough such that a sufficient accuracy in the pressure constraint is achieved as described in Remark~\ref{rem:choice_penalty}. The GL261 and LS174T case had a less dispersed distribution of the radius in connecting elements and, thus, of the permeability between larger and smaller vessels. For these topologies, our approach could estimate the mass transfer between larger 1D vessels and smaller homogenized vessels more accurately. The SW1222 case had a much higher variability of radius and flow between $\lbig$ and $\lsmall$. In this case, it could be advantageous to employ finite permeabilities to model the mass transfer and assign higher permeabilities to REVs or regions along the larger vessels where many branches go away from the main vessels. However, this would require additional parameterization of the model as well as additional data on regions where a lot of flow from larger into smaller vessels can be expected.

In addition, we have so far only employed very simple algorithms to optimize the parameters of the hybrid model. A much more powerful framework for coarse-graining physical models has been developed by Grigo et al.\cite{Grigo2019} and tested for flow through porous media. This could also be applied in our case to infer the optimal parameters of the hybrid model per REV. However, this would require much more microstructural features, such as tortuosity, blood vessel distances or radius data on the smaller homogenized blood vessels to calibrate the hybrid model. Again, it is questionable if this data can be acquired non-invasively and if these parameters are determining blood flow through tumors.
\section{Conclusion}
\label{sec:concl}
In this work, we have studied a hybrid embedded/homogenized model for computational modeling of solid tumor perfusion. Its guiding principle is that the complete morphology of vascular networks including blood vessel diameters and topology cannot be acquired with currently available \textit{in-vivo} imaging techniques. Thus, fully-resolved discrete models relying on this data cannot be applied in "real world" scenarios. If, however, the structure of larger vessels constituting the main branches of the vasculature is available, our hybrid approach where only these larger branches are completely resolved is a sensible alternative. By contrast, the smaller scale vessels are homogenized such that their exact structure is not required anymore. This results in a two-compartment or double porosity formulation where the larger vessels are still represented as one-dimensional embedded inclusions. The coupling between the resolved and homogenized part of the vasculature is realized via a line-based pressure constraint along the 1D larger vessels, which we enforce with a mortar-type formulation with penalty regularization. This also has the advantage that compared to previous hybrid models no additional parameter -- apart from the penalty parameter which has to be chosen large enough -- is required to couple the two distinct representations of the vasculature. 

The results of the hybrid model have been compared with reference solutions generated by a fully-resolved 1D-3D model. For that, three different network topologies extracted from three different tumor types grown in mice have been employed. These topologies consist of up to $\num{420000}$ vessel segments and have dimensions of up to $\SI{6}{\mm}\times\SI{8}{\mm}\times\SI{11}{\mm}$. To date, this is the largest and most challenging test case for a hybrid model, especially considering the abnormal and tortuous structure of the networks typical for the vasculature inside tumors. We have further shown how we artificially generate the hybrid from the fully-resolved model, define representative elementary volumes and assign boundary conditions. We are confident that the artificially created topologies of larger vessels are representative of real \textit{in-vivo} imaging data sets of larger vessels inside tumors such that they enable us to draw meaningful conclusions for more realistic scenarios where the full topology is not available such that a hybrid approach is the only option. 

For comparison of the results of the two models, we have defined several rigorous metrics involving the blood pressure in both resolved and homogenized vasculature, the pressure in the interstitial fluid and blood flow in the homogenized vasculature. These metrics have then been employed to obtain the optimal parameters for the hybrid model and to study its accuracy w.r.t.\ the fully-resolved one. We have obtained very good agreement in terms of blood pressure in the larger vessels and IF pressure. Larger deviations are present for blood pressure and flow in the homogenized vasculature. However, these limitations can be expected since the information on the smaller vessels is not retained in the hybrid model. Overall, the best correspondence has been achieved for the SW1222 case which also had the clearest vascular structure and distinction between larger and smaller vessels. All topologies showed a very good agreement in terms of REV IF pressure and REV blood pressure in smaller vessels with mean deviations in a range of $\SIrange[range-phrase = -,range-units=single]{20}{70}{\Pa}$ and $\SIrange[range-phrase = -,range-units=single]{40}{110}{\Pa}$ resp.\ $\SIrange[range-phrase = -,range-units=single]{0.7}{3.8}{\percent}$ and $\SIrange[range-phrase = -,range-units=single]{1.1}{2.9}{\percent}$. It is sufficient to resolve $\SIrange[range-phrase = -,range-units=single]{5}{10}{\percent}$ of all blood vessels segments by keeping them in the hybrid model since there is only a marginal improvement of the agreement with the fully-resolved model in terms of all investigated metrics when retaining a higher percentage $(\SIrange[range-phrase = -,range-units=single]{15}{20}{\percent})$ of blood vessels. Concerning the flow between smaller and larger vessels the error was mainly caused by the large variability of diameter and flow in the connectivity elements between large and small vessels for the SW1222 case. Possibly, this error could be reduced by allowing a varying permeability for coupling the two compartments. By including information about the blood vessel volume fraction of smaller blood vessels into the definition of their permeability tensor a better agreement with flow therein could be achieved for the LS174T case. Nevertheless, the abnormal vascular structure and blood flow patterns of tumor vasculature could impede this approach. 

Several other potential improvements have been discussed and remain subject to future work. Furthermore, the inclusion of species transport to simulate drug delivery or nutrient transport lies at hand. Species transport including the coupling between resolved and homogenized vasculature is possible within our hybrid multiphase tumor growth model\cite{Kremheller2019} and we have already studied nanoparticle delivery to solid tumors employing the homogenized compartment only.\cite{Wirthl2020} These models could ultimately enhance our understanding of the limitations of current drug delivery strategies and aid in devising more targeted therapies.

The next step towards a more realistic or even clinical usage of hybrid computational models for tissue perfusion is to devise a strategy which combines data which is available non-invasively.\cite{Li2020a} Faced with such a scenario, where a hybrid model is the only applicable option since the entire network topology is not known, the methods and metrics developed here could be applied in the following way:
\begin{enumerate}
\item Gather all physiological data, which can be accessed via \textit{in-vivo} measurements for the specific case. For instance, this could be tissue perfusion, hypoxic areas, REV or point-wise measurements of pressure or flow, volume fractions of homogenized blood vessels or their preferential direction and the transport of tracer molecules through the domain. 
\item Define the computational domain of interest as the embedded larger vessels and a surrounding domain of homogenized vasculature following the extent of the tumor. If available, include the information about volume fractions and preferential directions of smaller blood vessels in the definition of the permeability tensor.
\item Formulate an optimization problem similar to~\eqref{eq:optimization} to match the available information about transport, e.g., REV or point-wise flow and pressure data. However, not only the parameters of the homogenized vasculature would be part of the optimization as in this contribution but also the large majority of the (homogenized or resolved) pressure boundary conditions, which are additionally unknown. Note however, that far less boundary conditions compared to a fully-resolved setting need to be applied.
\item Employ the obtained flow state for \textit{in-silico} studies of drug delivery or the optimization of treatment strategies.
\end{enumerate}
\section*{Acknowledgments}
\label{sec:acknow}

The authors gratefully acknowledge the support of the Technical University of Munich -- Institute for Advanced Study, funded by the German Excellence Initiative and the T\"UV S\"UD Foundation. Research reported in this publication was supported by the National Cancer Institute of the National Institutes of Health under Award Number U54CA210181. The content is solely the responsibility of the authors and does not necessarily represent the official views of the National Institutes of Health. The software QUEENS was provided by the courtesy of AdCo Engineering\textsuperscript{GW} GmbH, which is gratefully acknowledged. We would also like to thank the authors of REANIMATE\cite{Desposito2018,Sweeney2019}, especially Paul W. Sweeney for sharing their data and making their code publicly available. In addition, we gratefully thank Barbara Wirthl and Anh-Tu Vuong for discussions about various aspects of the two models.

\appendix
\section{Segment-based line integration scheme for the evaluation of 1D-3D coupling terms}
\label{sec:app1}
\begin{figure}
\def\svgwidth{0.9\textwidth}
\centering
{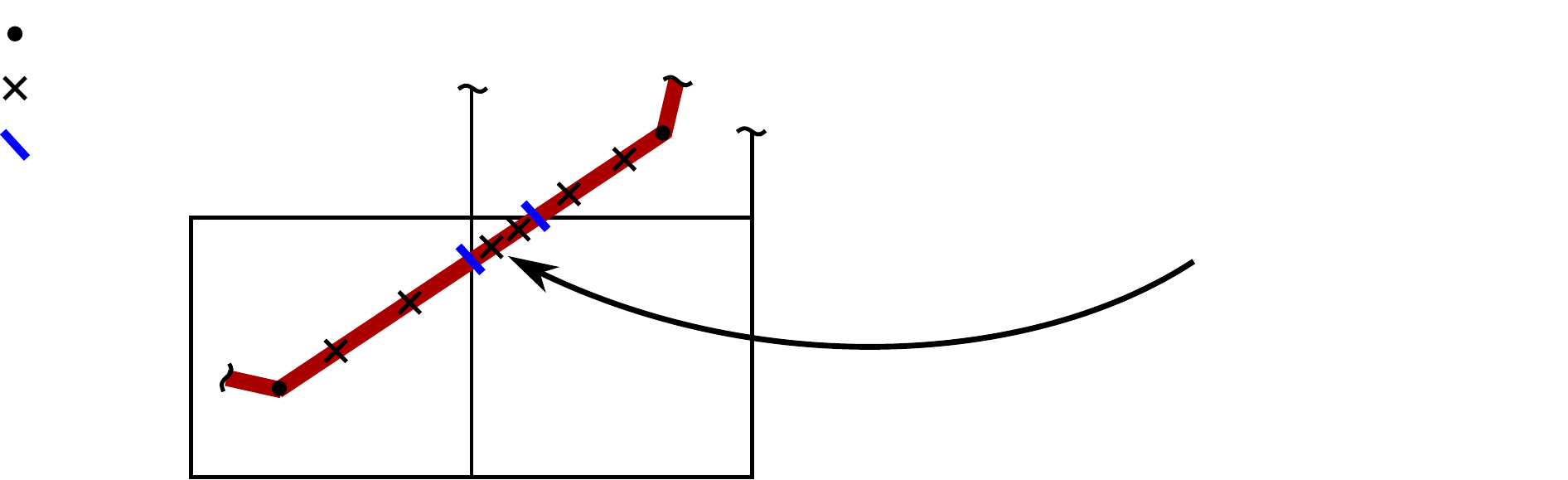}
\caption{Sketch of one-dimensional segment-based integration (in 2D) -- exemplarily two Gauss points per segment are defined in the one-dimensional parameter space and mapped onto the centerline where integration is performed. The 1D element here interacts with three 2D elements which defines three integration segments.}
\label{fig:segment_based_sketch}
\end{figure}
The numerical integration of the 1D-3D coupling terms is outlined in this appendix. These terms are the leakage terms from the 1D embedded vasculature into the IF occurring in the weak forms~\eqref{eq:full_weak_1D} and \eqref{eq:full_weak_3D} respectively~\eqref{eq:homo_weak_1D} and \eqref{eq:homo_weak_IF_3D} and the mortar coupling matrices $\bsd{D}$ and $\bsd{M}$ from~\eqref{eq:D} and~\eqref{eq:M}. After spatial discretization with appropriate shape functions, these terms comprise a line integral along the inclusion containing the product of shape functions either defined on the 1D or on the 3D domain.

A one-dimensional segment-based integration for these types of terms has been proposed in our previous publications.\cite{Kremheller2019,Steinbrecher2020} The goal of this procedure is to perform the integration in a general non-conforming scenario as sketched in Figure~\ref{fig:segment_based_sketch}. Numerically, the 1D integrals are evaluated with Gauss quadrature. However, at first each 1D element is segmented by finding its 2D or 3D interaction partners, i.e., those elements of the discretization of the surrounding domain it intersects. This yields 1D pieces interacting with a single 2D/3D element on which also the shape functions of the respective 2D/3D element are continuous. Then, GPs are defined in the single segments and mapped from the element parameter space on the 1D centerline. In addition, the spatial coordinate  of the respective GP $\bs{X}\left( S\left(\xi\right) \right)$, is required to obtain the shape function values of the respective 2D/3D element at this position. The integral for a specific 1D element then emerges as the sum over the integrals of all its segments and the respective contributions are assembled into the DOFs of the 1D and the interacting 2D/3D elements.

\end{document}